\documentclass[]{aa}
\usepackage{natbib}
\bibpunct{(}{)}{;}{a}{}{,}
\usepackage{graphicx}
\usepackage{txfonts}
\usepackage{subfigure}
\usepackage{longtable}
\usepackage{supertabular}
\begin{document}
   \title{Dust tori in radio galaxies}

   \subtitle{}

   \author{G.~van der Wolk
          \inst{1}
          \and
          P.D.~Barthel\inst{1}
	  \and
	  R.F.~Peletier\inst{1}
	  \and
	  J.W.~Pel\inst{1}
          }

   \offprints{G. van der Wolk, \email{G.van.der.Wolk@astro.rug.nl}}

   \institute{Kapteyn Astronomical Institute, University of Groningen, P.O.~Box~800, 9700AV Groningen, The Netherlands.
                       }

\date{Received 6 May 2009; accepted 20 November 2009}

\abstract
{}
{We investigate the quasar -- radio galaxy unification scenario and detect dust tori within radio galaxies of various types.} 
{Using VISIR on the VLT, we acquired sub-arcsecond ($\sim0.40\arcsec$) resolution N-band images, at a wavelength of 11.85~$\mu$m, of the nuclei of a sample of 27 radio galaxies of four types in the redshift range $z=0.006-0.156$. The sample consists of 8 edge-darkened, low-power Fanaroff-Riley class~I (FR-I) radio galaxies, 6 edge-brightened, class II (FR-II) radio galaxies displaying low-excitation optical emission, 7 FR-IIs displaying high-excitation optical emission, and 6 FR-II broad emission line radio galaxies. Out of the sample of 27 objects, 10 nuclei are detected and several have constraining non-detections at sensitivities of 7~mJy, the limiting flux a point source has when detected with a signal-to-noise ratio of 10 in one hour of source integration.}
{On the basis of the core spectral energy distributions of this sample we find clear indications that many FR-I and several low-excitation FR-II radio galaxies do not contain warm dust tori. At least $57\pm19$ percent of the high-excitation FR-IIs and almost all of the broad line radio galaxies exhibit excess infrared emission, which must be attributed to warm dust reradiating accretion activity. The FR-I and low-excitation FR-II galaxies are all of low efficiency, which is calculated as the ratio of bolometric to Eddington luminosity $L_{\mathrm{bol}}/L_{\mathrm{Edd}}<10^{-3}$. This suggests that thick tori are absent at low accretion rates and/or low efficiencies. The high-excitation FR-II galaxies are a mixed population with three types of nuclei: 1) low efficiency with dust torus, 2) low efficiency with weakly emitting dust torus, and 3) high efficiency with weak dust torus. We argue that the unification viewing angle range 0-45~degrees of quasars should be increased to $\sim$60~degrees, at least at lower luminosities.}
{}


\titlerunning{Dust tori in radio galaxies}
\authorrunning{G. van der Wolk et al.}

\maketitle
%

\section{Introduction}
Powerful radio galaxies are understood to host quasar-like nuclei that are obscured by circumnuclear dust tori perpendicular to the radio jets \citep{1989ApJ...336..606B, 1995PASP..107..803U}. If this unification scenario is correct, infrared observations of the nuclei of radio galaxies may act as calorimeters that indicate the levels of accretion activity. The dust tori, believed to be clumpy and to extend out to radii of tens to hundreds of parsecs \citep{1988ApJ...329..702K, 2002ApJ...570L...9N,1997ApJ...486..147G, 2006ApJ...653..127S}, reradiate the optical and ultraviolet emission originating from the accretion disks surrounding the supermassive black holes ($>~10^6~M_{\odot}$). However, only radio galaxies displaying high-excitation narrow-line optical emission, referred to as high-excitation radio galaxies (HEGs), are understood to host obscured broad line and bright continuum emission, and to be in the process of active black hole accretion \citep{1994ASPC...54..175B, 1994ASPC...54..201L}. Certain radio galaxies exhibit only low-excitation narrow-line emission \citep{1979MNRAS.188..111H}, referred to as low-excitation radio galaxies (LEGs). Accretion in these objects may be, temporarily or not, at a low level, the line emission being to some part due to hot star ionization. Infrared observations can help to confirm whether these radio galaxies indeed currently host inactive non-thermal nuclei.

A number of mid- and far-infrared surveys of radio galaxies have detected hidden quasars among powerful radio galaxies. These suggest that the scale-height of the torus increases, that is to say the inner radius increases but the outer radius does not change, for more luminous radio galaxies \citep{2004A&A...424..531H}. Starbursts are also infrequently found to be co-heating the dust \citep{2007ApJ...661L..13T}. On average, quasars are more powerful than radio galaxies in the mid-infrared because of the beamed emission of the former \citep{2007ApJ...660..117C}. Infrared surveys have furthermore identified a separate population among radio galaxies of Fanaroff-Riley type II (FR-II), which are mid-infrared weak and probably do not contain hidden quasars \citep{2001A&A...372..719M, 2006ApJ...647..161O}. These objects are most common at lower radio luminosities and among galaxies that exhibit low-excitation optical emission features. They are predominantly found at redshifts $z<0.3$, and have weaker jets. They may correspond to the excess number of radio galaxies relative to quasars at $z<0.5$ and explain their distributions of size \citep{1993MNRAS.262L..27S}. It is speculated that they form part of the parent population of variable active nuclei with a beamed component close to the line of sight \citep{1994ASPC...54..227L}.

This division between radio galaxy populations, according to the radiative properties of the active nucleus, is not linked one-to-one with the well-known radio-power dependent dichotomy in radio morphology \citep{1974MNRAS.167P..31F}. Low radio luminosity sources are edge-darkened and have radio jets that quickly decelerate and flare (Fanaroff-Riley type I sources, referred to as FR-Is). High luminosity sources are edge-brightened, with jets that remain relativistic over tens to hundreds of kpc, ending in bright hotspots (FR-IIs). The dividing line occurs around a luminosity $\log{\nu L_{178\mathrm{MHz}}} \sim 40.9~\mathrm{erg}~\mathrm{s}^{-1}$. Almost all FR-Is exhibit only low-excitation optical emission lines. FR-IIs have three kinds of excitation features: low-excitation or high-excitation narrow lines, and broad lines. LEGs have dominating low-excitation and/or weak/absent high-excitation emission. They are identified on the basis of one or more of the following criteria: emission-line ratios $[\ion{O}{III}]/\mathrm{H}\beta<5$ and $[\ion{O}{III}]/[\ion{O}{II}]<1$, equivalent widths $EW([\ion{O}{III}])<10~\AA$ or strengths $\log{\nu L_{[\ion{O}{III}]}}~<~40.7~\mathrm{erg}~\mathrm{s}^{-1}$ including for stellar absorption lines or the $4000~\AA$-break, while for HEGs these emission-line ratios, equivalent widths, and strengths are higher \citep{1992ApJ...389..208B, 1997MNRAS.286..241J, 1997MNRAS.284..541M}. Broad-line radio galaxies (BLRGs) are identified by having at least one broad emission-line ($v_{\mathrm{FWHM}}>2000~\mathrm{km}~\mathrm{s}^{-1}$) \citep{1997MNRAS.284..541M}. Since FR-IIs produce about ten to fifty times as much emission-line luminosity as FR-Is for the same radio core power, the engines of FR-Is and FR-IIs differ \citep{1995ApJ...448..521Z}. It could be that FR-I sources are produced when the central engine is fed by matter at a lower accretion rate, leading to the creation of a source in which the ratio of radiating to jet bulk kinetic energy is low, while FR-II sources are produced when the central engine is fed at a higher accretion rate, causing the central engine to deposit a higher fraction of its energy in radiating energy \citep{1995ApJ...451...88B}. Alternatively, \cite{2006ApJ...648L.101E} claim that the torus is produced by an outflow of material from the centre. For a low-luminosity, low-level activity nucleus, this outflow will transform into a radio jet, similar to what is believed to happen in X-ray binaries. This could explain why for the same emission-line activity the FR-Is produce more radio core power than the FR-IIs.

\begin{figure}
\centering
\includegraphics[width=8.8cm]{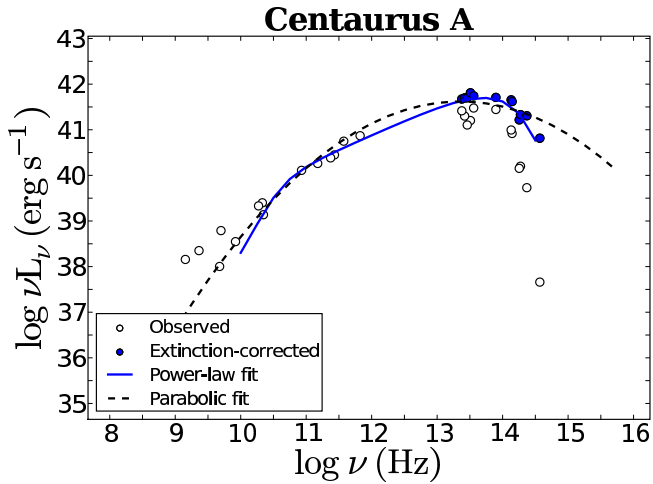}
\caption{The core spectral energy distribution of Centaurus~A, fitted by i) a straight power-law $\nu L_{\nu} \propto \nu^{0.64}$, which cuts off exponentially above $\nu_{\mathrm{c}} = 8 \times 10^{13}$~Hz and becomes optically thick below $\nu_{\mathrm{t}} = 5 \times 10^{10}$~Hz and ii) a parabolic function. The luminosities calculated from observed flux values are indicated with open circles. Filled circles indicate infrared to optical extinction-corrected values assuming a foreground extinction of $A_{\mathrm{V}}=$14~mag of a circumnuclear dust disk. See text for details.}
\label{censed}
\end{figure}

The thermal and/or non-thermal nature of the cores of low and high power radio galaxies, and their place in unification scenarios can be addressed using multiwavelength studies \citep{2000A&A...362...75P, 2003AJ....126.2677Q}. An infrared excess above a standard synchrotron spectrum fitted to high frequency radio core data is indicative of hot circumnuclear dust. On the other hand the absence of this excess or short-period flux variablity, is indicative of a purely non-thermal nucleus. On the basis of Hubble Space Telescope (HST) optical and ultraviolet data of the cores of FR-Is, \cite{1999A&A...349...77C, 2002ApJ...571..247C} and \cite{2002AJ....123.1334V} show that these nuclei follow a linear correlation in the radio-optical and radio-ultraviolet luminosity plane. This supports the hypothesis that the radio, optical, and ultraviolet emission have a common synchrotron origin. Within this picture, FR-I nuclei are not covered by thick dust tori and have spectral energy distributions (SEDs) that are dominated by non-thermal emission. This is confirmed for M87 \citep{2004ApJ...602..116W, 2007ApJ...663..808P} and for similar nuclei with low-ionization nuclear emission regions (LINERs) \citep{2005ApJ...625..699M}. At a one-parsec-level, the FR-I radio galaxy Centaurus~A is shown to be dominated by a synchrotron source, as inferred from mid-infrared interferometric observations with the VLT \citep{2007A&A...471..453M}. The synchrotron nucleus is argued to be surrounded by a small circumnuclear dust disk, which causes a visual foreground extinction of $A_V = 14$~mag. The overall, dereddened core spectrum of Cen~A is characterized by a $\nu L_{\nu} \propto \nu^{0.64}$ power-law that cuts off exponentially towards high frequencies at $\nu_{\mathrm{c}} = 8 \times 10^{13}$~Hz and becomes optically thick at $\nu_{\mathrm{t}} < 50$~GHz. In Fig.~\ref{censed}, the observed and extinction-corrected core luminosities, calculated from flux values listed in Table~\ref{coresed}, are plotted together with the synchrotron spectrum fit, expressed by 
\begin{equation}
\centering
\nu L_{\nu} \propto \nu \left(\frac{\nu}{\nu_{\mathrm{t}}}\right)^{\alpha_{1}} \left[ 1 - \exp{-(\frac{\nu_{\mathrm{t}}}{\nu})^{\alpha_{1}-\alpha_{2}}} \right] \exp{-\left(\frac{\nu}{\nu_{\mathrm{c}}}\right)} \,,
\label{ot}
\end{equation} 
where $\alpha_1=2.5$ and $\alpha_2=0.64$ are the optically thick and thin spectral indices \citep{2000A&A...362...75P}. This spectrum can also be fitted by a parabolic function, which is found to be a good approximation \citep{1986ApJ...308...78L,2002A&A...381..389A,2009ApJ...701..891L}. Both type of fits and the nuclear mid-infrared luminosity of $\log{\nu L_{\nu}}=41.8~\mathrm{erg}~\mathrm{s}^{-1}$ found by \cite{2004ApJ...602..116W} do not imply a large amount of thermal emission. Although spectral energy distributions seem to indicate that thick dust tori are absent in FR-I radio galaxy nuclei, dust disks, kiloparsecs in size, surrounding FR-I nuclei have been seen in HST images \citep{1999ApJS..122...81M}, and some contain high molecular gas masses as deduced from high resolution CO maps \citep{2005ApJ...620..673O,2005ApJ...629..757D}.

The multiwavelength properties of FR-II galaxy cores are more diverse than those of FR-Is. Broad-line radio galaxies (BLRGs) of FR-class II are very bright in the optical: their hosts are often of the N-type. Their nuclei show a large optical excess with respect to the radio-optical correlation found for FR-I galaxies. This excess could well represent unobscured thermal emission from an accretion disk. Broad lines are seen in objects spanning many orders of optical magnitude, from LINERs to quasars. The efficiency of the mass accretion onto the disks varies enormously \citep{2002A&A...394..791C}. The far-infrared emission of BLRGs, which is much brighter than that of LEGs and HEGs, indicates that they have a special character \citep{1995A&A...303....8H}. Their place in unification schemes is still unclear. While some may be considered to be weak quasars, others are viewed at grazing incidence to the torus and the jets are directed farther away from us than in the case of quasars \citep{2000A&A...364..501D}. The warm and cool dust emission indicate that BLRGs have possibly undergone a different merger evolution or have a relatively old active galactic nucleus \citep{2001A&A...379L..21V}.

Some HEGs show evidence of harboring a hidden quasar nucleus from optical excess in the radio-optical with respect to the radio-optical correlation found for FR-I galaxies. Others do not show this excess, but sometimes spectropolarimetric studies indicate a hidden quasar. Another group of HEGs show, similar to low-excitation radio galaxies (LEG) of type II, true unobscured FR-I-like nuclei \citep{2002A&A...394..791C}. It seems that LEGs and certain HEGs together constitute a distinct population, characterized by a low accretion rate and/or efficiency, weak or absent broad line emission, and lack of a significant nuclear absorbing structure. These LEGs and HEGs have probably undergone slower cosmic evolution \citep{2000MNRAS.316..449W} than the HEGs with hidden quasars and BLRGs. Alternatively, it has been argued that the torus covering fraction may increase with decreasing radio luminosity \citep{1991MNRAS.252..586L}.

To shed light on these issues, we obtained N-band (12~$\mu$m) VISIR observations of a sample of 8~FR-I, 6 low- and 7 high-excitation FR-II galaxies, and 6 broad-line radio galaxies. We address the thermal or non-thermal nature of their nuclei and their place in the unification scheme. The sample is described in Sect.~2 and the observations in Sect.~3. Core spectral energy distributions and other results are presented in Sect.~4. Discussion and conclusions are presented in Sect.~5 and 6. Throughout this paper, we assume a $\Lambda$ cold dark matter ($\Lambda$CDM) cosmology with a Hubble constant $H_0~=~74~\mathrm{km~s}^{-1}~\mathrm{Mpc}^{-1}$ and a critical matter and cosmological constant density parameter of $\Omega_{m}=0.24$ and $\Omega_{\Lambda}=0.76$ \citep{2007ApJS..170..377S}. All luminosities are calculated in the emitted frame, using
\begin{equation}
\centering
\nu L_{\nu} = \nu \frac{4\pi d^2 F_{\nu}}{1+z}~\mbox{erg~s}^{-1} \,,
\end{equation}
where $\nu$ is the observed frequency, $d$ is the distance calculated for this cosmology and redshift, $z$, and $F_{\nu}$ is the observed flux.

\begin{table*}
\caption{Basic parameters of the observed sample.}
\label{table1}
\centering
\begin{tabular}{lllccrccrrrcl}
\hline \hline
Source & $z$  & Class & Ref. & $\log{\nu L^{\mathrm{total}}_{\mathrm{408MHz}}}$ & $\log{R}$ & Freq. & Ref. & $\frac{[\ion{O}{III}]}{[\ion{O}{II}]}$ & $\frac{[\ion{O}{III}]}{[\mathrm{H}\beta]}$ & $\log{\nu L_{[\ion{O}{III}]}}$ & Ref. & VISIR \\
&&&&&&&&&&&& Resolution \\
&&&&(erg s$^{-1}$)&&(GHz)&&&&(erg s$^{-1}$)&&(parsec) \\
(1)&(2)&(3)&(4)&(5)&(6)&(7)&(8)&(9)&(10)&(11)&(12)&(13) \\
\hline
4C12.03      & 0.15600 & LEG    & 5     & 41.98 & $ $ $-$2.66 &   1.5 &  5      &  $\dots$ &  $\dots$ & $ $ 40.9 & 7      & 1400 \\
3C015        & 0.07300 & LEG    & 8     & 41.64 & $ $ $-$2.02 &   2.3 &  6      &  $\dots$ &  3.7     & $ $ 40.4 & 6      &  610 \\
3C029        & 0.04503 & FR-I   & 8     & 41.26 & $ $ $-$1.79 &   2.3 &  6      &  $\dots$ &  1.7     & $ $ 39.9 & 6      &  370 \\
3C040        & 0.01800 & FR-I   & 8     & 40.63 & $ $ $-$1.61 &   2.3 &  6      &  $\dots$ &  $\dots$ & $<$ 38.9 & 6      &  140 \\
3C078        & 0.02865 & FR-I   & 8     & 40.94 & $ $ $-$0.92 &   2.3 &  6      &  $\dots$ &  $\dots$ & $<$ 39.1 & 6      &  230 \\
Fornax~A     & 0.00587 & FR-I   & 8     & 40.72 & $ $ $-$3.40 &   2.3 &  6      &  $\dots$ &  $\dots$ & $<$ 38.6 & 6      &   40 \\
3C093        & 0.35700 & BLRG   & 6     & 43.02 & $ $ $-$2.38 &   5.0 &  2,7    &  3.9     &  $\dots$ & $\dots$  & 3      & 3500 \\
3C098        & 0.03045 & HEG    & 5     & 41.31 & $ $ $-$2.70 &   8.3 &  4      &  2.9     & 15       & $ $ 40.9 & 1      &  240 \\
3C105        & 0.08900 & HEG    & 8     & 41.80 & $ $ $-$2.27 &   2.3 &  6      &  $\dots$ & 16       & $ $ 40.8 & 6      &  750 \\
3C120        & 0.03301 & BLRG   & 8     & 40.73 & $ $    1.01 &   2.3 &  6      &  5.7     &  $\dots$ & $ $ 41.8 & 6      &  270 \\
3C135        & 0.12738 & HEG    & 5     & 42.05 & $ $ $-$2.71 &   8.3 &  3      &  $\dots$ &  $\dots$ & 42.0  & 9         & 1100 \\
Pictor~A     & 0.03506 & BLRG   & 8     & 42.22 & $ $ $-$1.57 &   2.3 &  6      &  1.8     &  4.2     & $ $ 41.2 & 6      &  280 \\
3C198        & 0.08147 & HEG    & 3     & 41.56 & $ $ $-$2.20 &   5.0 &  1,7    &  1.2     &  2.1     & $ $ 41.1 & 8      &  680 \\
3C227        & 0.08583 & BLRG   & 8     & 42.13 & $ $ $-$2.55 &   2.3 &  6      & 13       &  9.0     & $ $ 41.8 & 6      &  710 \\
3C270        & 0.00746 & FR-I   & 8     & 40.21 & $ $ $-$1.71 &   2.3 &  6      &  $\dots$ &  4.7     & $ $ 38.9 & 2,10   &   55 \\
Centaurus~A  & 0.00095 & FR-I   & 8     & 40.30 & $ $ $-$1.35 &   2.3 &  6      &  2.5     &  6.3     & $ $ 38.3 & 5,6    &    7 \\
PKS1417$-$19 & 0.12000 & BLRG   & 4     & 41.79 & $ $ $-$2.20 &   5.0 &  7,8    & 22       &  2.2     &  $\dots$ & 4      & 1000 \\
3C317        & 0.03446 & FR-I   & 8     & 41.38 & $ $ $-$0.79 &   2.3 &  6      &  0.37    &  3.1     & $ $ 40.0 & 6      &  280 \\
3C327        & 0.10480 & HEG    & 8     & 42.17 & $ $ $-$2.19 &   2.3 &  6      & 12       & 15       & $ $ 42.1 & 6      &  890 \\
3C353        & 0.03042 & LEG    & 8     & 42.01 & $ $ $-$2.64 &   2.3 &  6      &  $\dots$ &  0.92    & $ $ 39.3 & 6      &  240 \\
PKS1839$-$48 & 0.11120 & LEG    & 8     & 41.99 & $ $ $-$1.16 &   2.3 &  6      &  $\dots$ &  $\dots$ & $<$ 39.2 & 6      &  960 \\
3C403        & 0.05900 & HEG    & 8     & 41.59 & $ $ $-$2.18 &   2.3 &  6      &  8.4     & 20       & $ $ 41.6 & 6      &  490 \\
3C403.1      & 0.05540 & LEG    & 7     & 41.04 &  $\dots$    &$\dots$&  $\dots$& $\dots$  &  $\dots$ & $\dots$  & $\dots$&  450 \\
3C424        & 0.12699 & LEG    & 5     & 42.03 & $ $ $-$1.70 &   8.3 &  3      &  $\dots$ &  $\dots$ & $ $ 40.7 & 9      & 1100 \\
PKS2158$-$380& 0.03330 & HEG    & 2     & 40.56 & $<$ $-$1.92 &   5.0 &  7,9    & $\dots$  &  $\dots$ &  $\dots$ & $\dots$&  270 \\
3C445        & 0.05620 & BLRG   & 8     & 41.66 & $ $ $-$1.57 &   2.3 &  8      & 14       & 13       & $ $ 42.0 & 6      &  460 \\
PKS2354$-$35 & 0.04905 & FR-I   & 1     & 41.24 & $ $ $-$1.16 &   5.0 &  7,10   & $\dots$  &  $\dots$ &  $\dots$ & $\dots$&  400 \\
\hline
\end{tabular}
\begin{list}{}{}
\item[]\textit{Column (2)} Redshifts are from the NED catalogue, except for Centaurus~A, which is calculated from the distance from \cite{2004A&A...413..903R}. \textit{Column (3)} Classifications: FR-I stands for Fanaroff-Riley class I radio galaxy, LEG and HEG for low- and high-excitation Fanaroff-Riley class II radio galaxy and BLRG for broad-line radio galaxy. \textit{Column (4)} References to classifications: 1=\cite{2006AJ....131..100B}, 2=\cite{1992ApJ...389..208B}, 3=\cite{1992MNRAS.256..186B}, 4=\cite{1978ApJ...220..783G}, 5=\cite{1998MNRAS.296..445H}, 6=\cite{1996A&A...313..423H}, 7=\cite{1997MNRAS.286..241J}, and 8=\cite{1997MNRAS.284..541M}. \textit{Column (5)} Integrated radio source luminosities, at rest 408~MHz from \cite{1996yCat.8015....0W}. \textit{Columns (6-8)} Radio core fraction, i.e. the ratio of the core to the extended radio flux at 1.5, 2.3, 5.0, or 8.0 GHz frequencies and references: 1=\cite{2003yCat.8071....0B}, 2=\cite{1994A&AS..105...91B}, 3=\cite{1998MNRAS.296..445H}, 4=\cite{1999MNRAS.304..135H}, 5=\cite{1991AJ....102..537L}, 6=\cite{1997MNRAS.284..541M}, 7=\cite{1996yCat.8015....0W}, 8=\cite{1995ApJ...448..521Z}, 9=\cite{1982MNRAS.201..991F}, and 10=\cite{1991ApJ...376..424S}.  \textit{Columns (9-12)} Optical emission-line ratios of $[\ion{O}{III}]$ to $[\ion{O}{II}]$, $[\ion{O}{III}]$ to H$\beta$, and optical $[\ion{O}{III}]$ emission-line luminosities calculated from flux values found in the literature. References to emission-line ratios and fluxes: 1=\cite{1977ApJ...211..675C}, 2=\cite{1996ApJ...470..444F}, 3=\cite{2003ApJ...599..886E}, 4=\cite{1978ApJ...220..783G}, 5=\cite{2007ApJ...670L..81H}, 6=\cite{1993MNRAS.263..999T}, 7=\cite{2007A&A...470..531W}, 8=\cite{1996MNRAS.281..509S}, 9=\cite{2009A&A...495.1033B}, and 10=\cite{1997ApJS..112..315H}. \textit{Column (13)} Scale of the VISIR mid-infrared diffraction-limit of $0.40\arcsec$ in parsecs.
\end{list}
\end{table*}

\begin{table*}
\caption{Observation parameters}
\label{tablevisir}
\centering
\begin{tabular}{llllrllrrrlll}
\hline \hline
PKS &  Alias & Date & Dist. &  $F_{11.85~\mu\mathrm{m}}^{\mathrm{core}}$  &  Lit. & Ref. &  $\log{\nu L_{11.85~\mu\mathrm{m}}^{\mathrm{core}}}$  &  $\log{\nu L_{5\mathrm{GHz}}^{\mathrm{core}}}$ &  $\log{\nu L_{12~\mu\mathrm{m}}^{\mathrm{total}}}$ & $R50$ & \multicolumn{2}{c}{FWHM} \\
&&&&&&&&&&& Obj. & St. \\
 & & & Mpc & (mJy) & (mJy) & & ($\mathrm{erg}~\mathrm{s}^{-1}$) & ($\mathrm{erg}~\mathrm{s}^{-1}$) & ($\mathrm{erg}~\mathrm{s}^{-1}$) & (pc) & ($\arcsec$) & ($\arcsec$) \\
(1)&(2)&(3)&(4)&(5)&(6)&(7)&(8)&(9)&(10)&(11)&(12)&(13)\\
\hline
0007+12        & \object{4C12.03}      & 2006-06-18   &   710 & $<$    3            &    $\dots$  & $\dots$     & $<$ 43.58      & 40.28   &    $\dots$ &     $\dots$ &  $\dots$ &  $\dots$ \\
0034$-$01      & \object{3C015}        & 2006-06-18   &   310 & $<$    3            &  6 & 4  & $<$ 42.93      & 41.21   &    $\dots$ &     $\dots$ &  $\dots$ &  $\dots$ \\
0055$-$01      & \object{3C029}        & 2006-12-27   &   190 & $<$    4            &  2 & 4  & $<$ 42.56      & 40.28   &    $\dots$ &     $\dots$ &  $\dots$ &  $\dots$ \\
0123$-$01      & \object{3C040}        & 2006-12-27   &    70 & $<$    4            &    $\dots$  & $\dots$     & $<$ 41.76      & 39.51   &$<$ 43.27   &     $\dots$ &  $\dots$ &  $\dots$ \\
0305+03        & \object{3C078}        & 2006-12-27   &   120 & $<$    4            &    $\dots$  & $\dots$     & $<$ 42.18      & 40.78   &$<$ 43.57   &     $\dots$ &  $\dots$ &  $\dots$ \\
0320$-$37      & \object{Fornax~A}     & 2006-12-27   &    20 & $<$    9            &    $\dots$  & $\dots$     & $<$ 41.03      & 37.81   &    42.44   &     $\dots$ &  $\dots$ &  $\dots$ \\
0340+04        & \object{3C093 }       & 2006-12-27   &  1800 & $<$    5            &    $\dots$  & $\dots$     & $<$ 44.56      & 40.75   &    $\dots$ &     $\dots$ &  $\dots$ &  $\dots$ \\
0356+10        & \object{3C098}        & 2005-12-18   &   130 &       20$\pm$1.4    &     24      & 2           & 42.93$\pm$0.04 & 38.91   &$<$ 42.95   &   180       &  0.47    &  0.37 \\
0404+03        & \object{3C105}        & 2005-12-18   &   390 & $<$    2            &    $\dots$  & $\dots$     & $<$ 42.93      & 40.06   &    $\dots$ &     $\dots$ &  $\dots$ &  $\dots$ \\
0430+05        & \object{3C120}        & 2006-12-27   &   140 &      240$\pm$15     &    109      & 1           & 44.10$\pm$0.03 & 41.58   &    44.20   &   170       &  0.36    &  0.33 \\
0511+00        & \object{3C135}        & 2006-12-27   &   570 & $<$    4            &    $\dots$  & $\dots$     & $<$ 43.51      & 39.89   &    $\dots$ &     $\dots$ &  $\dots$ &  $\dots$ \\
0518$-$45      & \object{Pictor~A}     & 2006-12-27   &   150 &       68$\pm$4.3    &    $\dots$  & $\dots$     & 43.59$\pm$0.04 & 41.09   &    43.82   &   200       &  0.34    &  0.33 \\
0819+06        & \object{3C198 }       & 2006-12-27   &   350 & $<$    4            &    $\dots$  & $\dots$     & $<$ 43.13      & 39.13   &    $\dots$ &     $\dots$ &  $\dots$ &  $\dots$ \\
0945+07        & \object{3C227}        & 2006-12-27   &   370 &       29$\pm$3.1    &    $\dots$  & $\dots$     & 44.01$\pm$0.05 & 40.38   & $<$44.96   &   410       &  0.36    &  0.33 \\
1216+06        & \object{3C270}        & 2006-06-18   &    30 & $<$    2            &       27    & 4           & $<$ 40.72      & 39.17   & $<$42.44   &     $\dots$ &  $\dots$ &  $\dots$ \\
1322$-$42      & \object{Centaurus~A}  & 2006-12-27   &   3.8 &     1200$\pm$47     &    1800     & 3           & 41.69$\pm$0.03 & 38.79   &    42.69   &     5       &  0.37    &  0.33 \\
1417$-$19      & \object{PKS1417$-$19} & 2006-12-27   &   530 &       26$\pm$4.3    &    $\dots$  & $\dots$     & 44.27$\pm$0.08 & 39.90   &    $\dots$ &   770       &  0.44    &  0.33 \\
1514+07        & \object{3C317}        & 2006-06-18   &   142 & $<$    3            &  3 & 4  & $<$ 42.24      & 40.66   & $<$43.67   &     $\dots$ &  $\dots$ &  $\dots$ \\
1559+02        & \object{3C327}        & 2005-07-02   &   460 &       78$\pm$7.5    &    $\dots$  & $\dots$     & 44.62$\pm$0.05 & 40.50   & $<$44.68   &  1300       &  1.25    &  0.42 \\
1717$-$00      & \object{3C353}        & 2005-07-02   &   125 & $<$    4            &    $\dots$  & $\dots$     & $<$ 42.28      & 40.02   &   $\dots$  &     $\dots$ &  $\dots$ &  $\dots$ \\
1839$-$48      & \object{PKS1839$-$48} & 2006-06-18   &   490 & $<$    4            &    $\dots$  & $\dots$     & $<$ 43.40      & 41.33   &   $\dots$  &     $\dots$ &  $\dots$ &  $\dots$ \\
1949+02        & \object{3C403}        & 2006-06-18   &   250 &       89$\pm$6.8    &    $\dots$  & $\dots$     & 44.17$\pm$0.04 & 39.55   &     44.50  &   330       &  0.34    &  0.37 \\
1949$-$01      & \object{3C403.1 }     & 2006-06-18   &   230 & $<$    4            &    $\dots$  & $\dots$     & $<$ 42.78      & $\dots$ &    $\dots$ &     $\dots$ &  $\dots$ &  $\dots$ \\
2045+06        & \object{3C424  }      & 2006-06-18   &   560 & $<$    4            &  2 & 4  & $<$ 43.46      & 40.48   &    $\dots$ &     $\dots$ &  $\dots$ &  $\dots$ \\
2158$-$380     & \object{PKS2158$-$380}& 2006-06-18   &   140 &       17$\pm$2.8    &    $\dots$  & $\dots$     & 42.94$\pm$0.07 &$<$38.88 & $<$ 43.68  &   200       &  0.37    &  0.33 \\
2221$-$02      & \object{3C445   }     & 2005-07-02   &   240 &      190$\pm$16     &    210      &  2          & 44.46$\pm$0.04 & 40.44   &     44.44  &   390       &  0.42    &  0.42 \\
2354$-$35      & \object{PKS2354$-$35} & 2006-06-18   &   210 & $<$    3            &    $\dots$  & $\dots$     & $<$ 42.55      & 39.08   &    $\dots$ &     $\dots$ &  $\dots$ &  $\dots$ \\
\hline
\end{tabular}
\begin{list}{}{}
\item[] \textit{Column (1)} Parkes Radio catalogue reference. \textit{Column (3)} Date of observation. \textit{Column (4)} Distances under the chosen cosmology, determined from the redshifts. \textit{Column (5-7)} VISIR mid-infrared core flux and 3-$\sigma$ upper limits, at the SiC-band (11.9~$\mu$m), compared to literature values: 1=\cite{2004ApJ...605..156G} (TIMMI, 1.5~$\arcsec$, 10.8~$\mu$m), 2=\cite{2004A&A...421..129S} (ISOCAM, 5~$\arcsec$, 14.3~$\mu$m), 3=\cite{2004ApJ...602..116W} (Keck, 0.3~$\arcsec$, 11.7~$\mu$m) and 4=\cite{2009ApJ...701..891L} (Spitzer IRS, 15~$\mu$m). \textit{Column (8)} VISIR mid-infrared core luminosity at rest wavelength. \textit{Column (9)} Radio (5~GHz) core luminosity at rest wavelength from fluxes tabulated in Table~\ref{coresed}. \textit{Column (10)} IRAS mid-infrared galaxy luminosity at rest wavelength derived for most sources from flux from \cite{1988AJ.....95...26G}, for 3C227 from \cite{1995A&A...303....8H}, and for PKS2158$-$380 from \cite{1990IRASF.C......0M}. \textit{Columns (11-13)} The radius at $50\%$ of the flux, the full width half maximum (FWHM) of the flux of the source and of the standard star.
\end{list}
\end{table*}

\section{Sample selection}
From the well studied 3CR/4C and PKS radio galaxy samples, we extracted subsamples of nearby double-lobed FR-I galaxies, low- and high-excitation, and broad-line FR-II radio galaxies visible from Chile. The sources range in redshift from $z=0.006$ to $0.156$, with two outliers: Centaurus~A at $z=0.00095$ and 3C93 at $z=0.35700$. The subsamples were chosen to span three orders of magnitude in radio core as well as in total power, to search for trends with radio power. Southern and equatorial objects were selected from three different studies: 1) FR-IIs from a $z<0.15$ high-resolution 3.6~cm VLA study \citep{1992MNRAS.256..186B}, 2) FR-Is and FR-IIs from the Hubble Space Telescope snapshot survey of a subsample of $z<0.1$ 3CR sources \citep{1999ApJS..122...81M}, and 3) FR-Is and FR-IIs from a $z<0.7$, 2.3~GHz radio study \citep{1997MNRAS.284..541M} of a subsample of radio galaxies complete down to a flux density of $>2$~Jy at $2.7$~GHz \citep{1985MNRAS.216..173W}. High resolution radio images for all but one object, 3C403.1, were available. All relevant optical spectroscopic classifications are extracted from \cite{1997MNRAS.286..241J}, \cite{1998MNRAS.296..445H, 1999MNRAS.304..135H}, and \cite{1997MNRAS.284..541M}. We note that BLRG 3C120 differs from the rest of the selected BLRG subsample, since it is strongly core-dominated and has a FR-I morphology. 

The basic parameters of the radio galaxy sample - that is to say the unbiased subset as eventually observed with VISIR - are given in Table~\ref{table1}. The original division between edge-darkened, low-radio-power FR-Is and edge-brightened, radio-powerful FR-IIs was established by \cite{1974MNRAS.167P..31F}, where FR-IIs have radio luminosities $\log{\nu L_{178\mathrm{MHz}}}>40.9~\mathrm{erg}~\mathrm{s}^{-1}$. In our sample, some galaxies classified as FR-Is have FR-II-like radio power close to the FR-I/FR-II division line. We here consider the radio morphology to be the defining classification property and not the radio power. Not surprisingly, we find some radio luminosity overlap between FR-Is and FR-IIs. The objects span a broad range in radio core fraction $R=S_{\rm{core}}/S_{\rm{extended}}$, where $S_{\rm{core}}$ is the radio emission of the core and $S_{\rm{extended}}$ of the lobes: we find typical values of $R$ at GHz-frequencies of $\sim1\times10^{-2}$ and note that the unusually high value for 3C120 is most likely caused by beamed core emission. Depending on their distance, the objects permit linear resolution of approximately 10-1000 parsec, for the diffraction-limit in the mid-infrared of $0.40\arcsec$, which is achievable with VISIR. Table~\ref{tablemedian} lists the median values of important physical parameters for the four different classes. In general, our sample constitutes fairly low-luminosity sources with $\log{\nu L_{\mathrm{408MHz}}^{\mathrm{total}}} \mathrm{(median)}= 41.6~\mathrm{erg}~\mathrm{s}^{-1}$ compared to the sources detected by mid-infrared studies at intermediate redshift \citep{2006ApJ...647..161O}.

\section{VISIR observations and data reduction}
We observed the nuclei of the 27 radio galaxies with VISIR, the Very Large Telescope imager and spectrometer for the mid-infrared \citep{2004Msngr.117...12L, 2005Msngr.119...25P}, mounted on the Cassegrain focus of the VLT unit telescope 3, Melipal, in Cerro Paranal, Chile. The imaging data were obtained in guaranteed time during three nights in total in July and December 2005 and June and December 2006, through the SiC filter ($11.85~\pm~2.34~\mathrm{\mu m}$) in the N-band. The pixel scale is $0.075\arcsec\mathrm{/pixel}$ resulting in a $19\arcsec.2$ field of view. To help remove the sky background, secondary mirror chopping was performed in the north-south direction with an amplitude of $8\arcsec$ at a frequency of $0.2~\mathrm{Hz}$. This rate is sufficiently higher than the rate of the background fluctuations. Nodding was applied every 30 seconds using telescope offsets of $8\arcsec$ in the east-west direction to eliminate optical path residuals that remain after chopping. The detector integration time was $20$~ms. Total source integration times were in the range of $2$ to $60~\mathrm{minutes}$.

With the SiC filter, the limiting flux of a point source detected at a signal-to-noise-ratio of 10 in one hour of on-source integration is of the order of 7~mJy as determined from long-term observations of calibration standard stars (VISIR user manual). Mid-infrared standard stars \citep{1999AJ....117.1864C} were observed frequently during the nights. Their FWHMs (Table~\ref{tablevisir}) show that with this filter on average a diffraction-limited resolution of 0.36$\arcsec$ is reached. This is an order of magnitude higher than can currently be obtained with infrared space observatories.

The standard star and program source data are reduced and analyzed in the same way. Stripes caused by variations in the detector temperature are removed, bad pixels blocked, the nod-images are co-added into one and finally the four beams are shifted and added into one. Flux and noise levels are extracted using multi-aperture photometry and the curve-of-growth method. The sky annulus is fixed to a size of 0.5$\arcsec$ and begins where the source aperture radius ends, varying from 1 to 1.5$\arcsec$. Of the 27 sources, we detect 10 with high signal-to-noise ratios; these span a substantial range in flux, from 17 to 1200~mJy and in distance, from 4 to 530~Mpc. Comparing the VISIR flux values to the wide beam IRAS 12~$\mu$m values, we find mid-infrared core concentrations ranging from 4 -- 100 percent. For the 17 non-detected sources, mostly FR-Is and LEG/FR-IIs, we are able to place three-sigma, three standard deviations, level upper limits using the sensitivies determined from the standard stars. The resulting values and derived luminosities are presented in Table~\ref{tablevisir}.

Certain sample sources were observed previously by other investigators, at comparable infrared-wavelengths but somewhat lower resolution. The nucleus of FR-I Centaurus~A is estimated to have a flux of 1.8~Jy \citep{2004ApJ...602..116W} for a 11.7~$\mu$m filter with a 2.4~~$\mu$m bandwidth. Both this and our measurement are in agreement with the 4$\arcsec$ ISOCAM 10-20~$\mu$m spectrum of the nucleus (see Fig.~2 of \cite{1999A&A...341..667M}). Our flux measurement of 3C120, 240~mJy, is notably higher than that measured by \cite{2004ApJ...605..156G}, which was 109~mJy and derived from a lower $1.5\arcsec$ resolution image. This demonstrates the most likely variable, beamed, and non-thermal nature of the infrared-emission. This conclusion is also supported by the lack of an infrared excess above the synchrotron fit in the core spectral energy distribution (Fig.~\ref{allsedsBLRG}). The 10-20~$\mu$m ISOCAM observations of HEG 3C98 and BLRG 3C445 \citep{2004A&A...421..129S} infer the same fluxes as our measurements. On the other hand, our non-detection and the detection with the Spitzer IRS (3.7$\arcsec$ slit width) \citep{2009ApJ...701..891L} show that the mid-infrared emission in FR-I 3C270 must be of an extended nature compared to the VLT/VISIR 0.36\arcsec~beam. Possibly inconsistent with this conclusion is that this extended emission probably originates from star formation and in situ heating, yet PAHs are weak or absent in the high-quality Spitzer spectrum. Galaxy 3C15 is detected with the Spitzer IRS, but not with VISIR: the IRS spectrum includes the radiation of the arcsecond scale jet, seen in the optical, but shows no sign of star formation. Three other galaxies, 3C29, 3C317, and 3C424 with Spitzer IRS spectra have flux values consistent with our VISIR upper limits. Finally, published VISIR observations of Centaurus~A and 3C445 in three other N-band-filters, centred on 10.49, 11.25, and 12.81~$\mu$m \citep{2009A&A...495..137H} are consistent with our data.

\begin{figure*}
\centering
\begin{displaymath}
\begin{array}{cccccc}
\includegraphics[width=2.8cm]{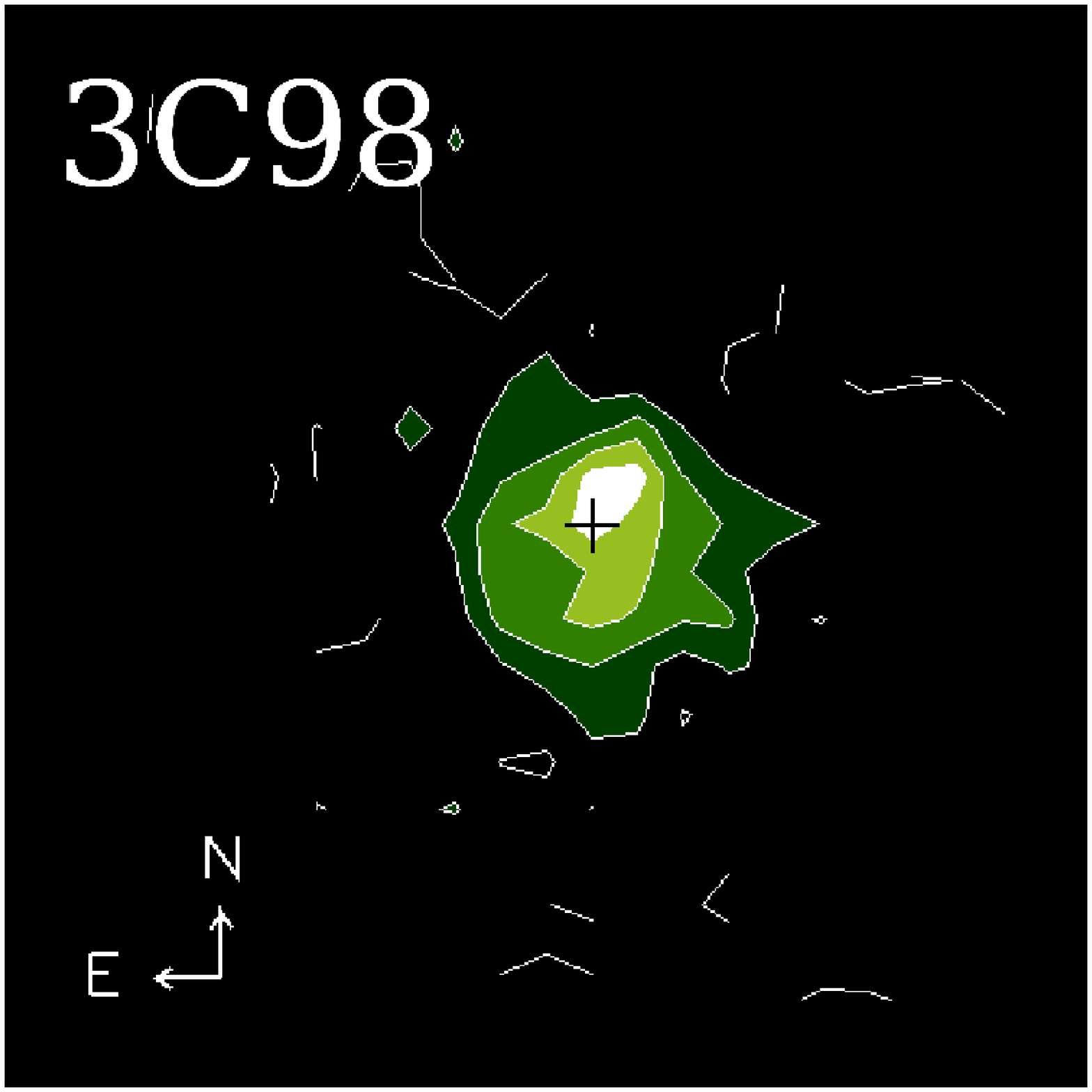} &
\includegraphics[width=2.8cm]{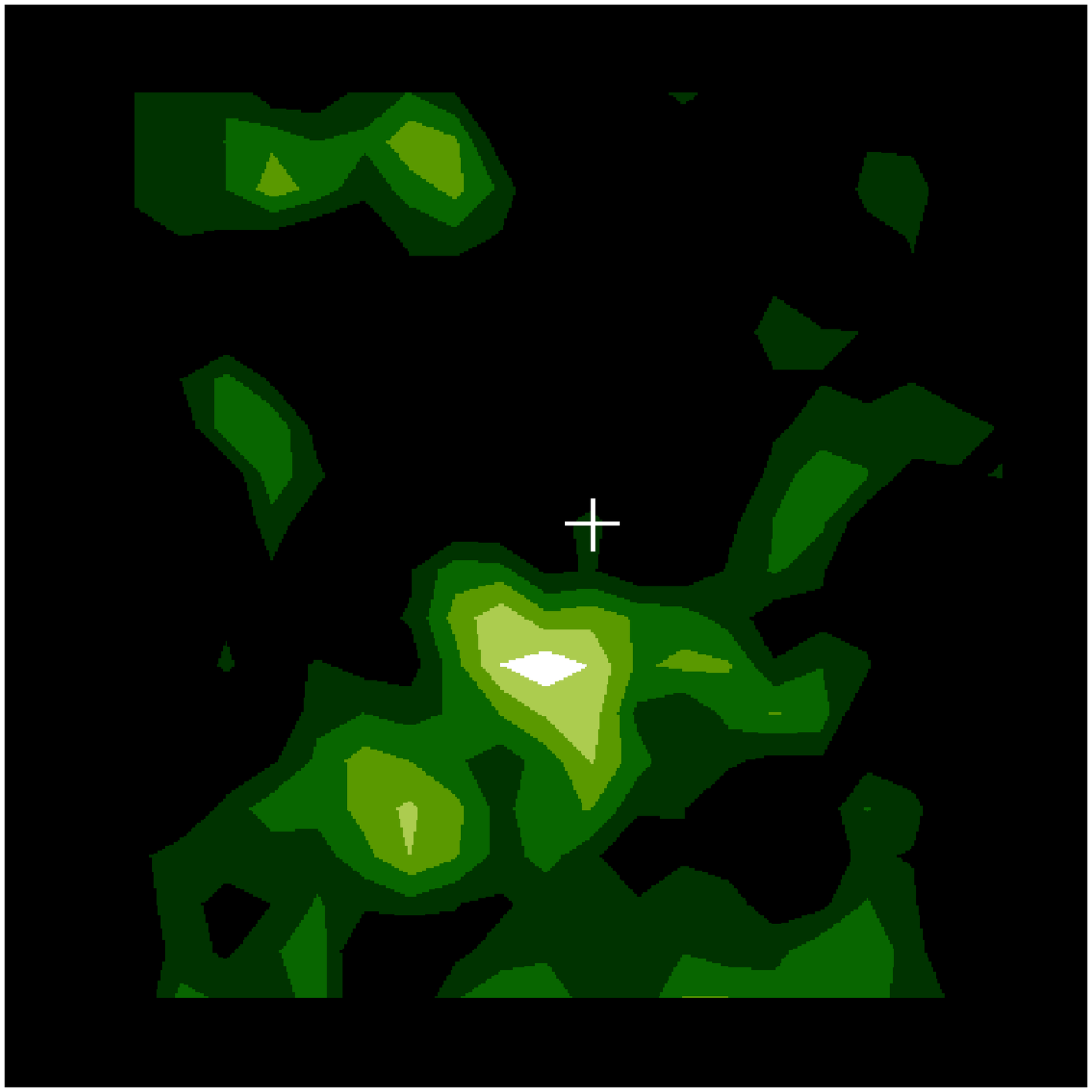} &
\includegraphics[width=3.4cm]{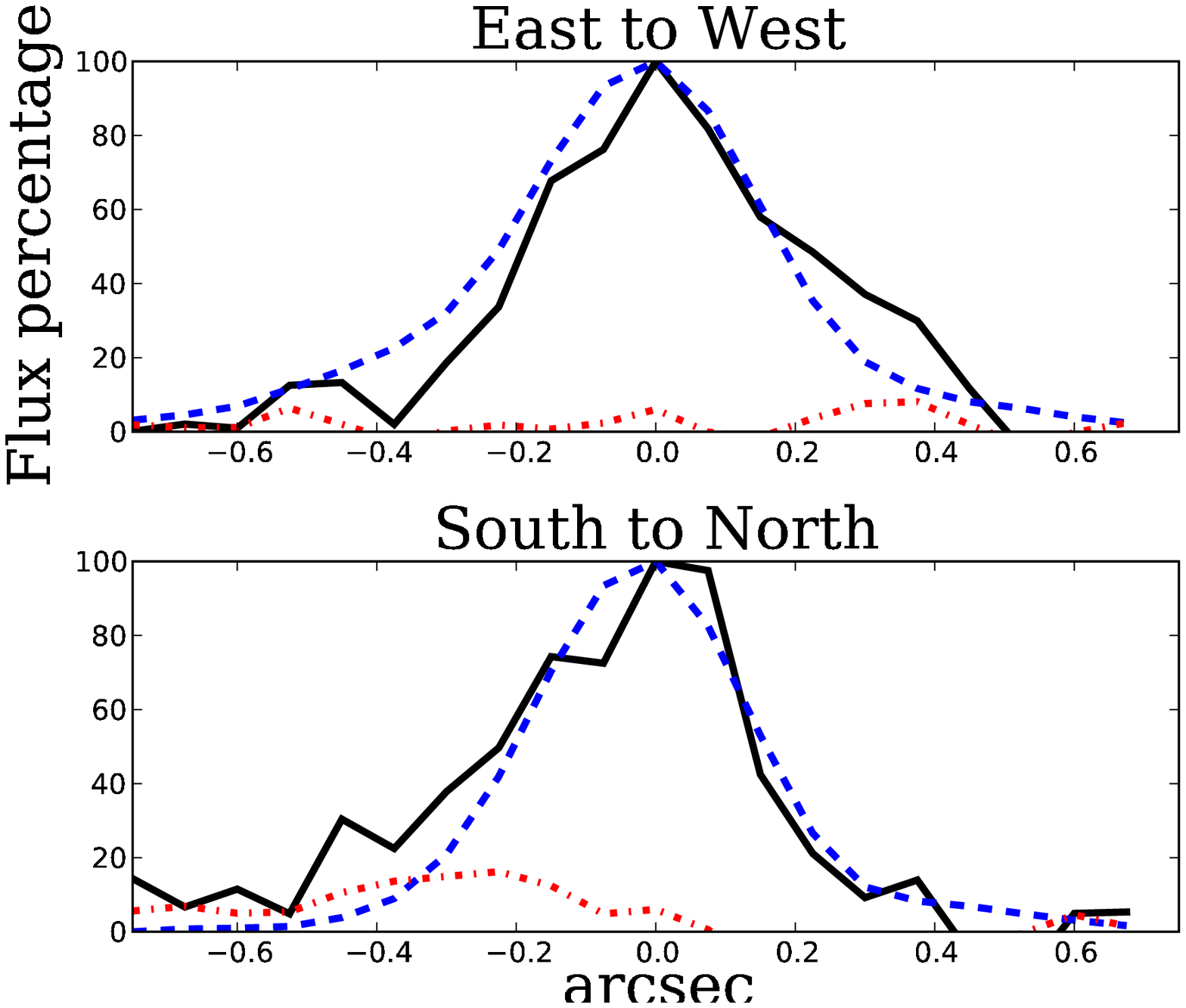} &
\includegraphics[width=2.8cm]{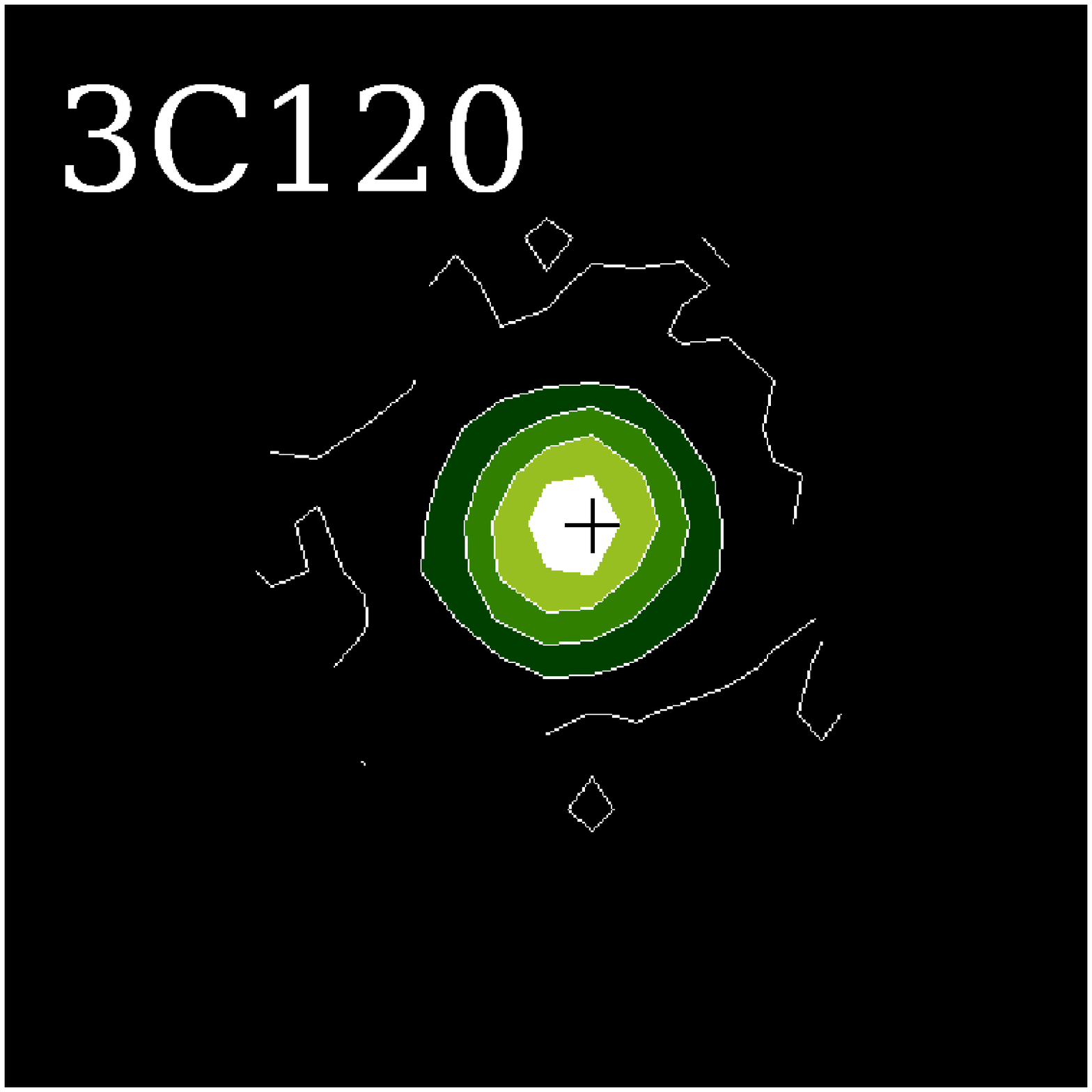} &
\includegraphics[width=2.8cm]{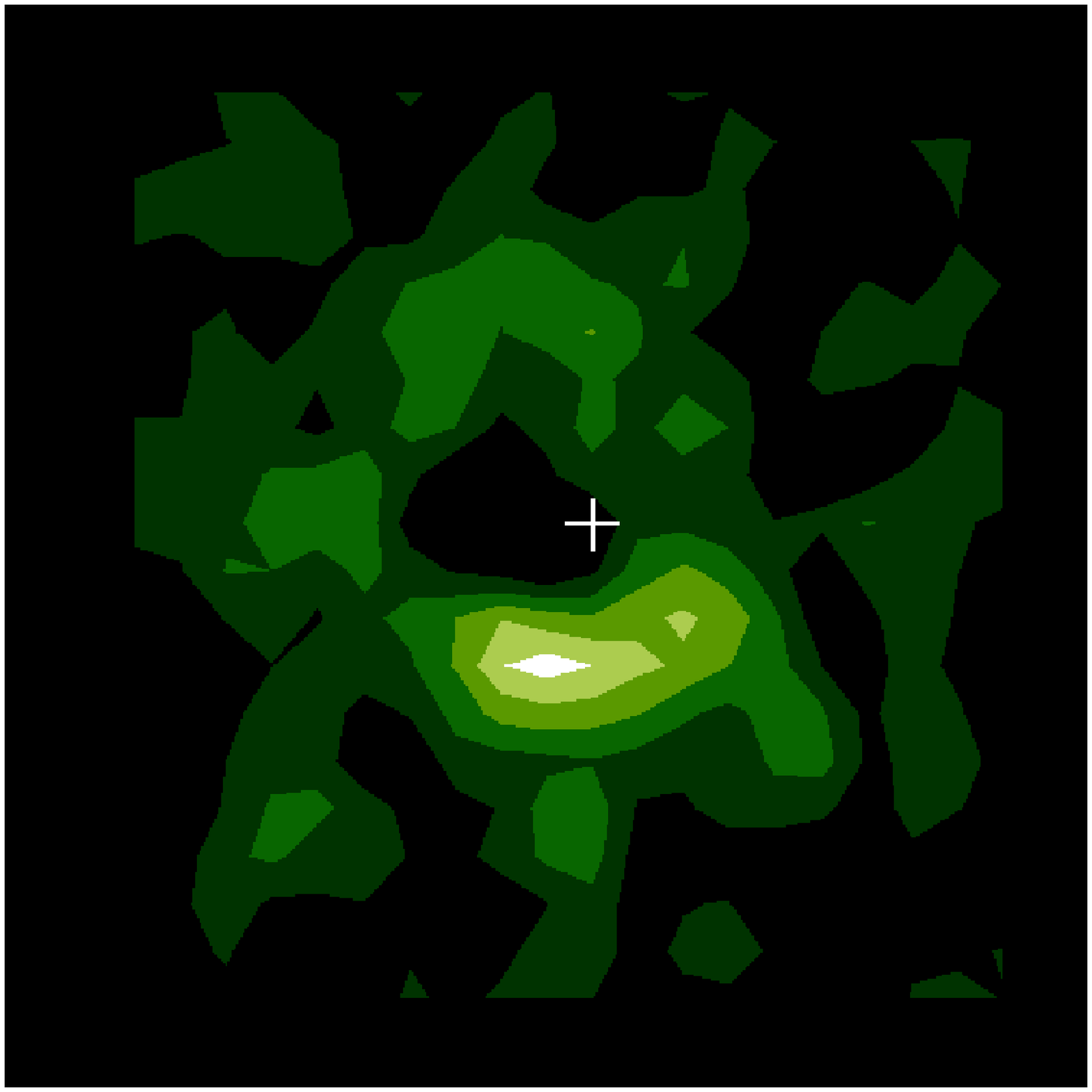} &
\includegraphics[width=3.4cm]{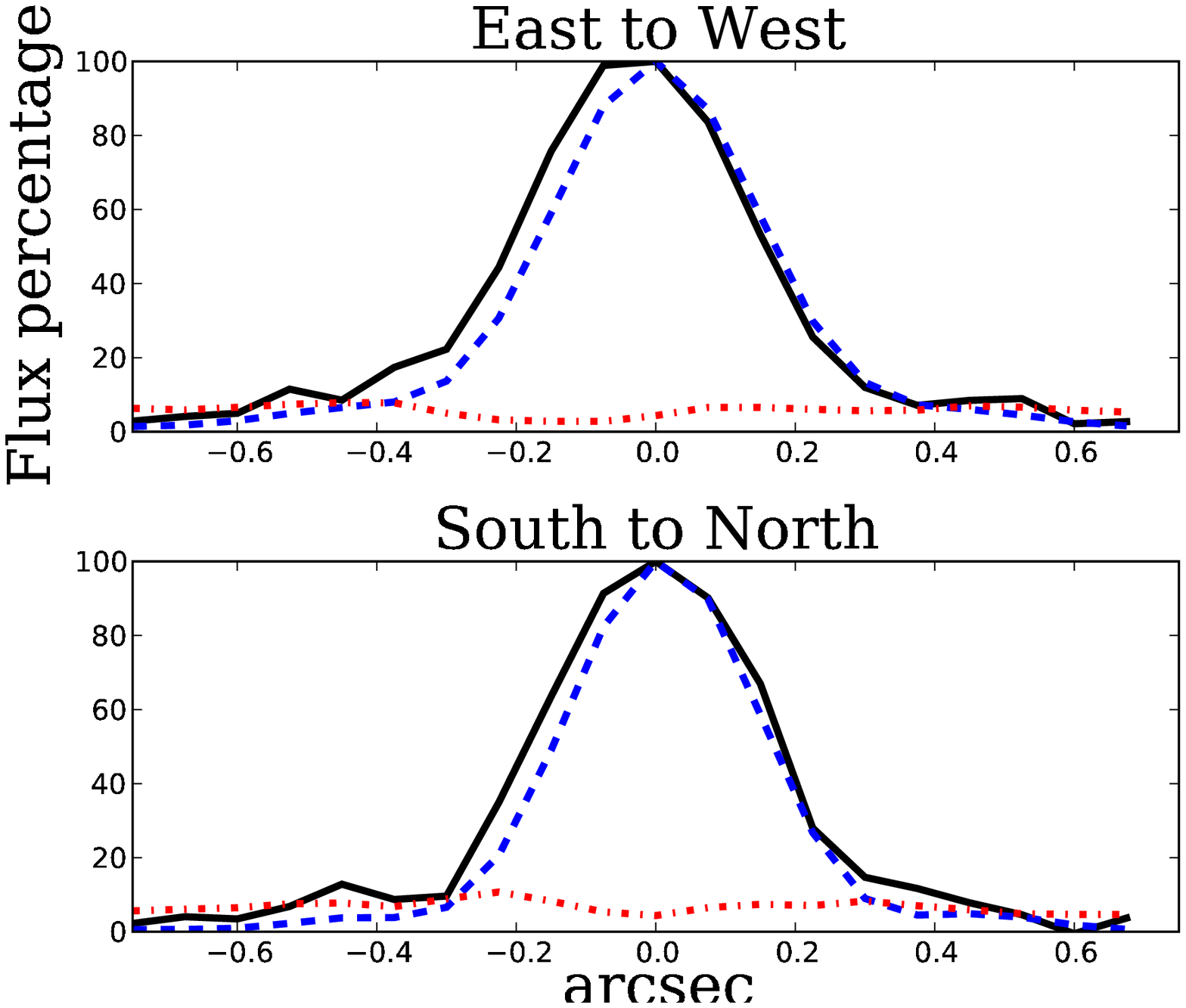} \\
\includegraphics[width=2.8cm]{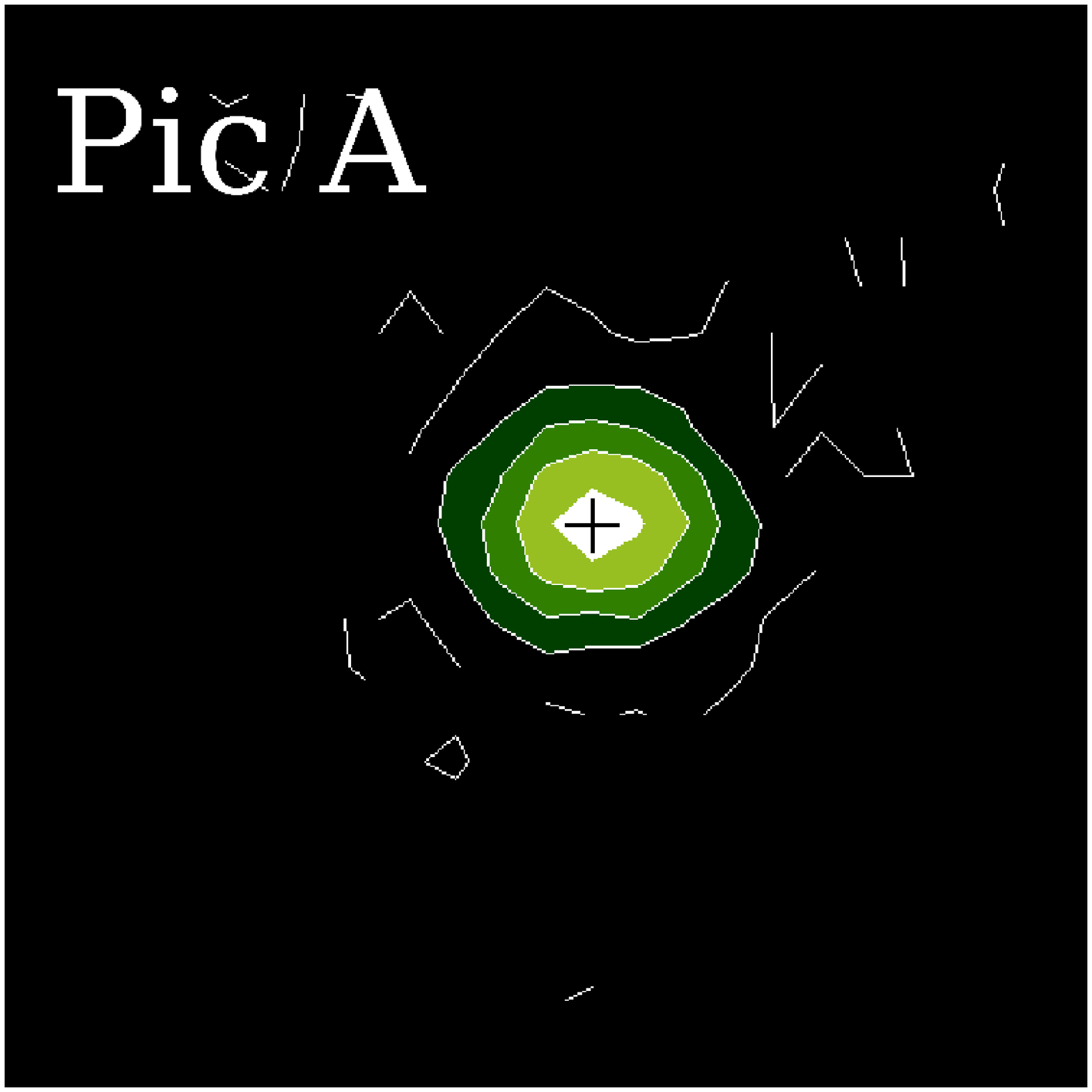} &
\includegraphics[width=2.8cm]{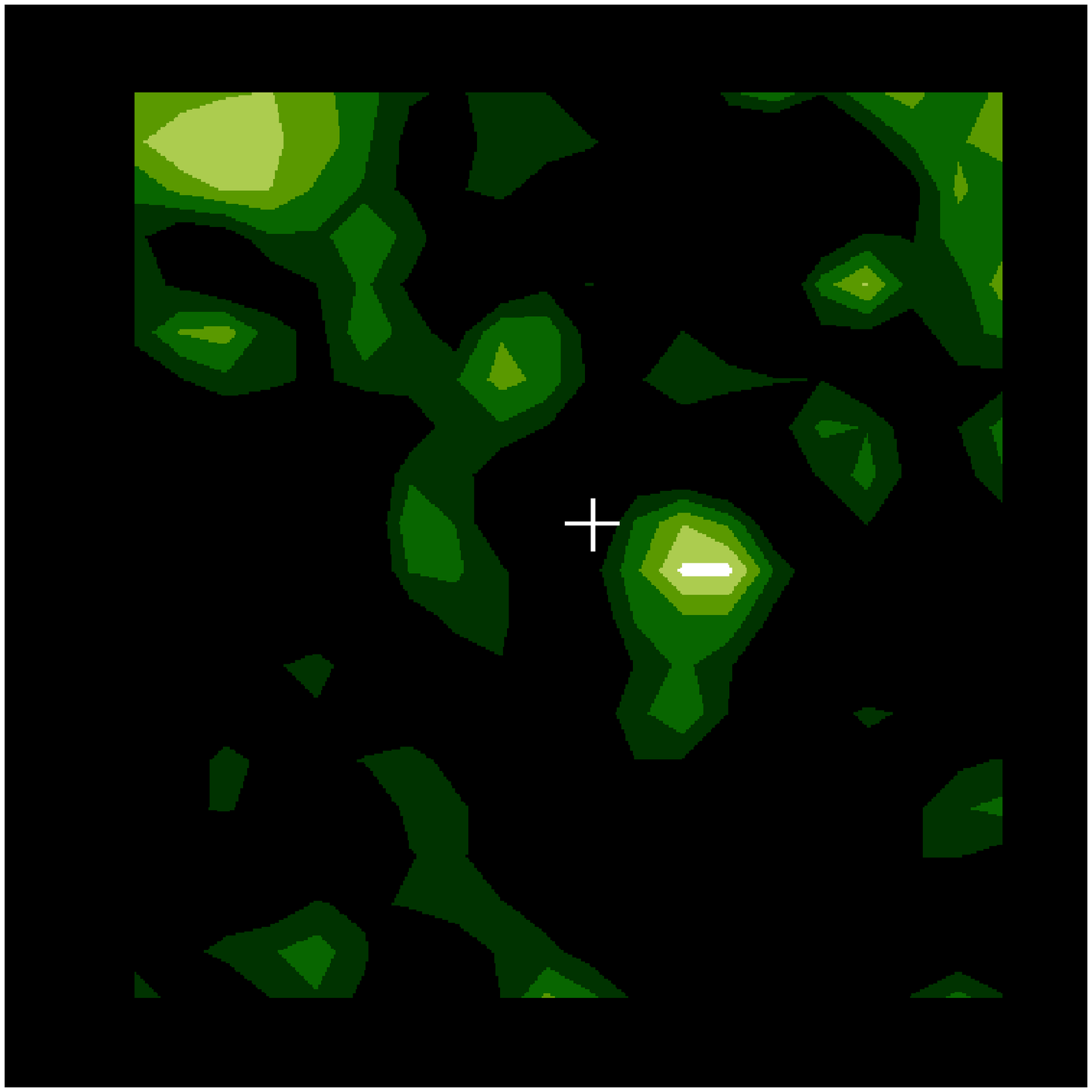} &
\includegraphics[width=3.4cm]{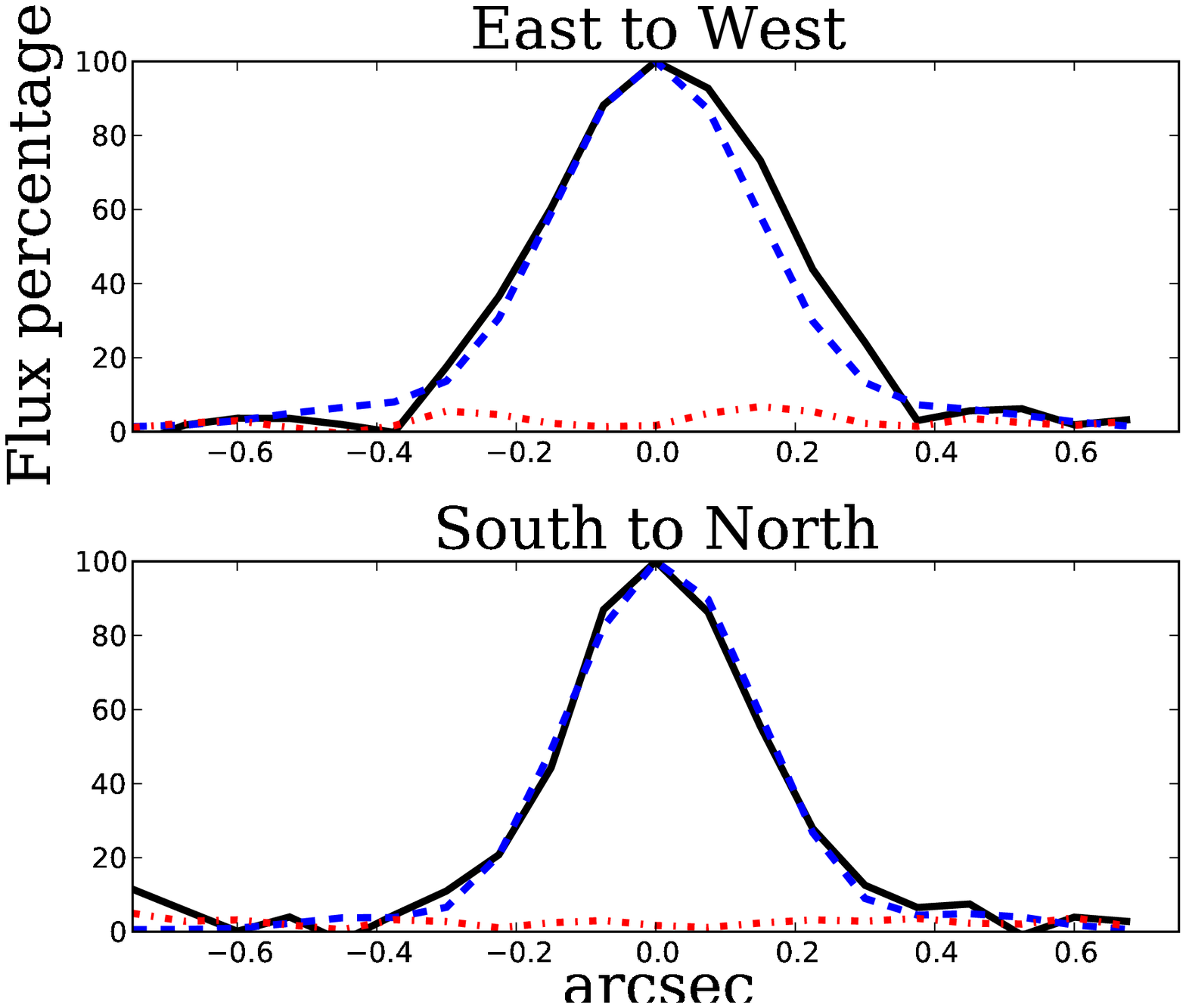} &
\includegraphics[width=2.8cm]{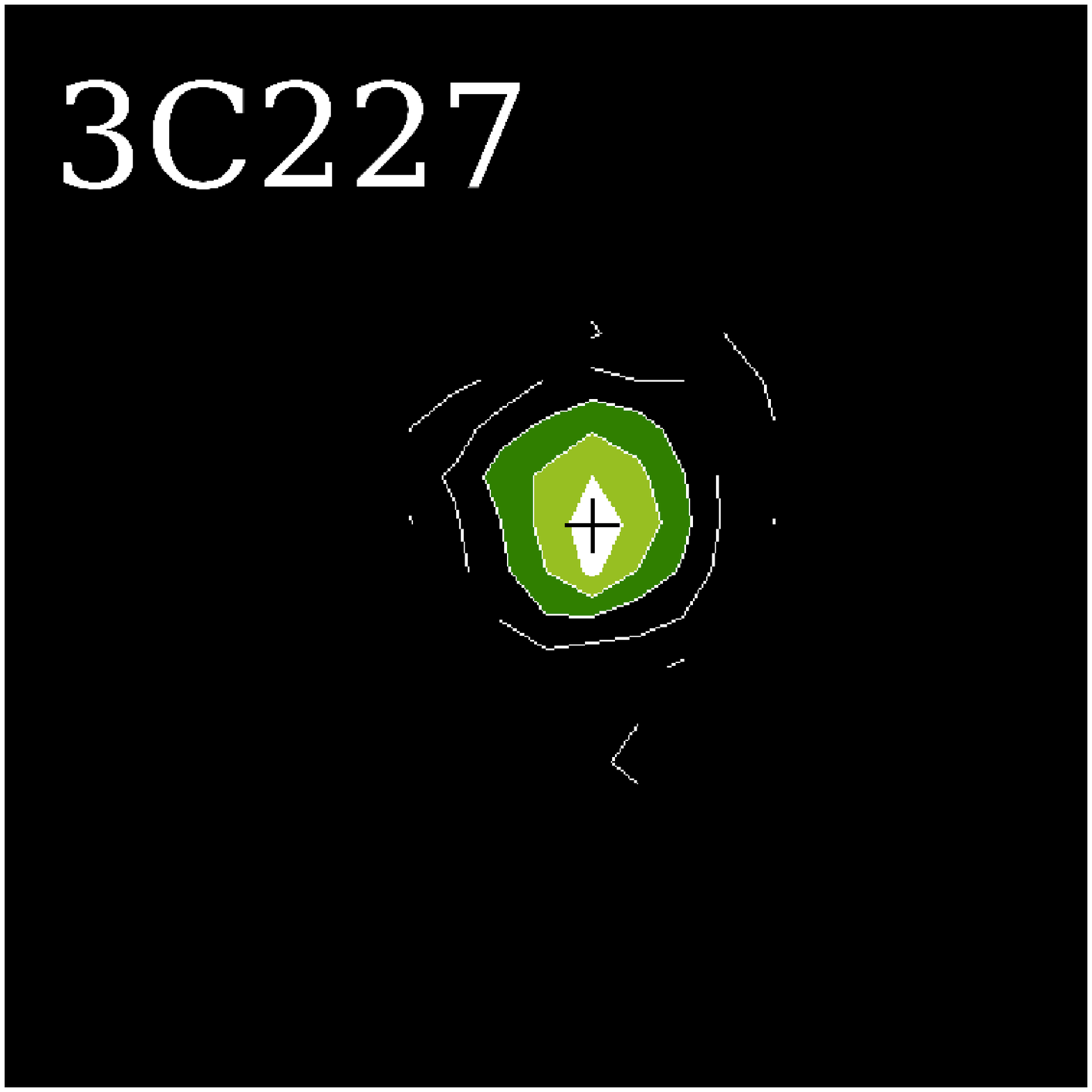} &
\includegraphics[width=2.8cm]{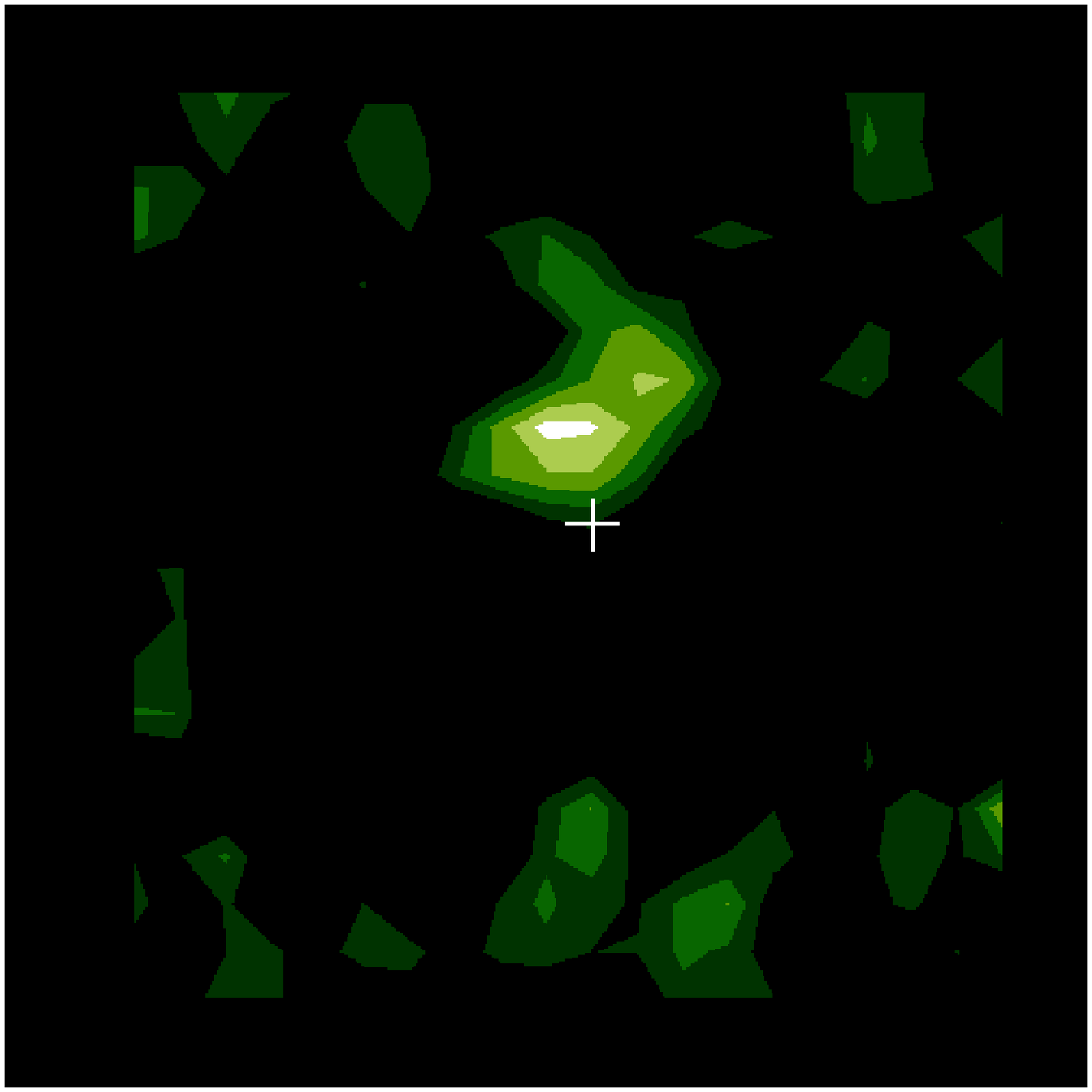} &
\includegraphics[width=3.4cm]{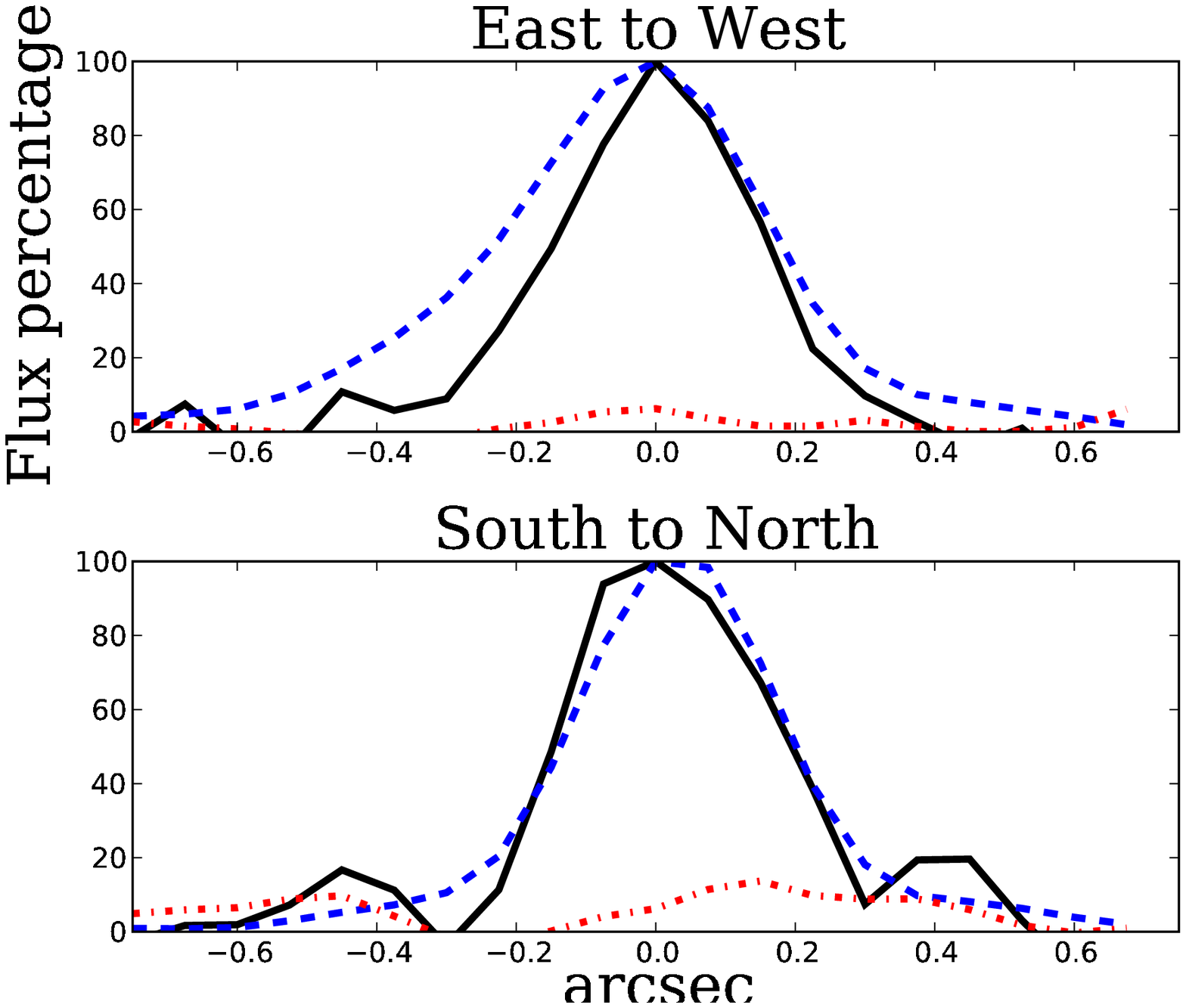} \\
\includegraphics[width=2.8cm]{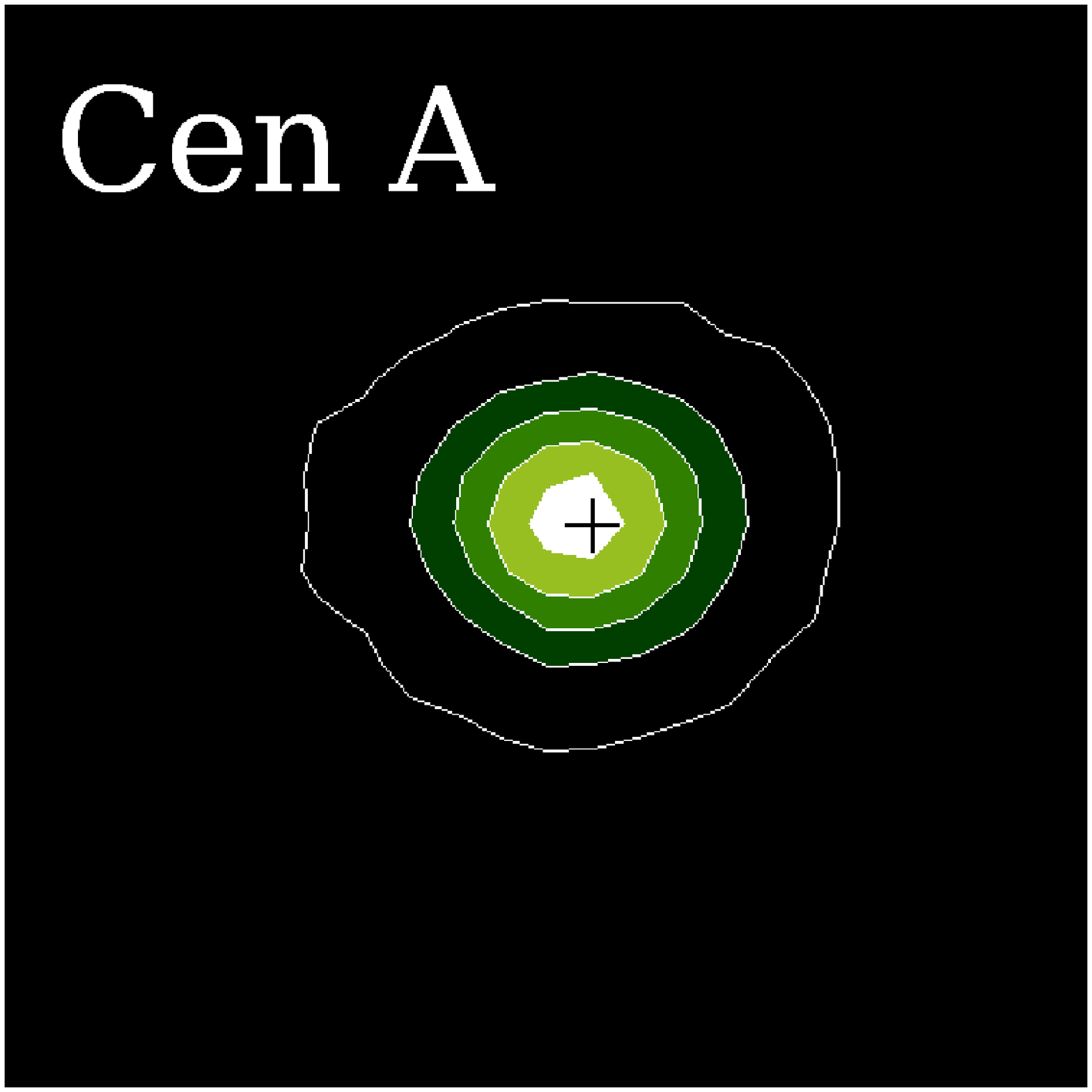} &
\includegraphics[width=2.8cm]{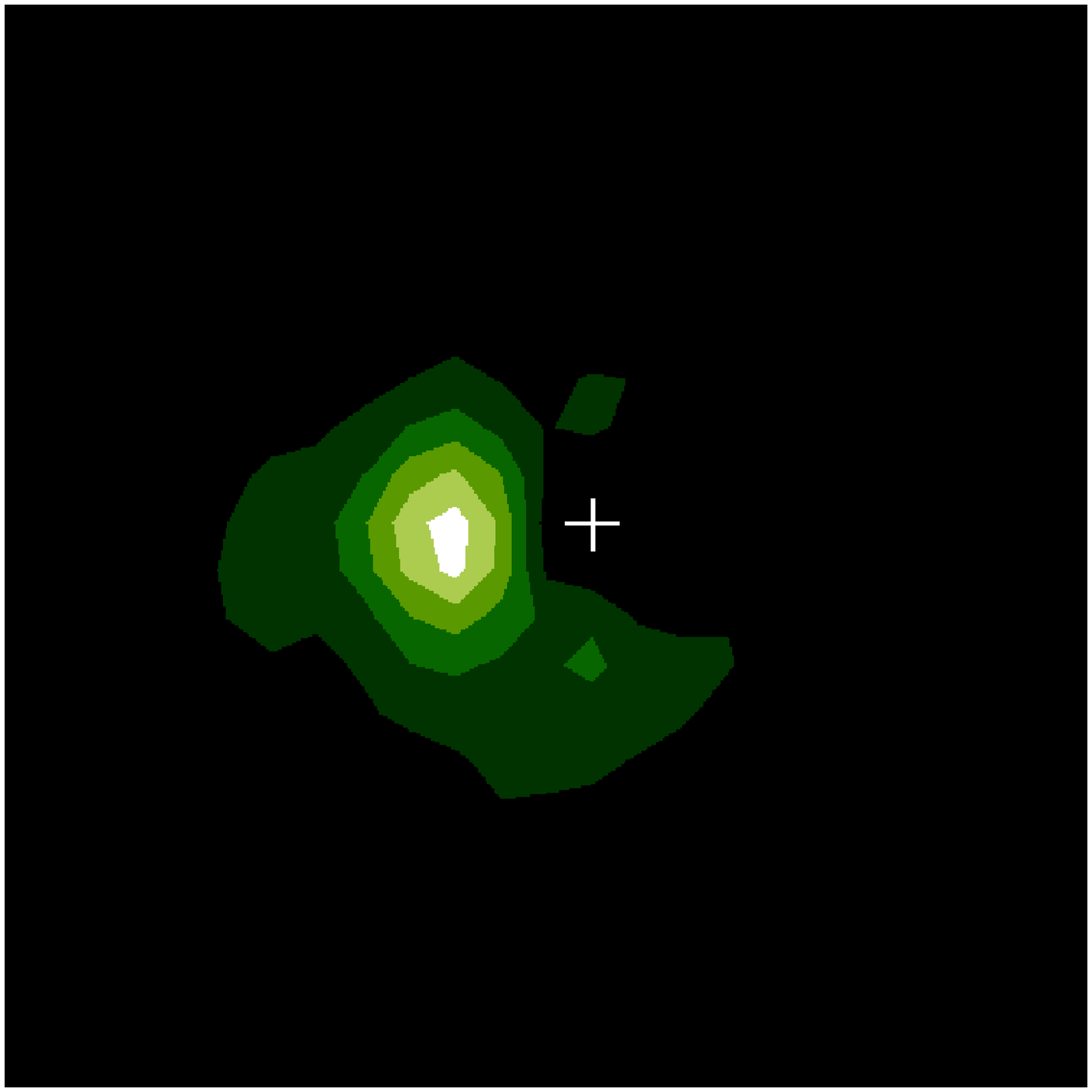} &
\includegraphics[width=3.4cm]{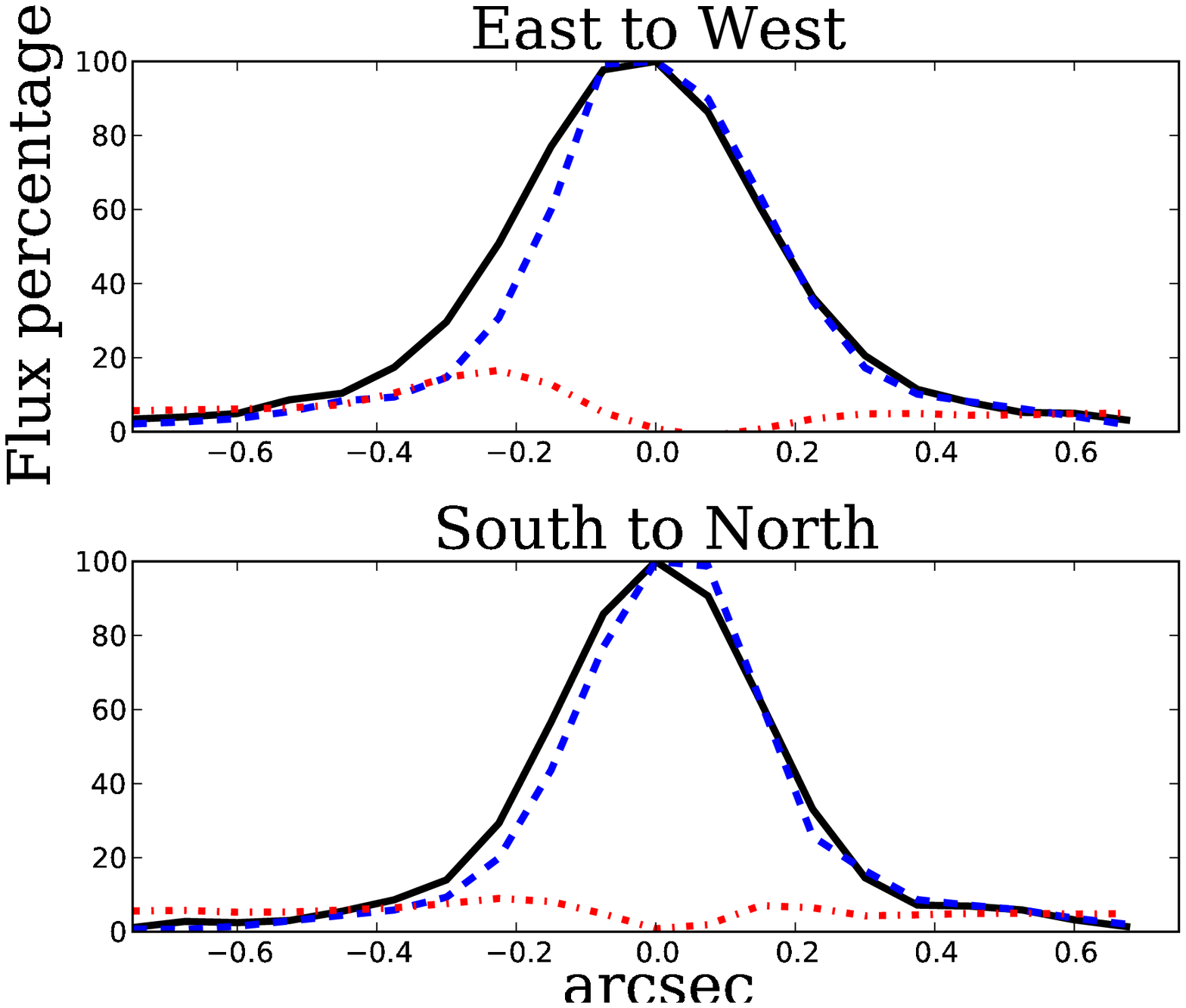} &
\includegraphics[width=2.8cm]{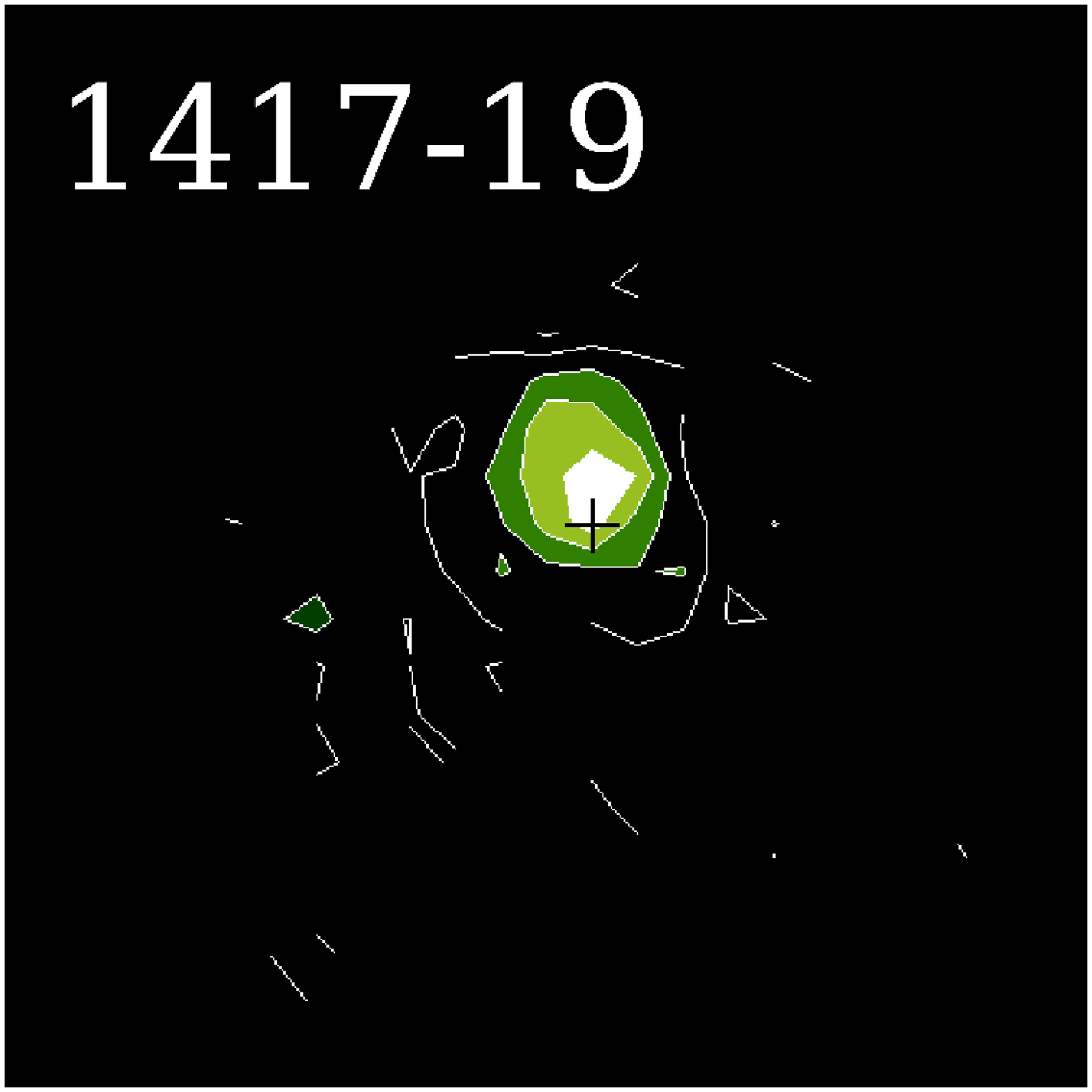} &
\includegraphics[width=2.8cm]{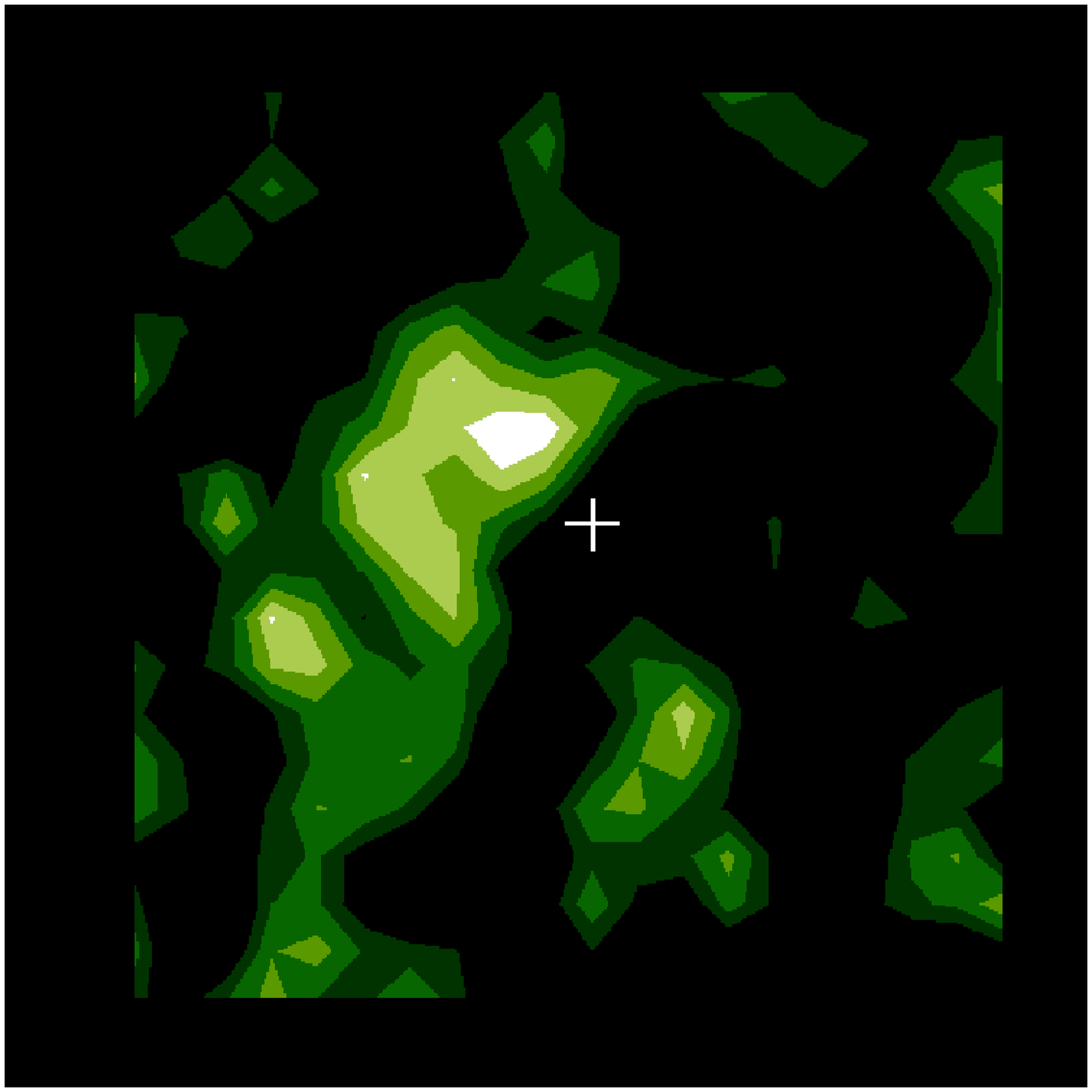} &
\includegraphics[width=3.4cm]{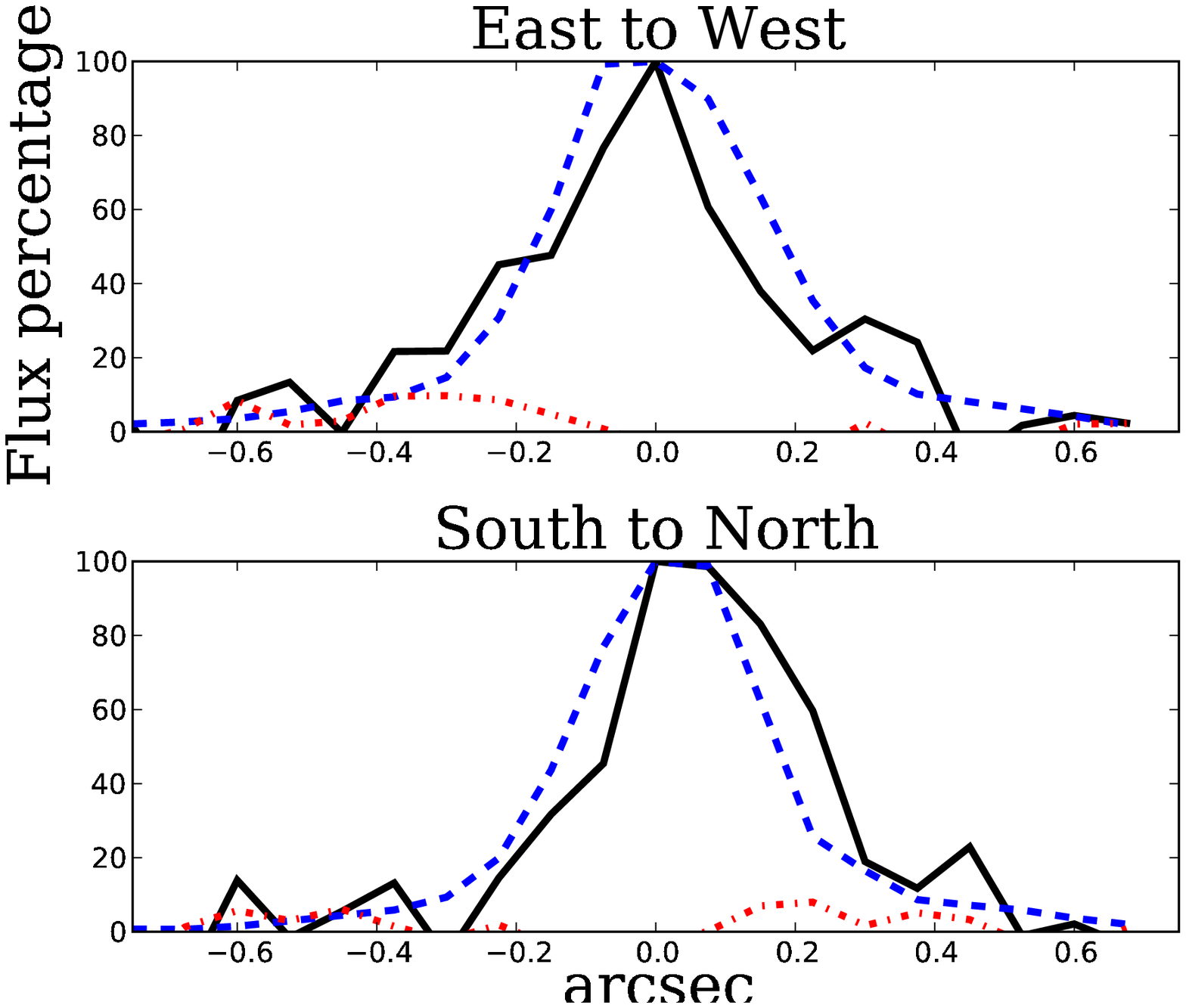} \\
\includegraphics[width=2.8cm]{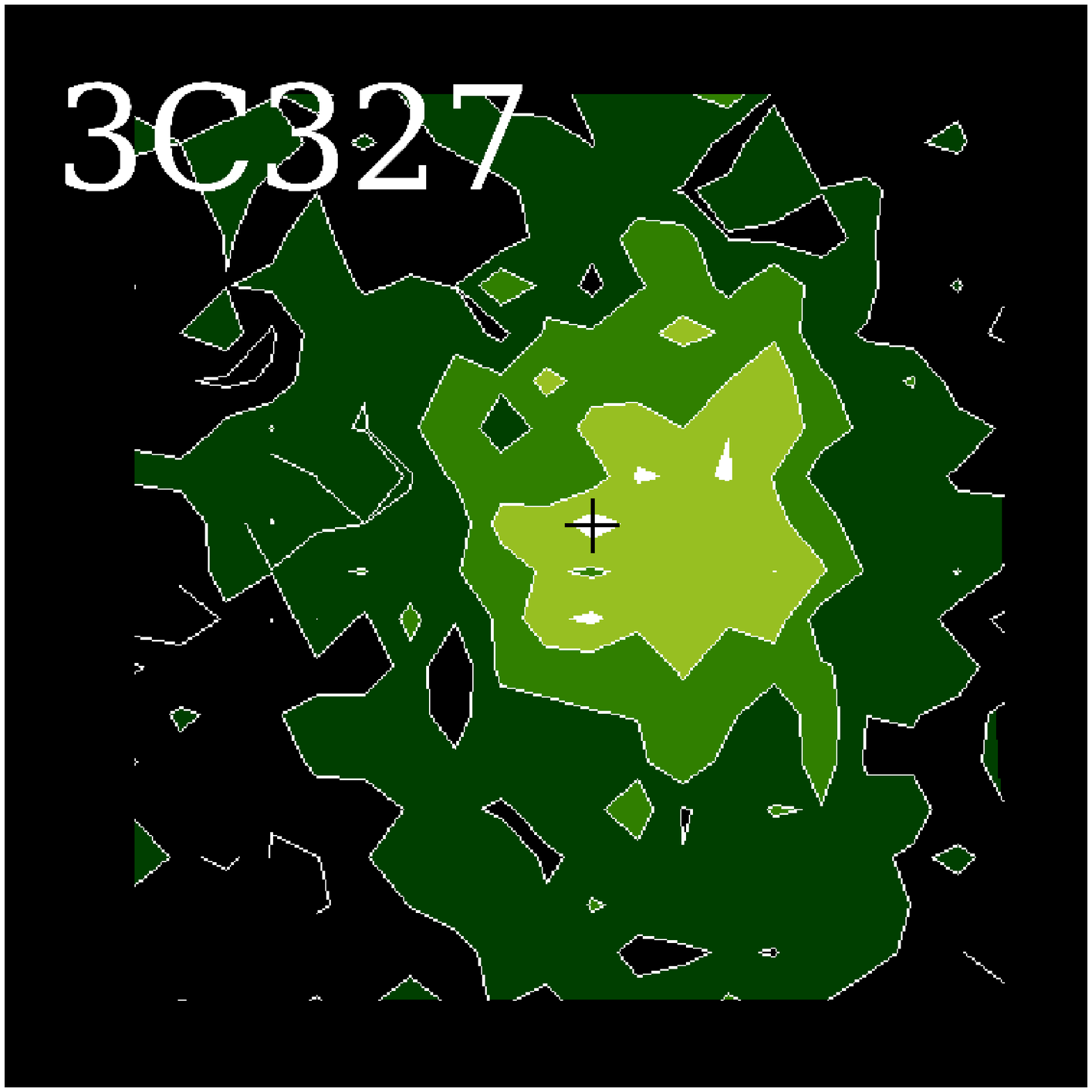} &
\includegraphics[width=2.8cm]{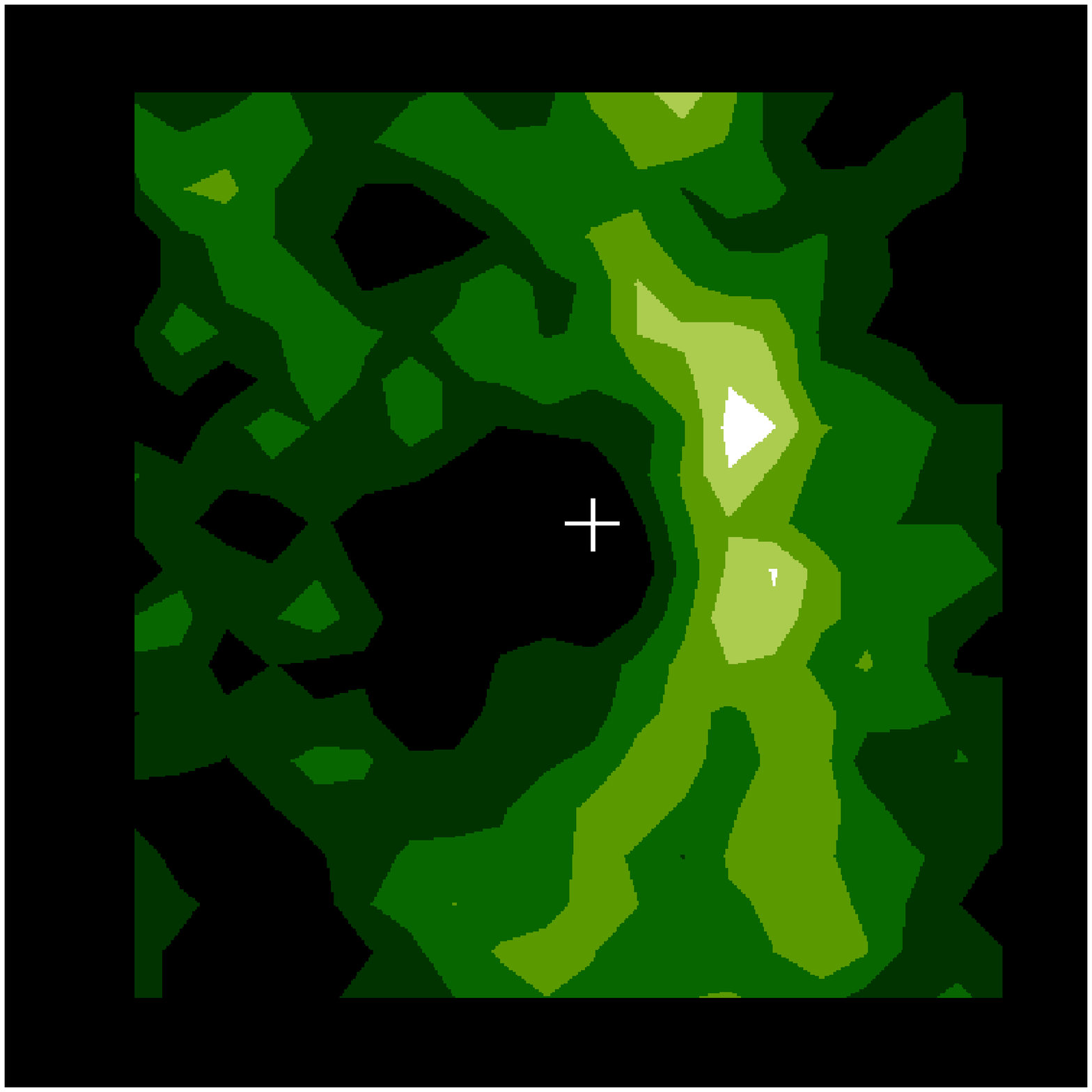} &
\includegraphics[width=3.4cm]{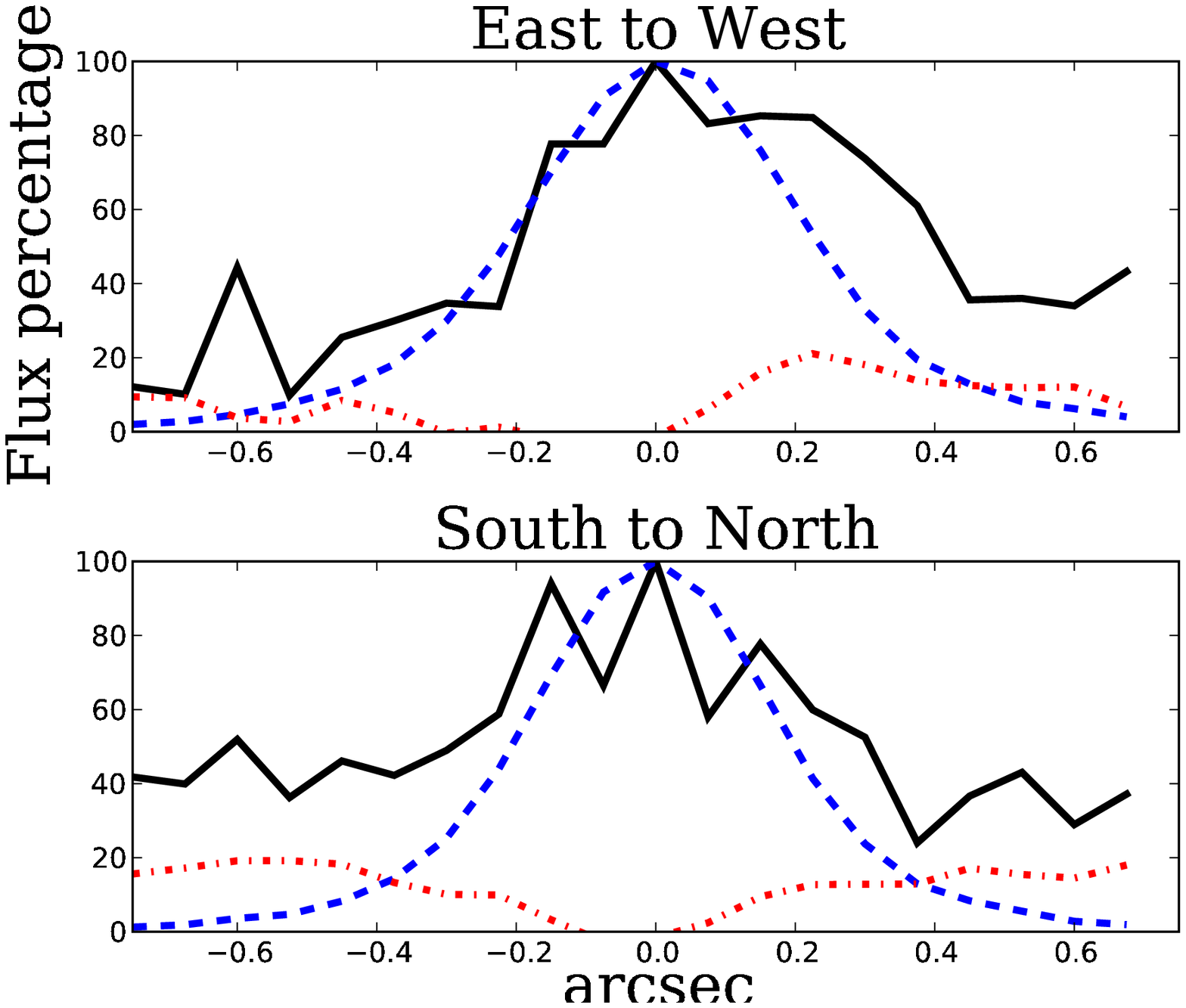} &
\includegraphics[width=2.8cm]{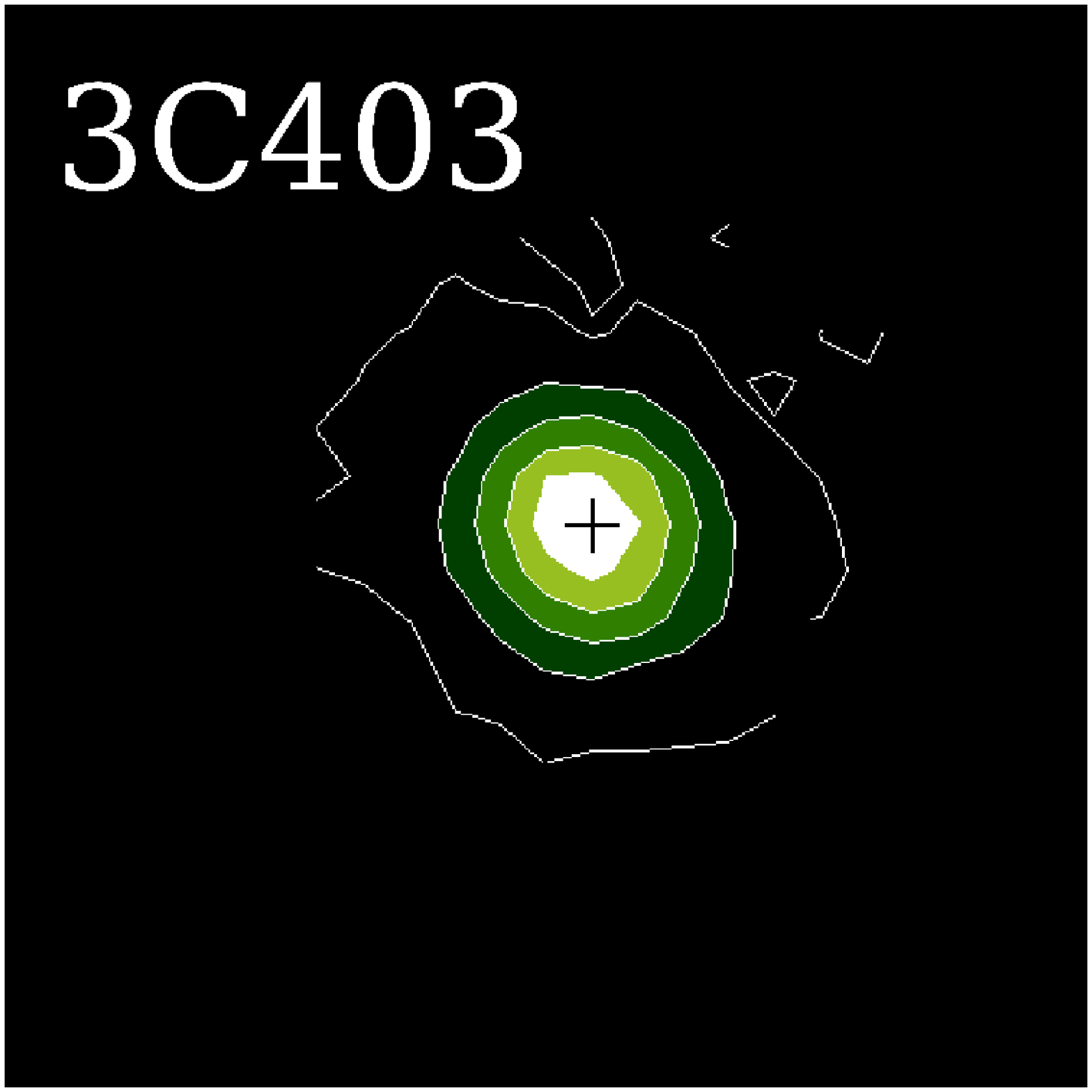} &
\includegraphics[width=2.8cm]{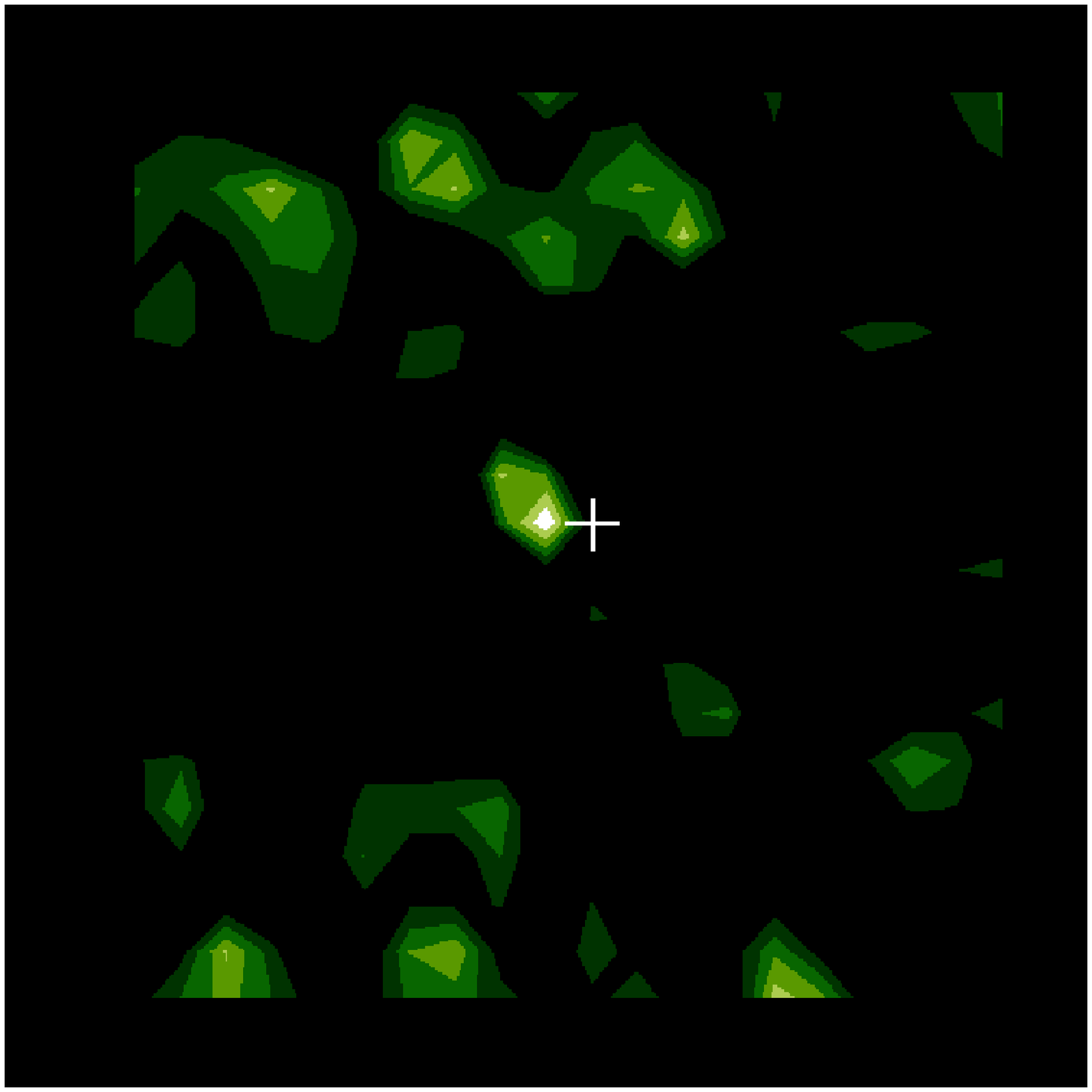} &
\includegraphics[width=3.4cm]{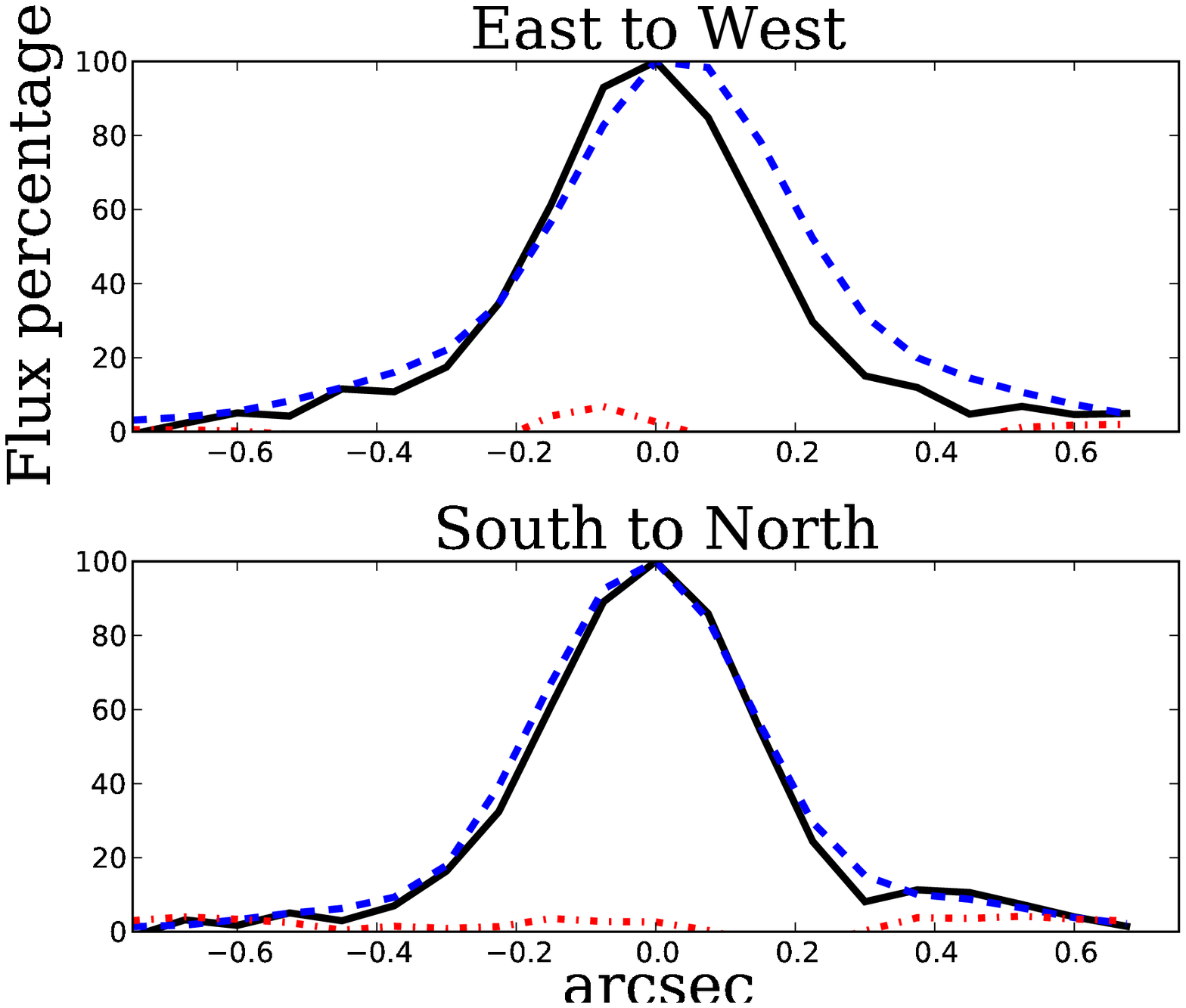} \\
\includegraphics[width=2.8cm]{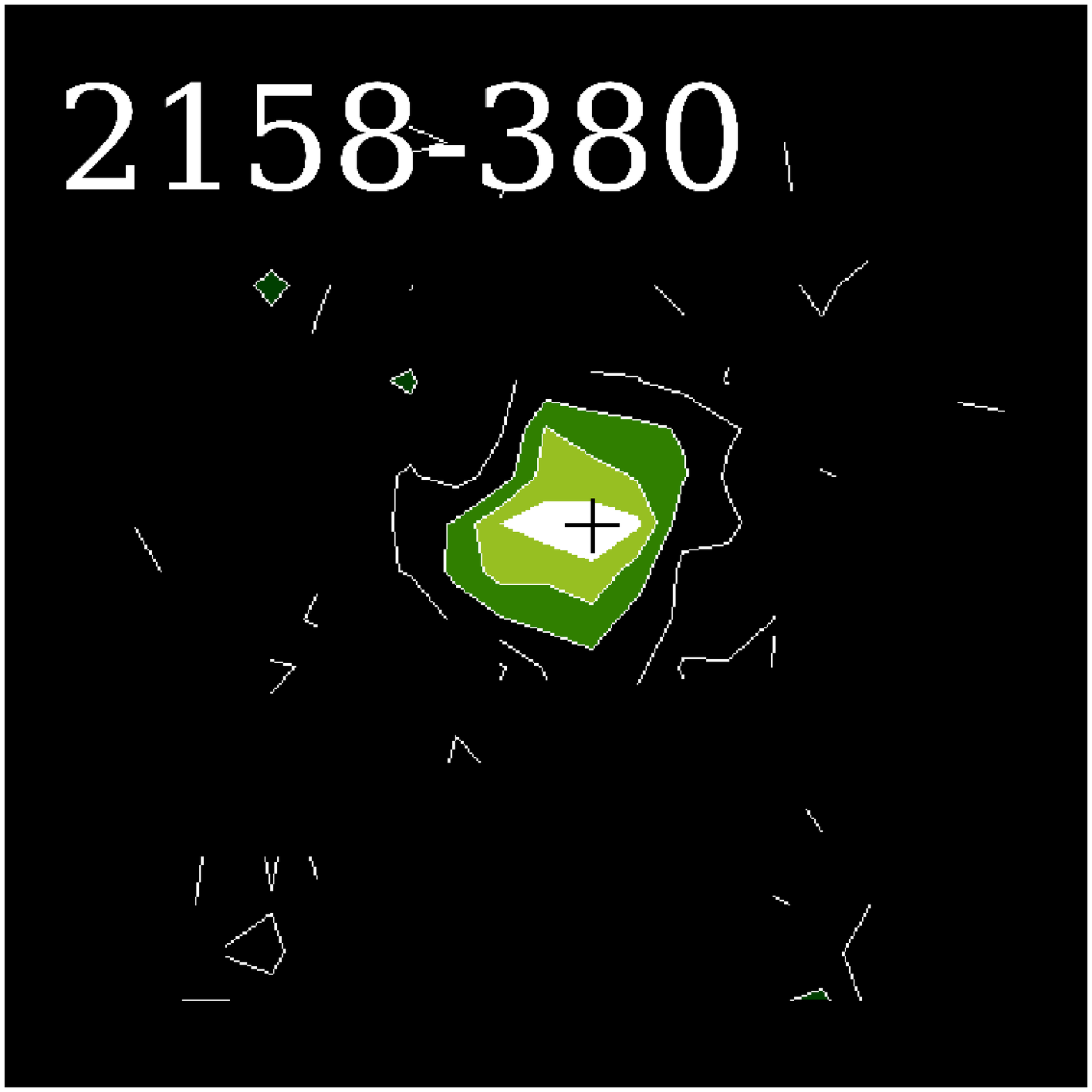} &
\includegraphics[width=2.8cm]{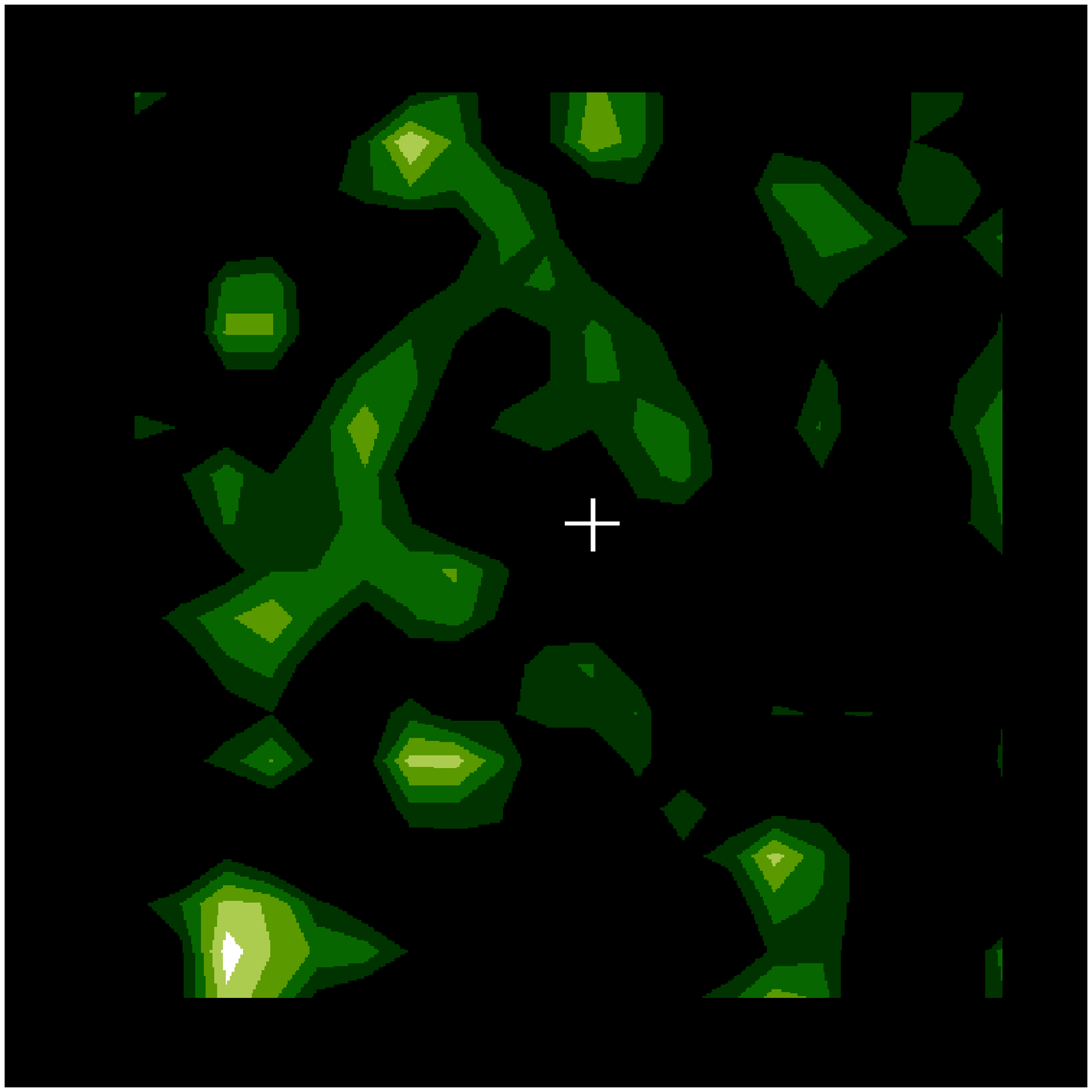} &
\includegraphics[width=3.4cm]{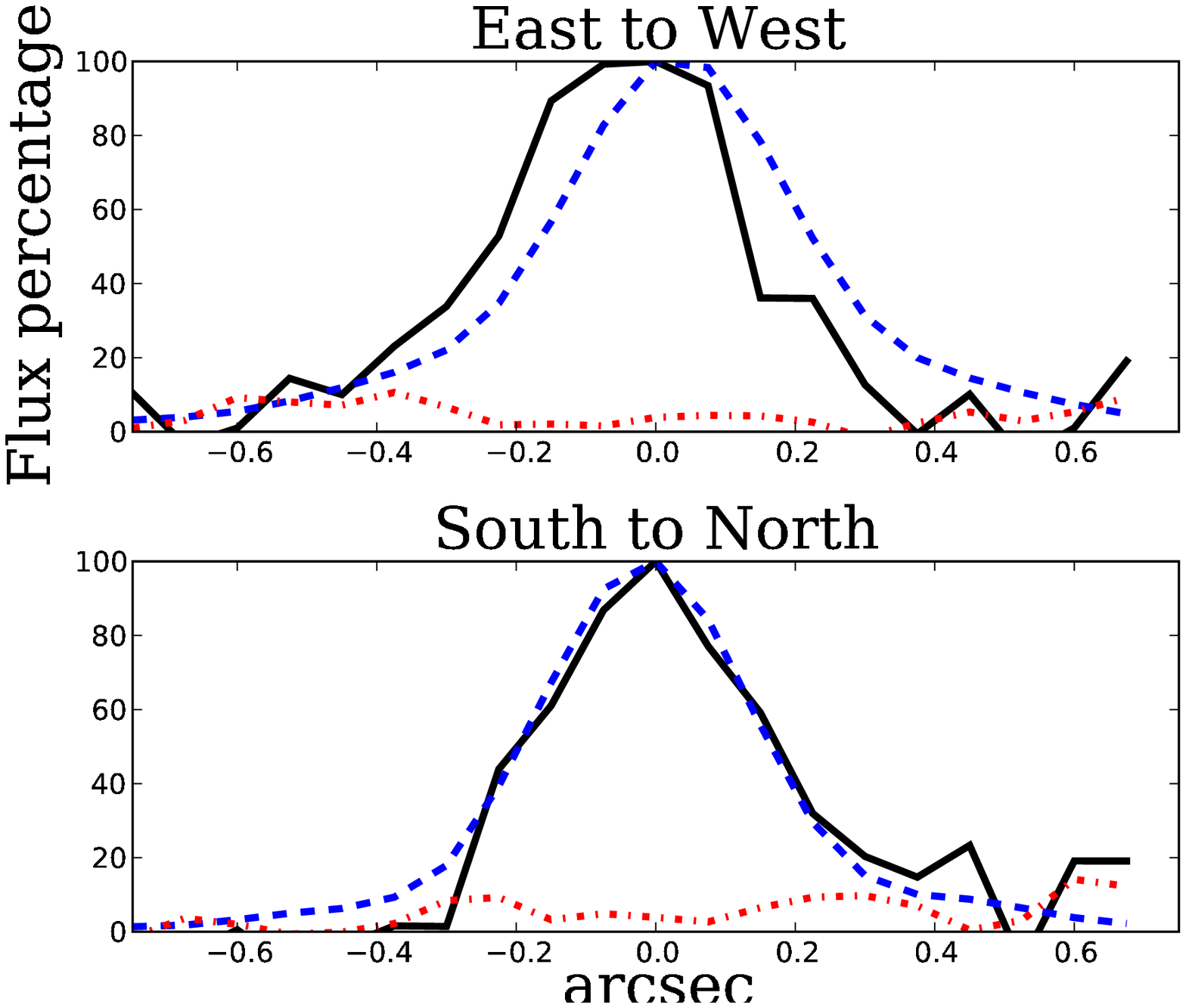} &
\includegraphics[width=2.8cm]{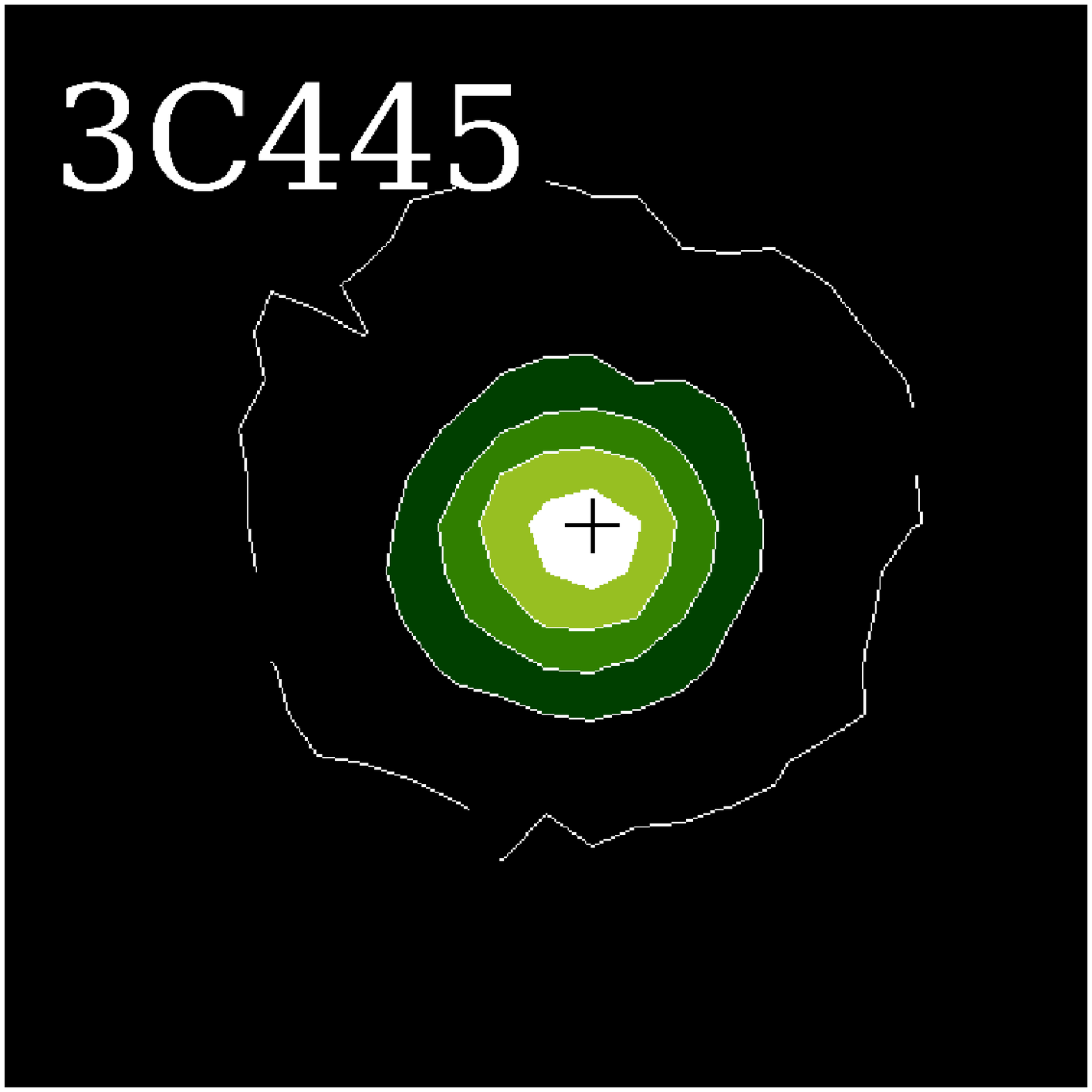} &
\includegraphics[width=2.8cm]{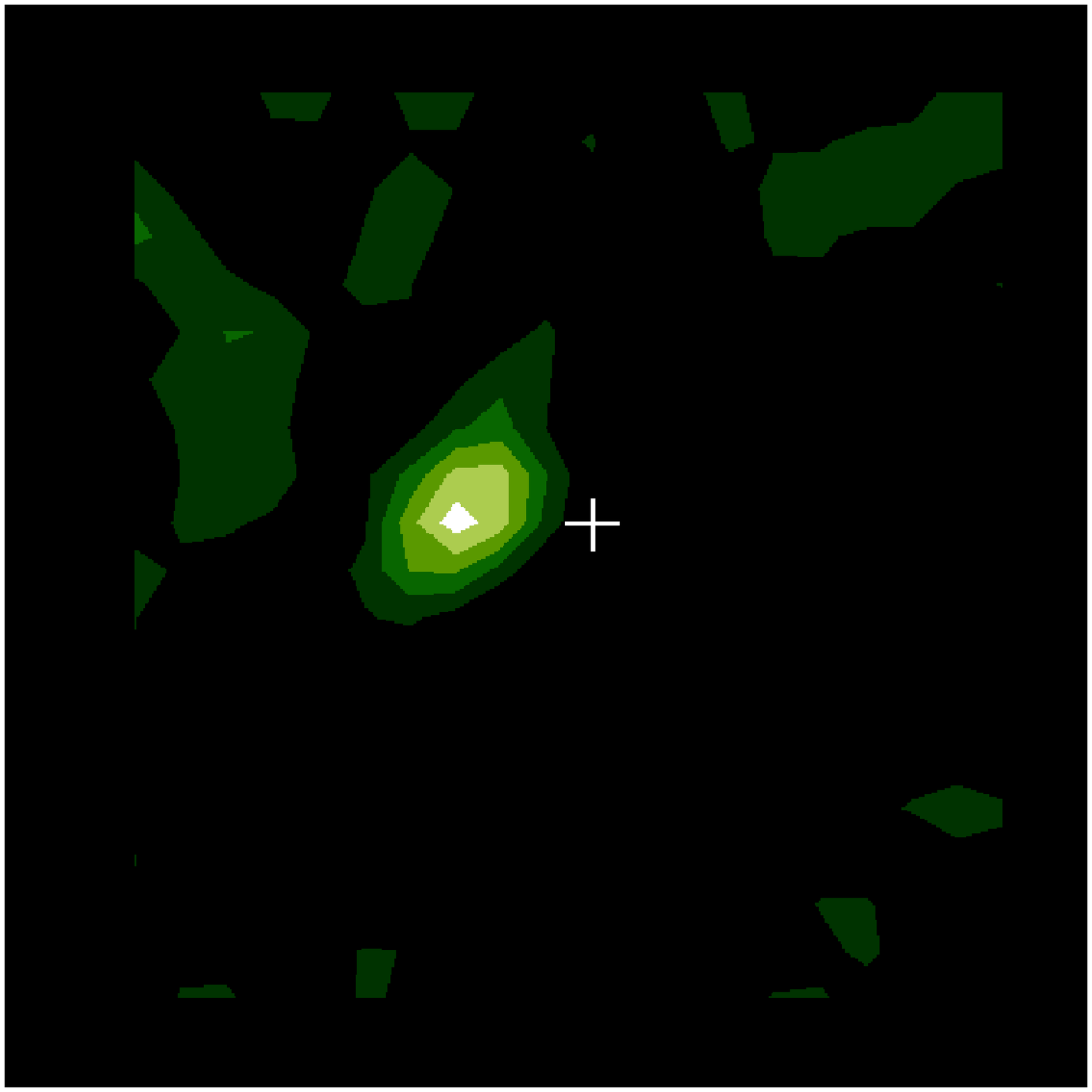} &
\includegraphics[width=3.4cm]{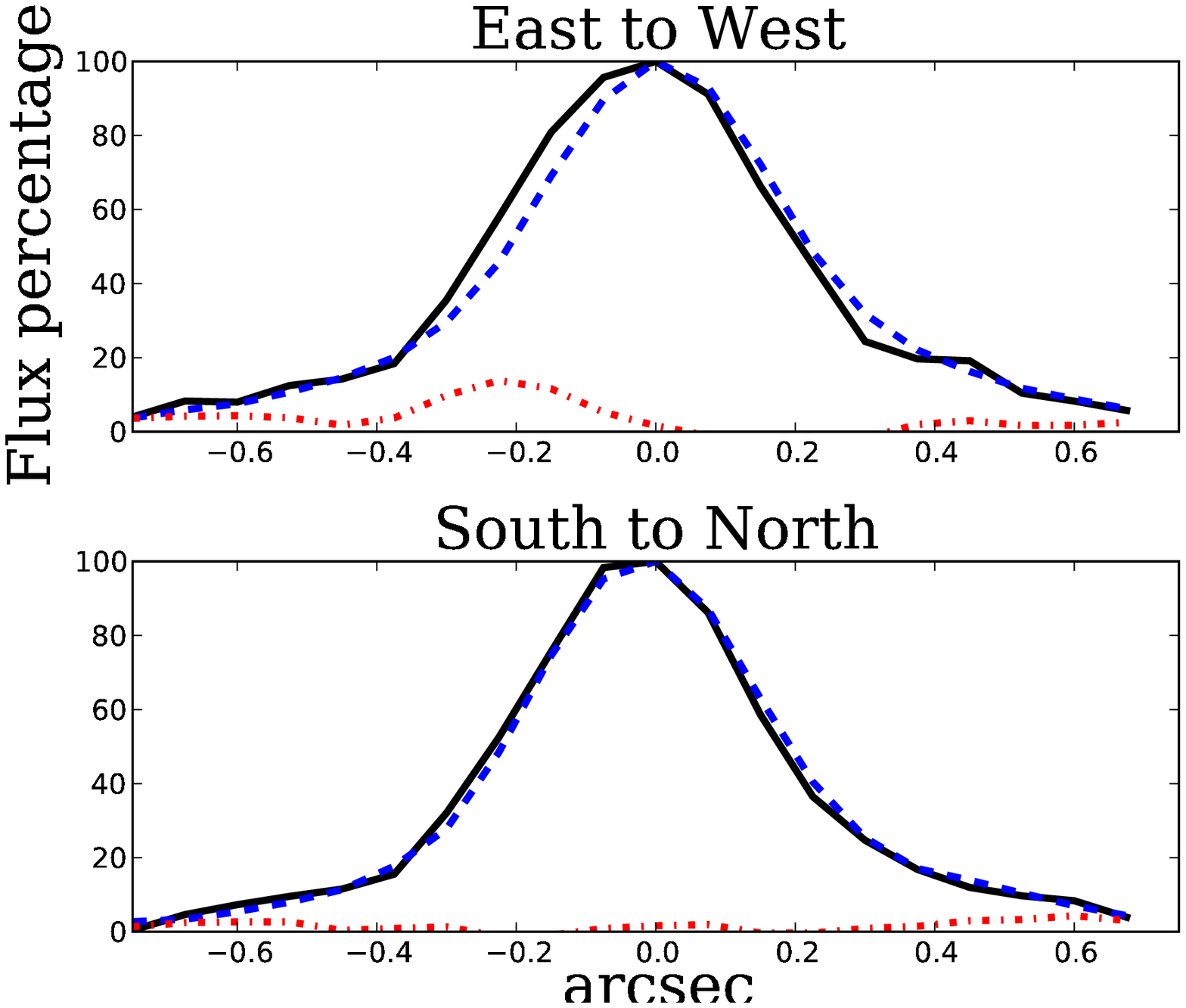}  \\
\includegraphics[width=2.8cm]{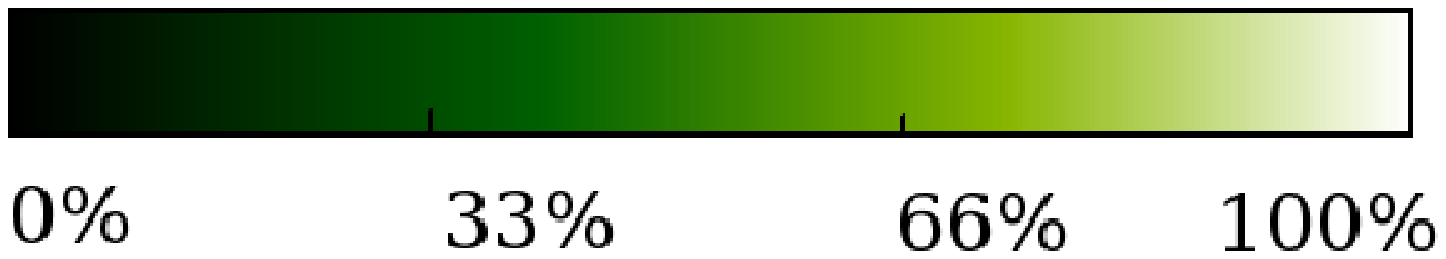} &
\includegraphics[width=2.8cm]{12435fg2ae.ps} &
\includegraphics[width=3.4cm]{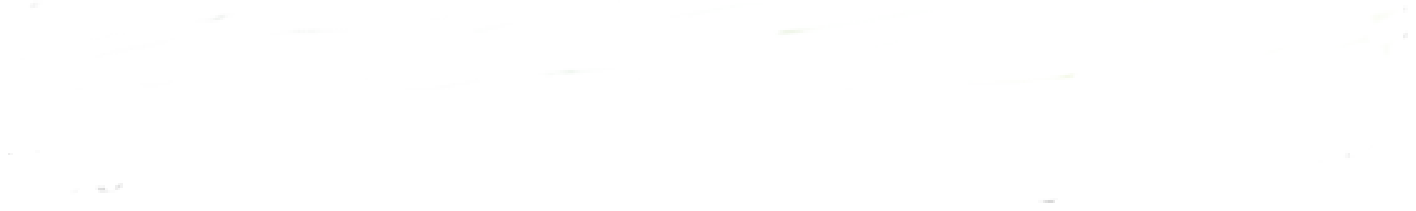} &
\includegraphics[width=2.8cm]{12435fg2ae.ps} &
\includegraphics[width=2.8cm]{12435fg2ae.ps} &
\includegraphics[width=3.4cm]{12435fg2af.ps} \\
\end{array}
\end{displaymath}
\caption{The raw (left) and PSF-subtracted (middle) VISIR N-band images and the radial profiles (right) from East to West and North to South of the object (black solid line), PSF (blue dashed line) and the residual emission (red dash-dotted line). The images are 1.5~\arcsec~$\times$~1.5\arcsec~in size, North is up, East is left. Contour levels are in steps of 17 and 20~percent of the peak flux in respectively the PSF-subtracted and raw image.}
\label{psfsubtraction}
\end{figure*}

To search for extended dust structures in the detected sources, we subtracted the PSF obtained by averaging standard stars observed close in time to the sources. The raw images, extended dust structure maps, and radial profiles are shown in Fig.~\ref{psfsubtraction}. Our PSF-subtraction method subtracts the PSF from the photometric center of the sources. Extended structures contributing up to 10~percent of the flux are detected. These prominent extended structures are found mostly in the BLRGs.

\begin{figure*}
\label{allsedsFR-I}
\centering
\begin{displaymath}
\begin{array}{ccc}
\includegraphics[width=6cm]{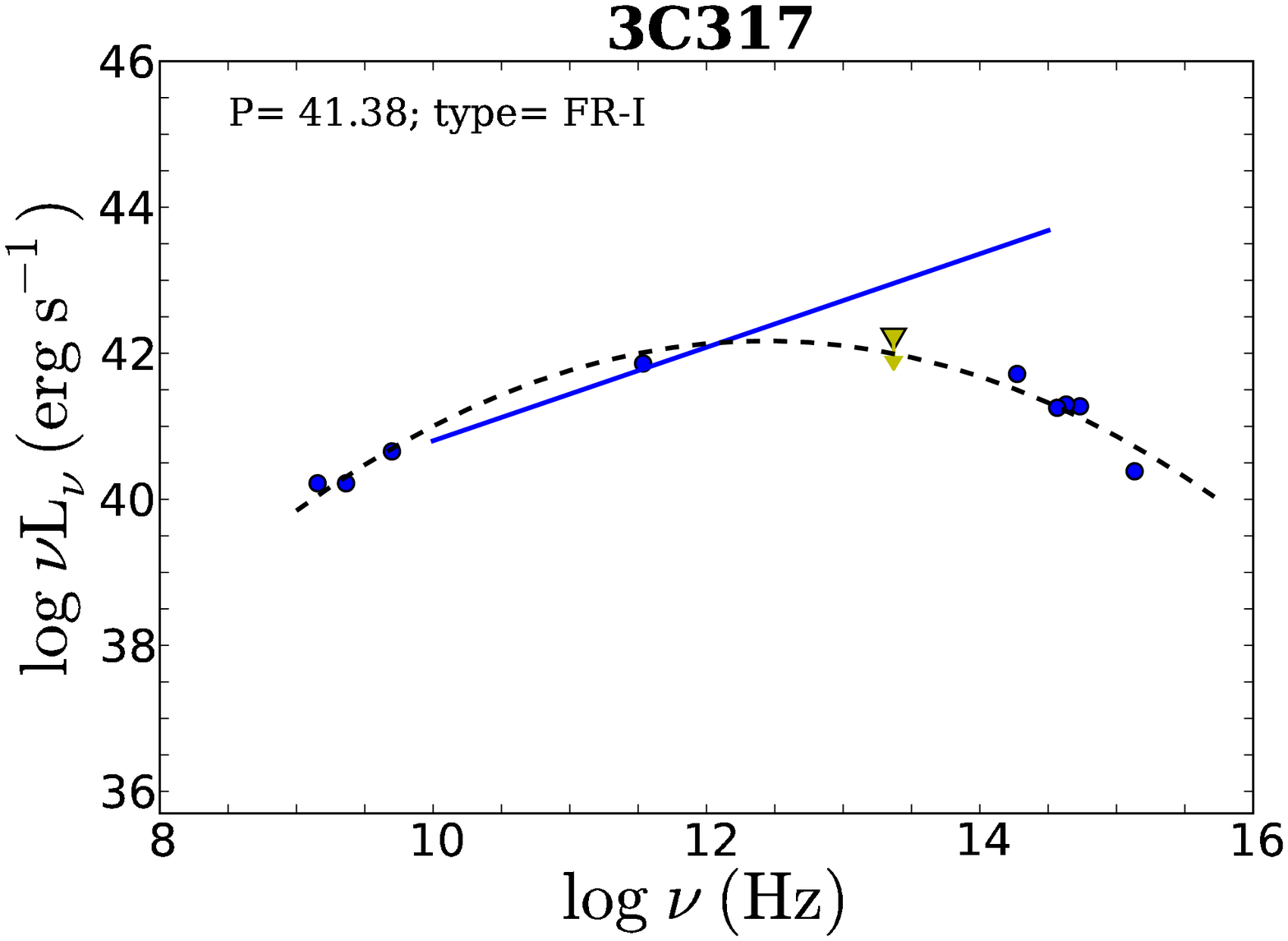} &
\includegraphics[width=6cm]{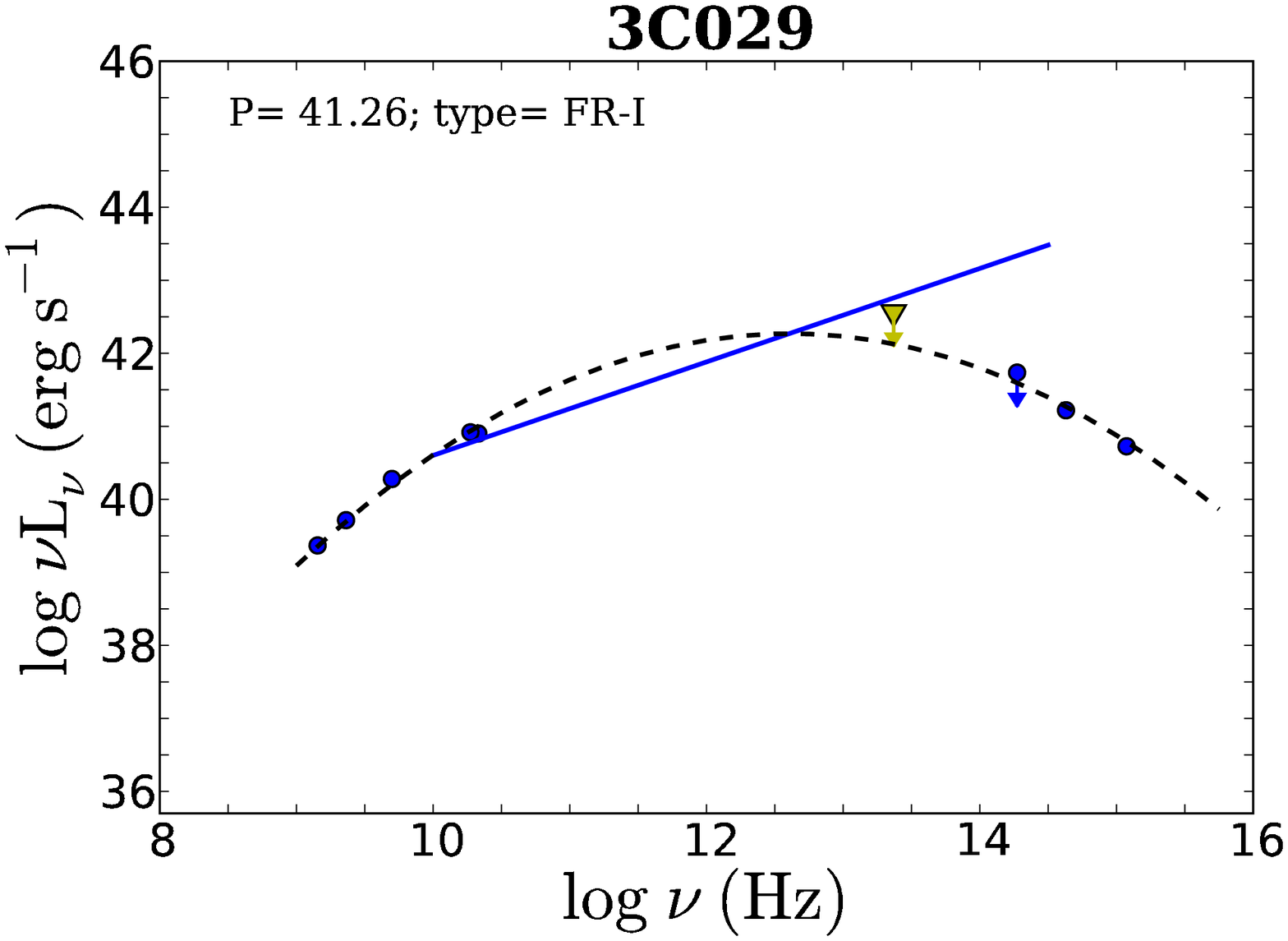} &
\includegraphics[width=6cm]{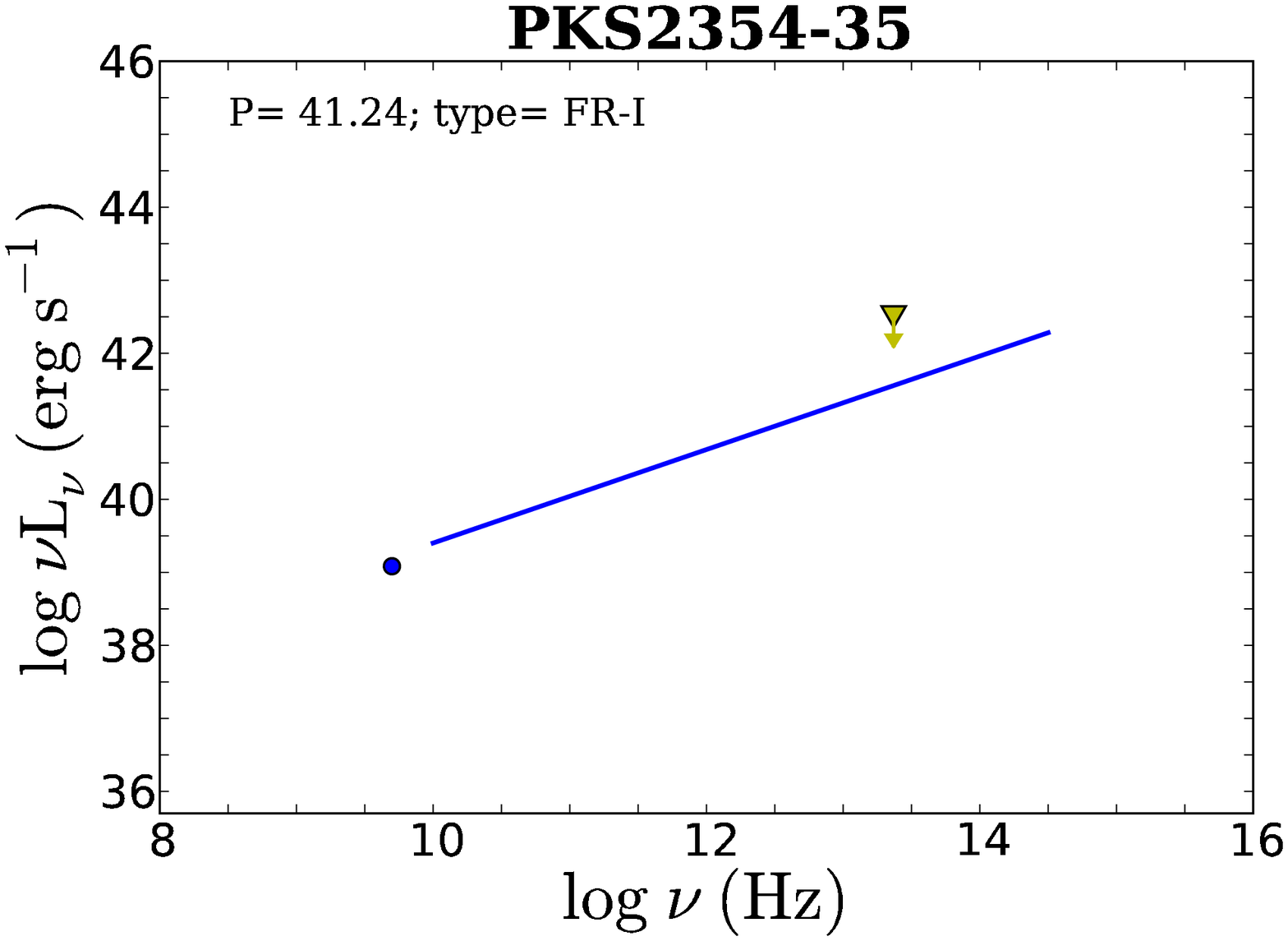} \\
\includegraphics[width=6cm]{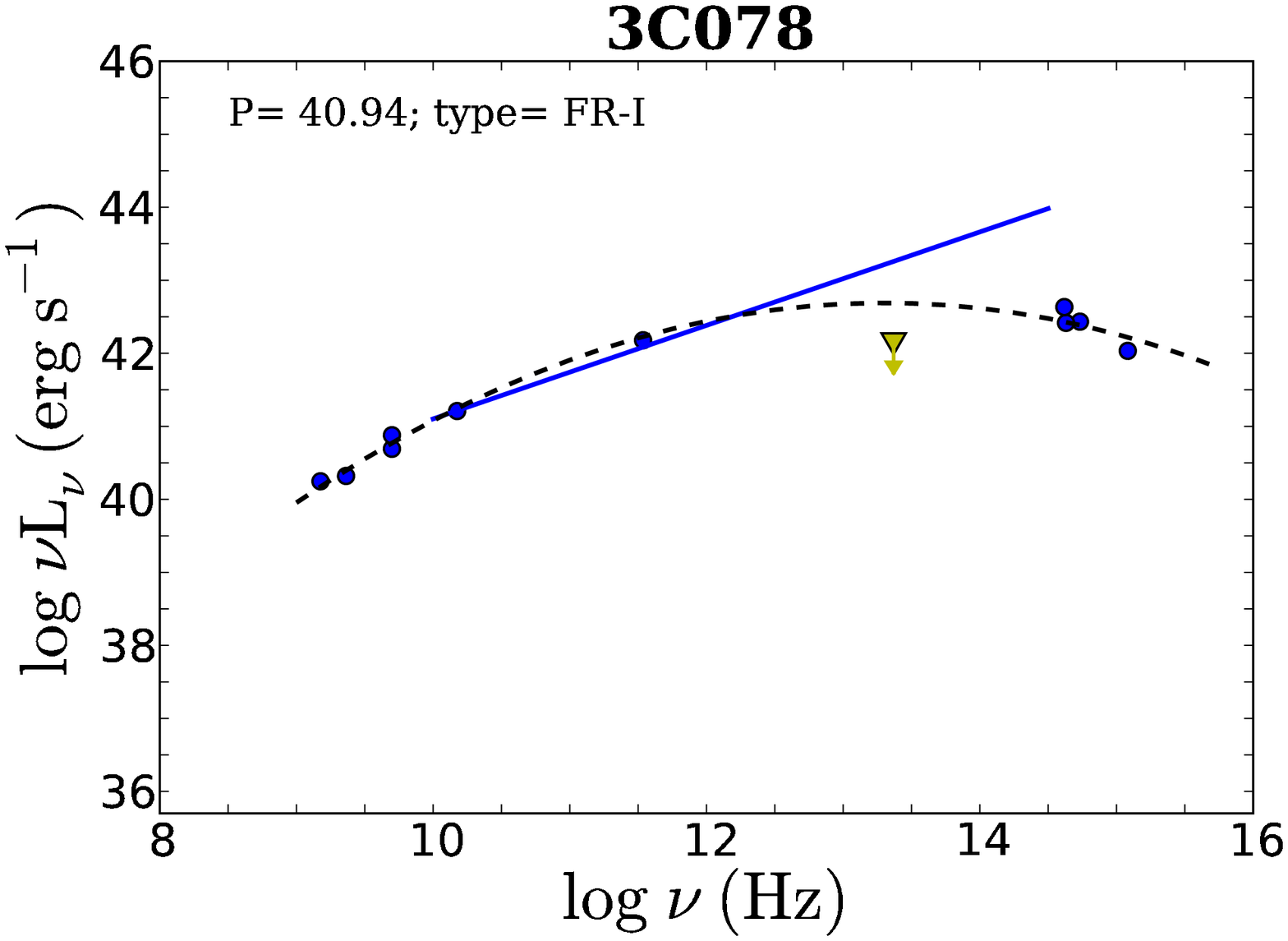} &
\includegraphics[width=6cm]{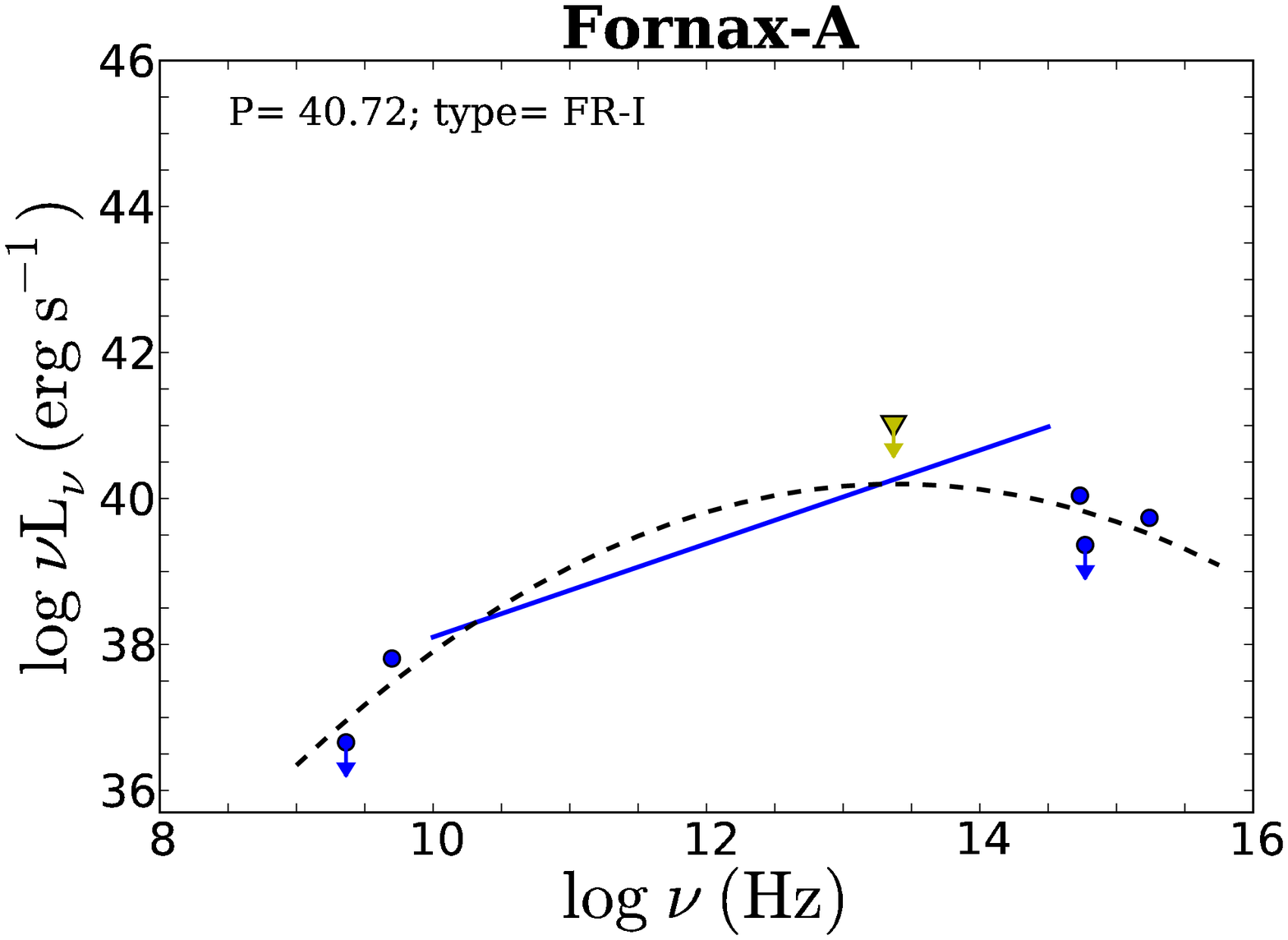} &
\includegraphics[width=6cm]{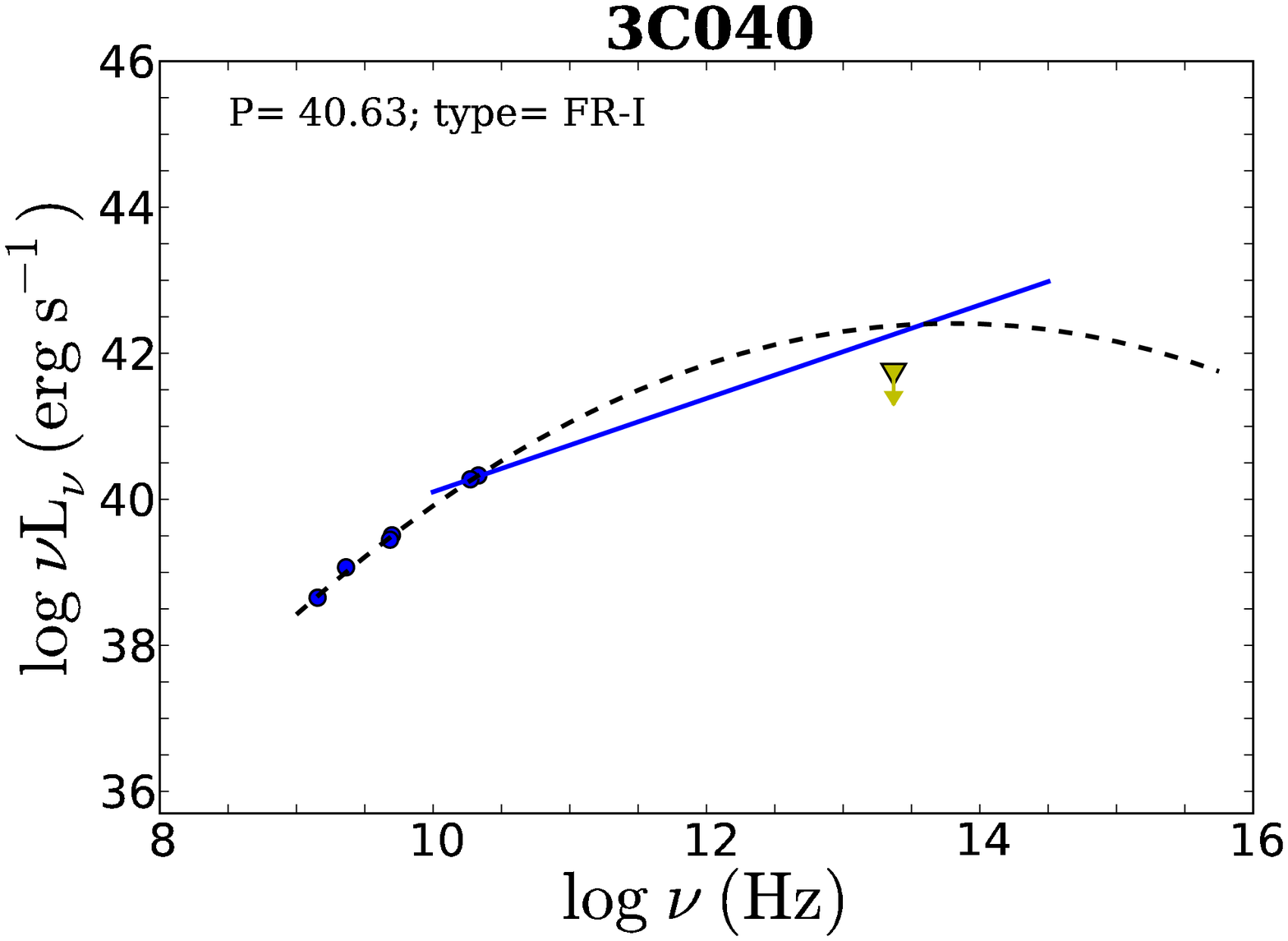} \\
\includegraphics[width=6cm]{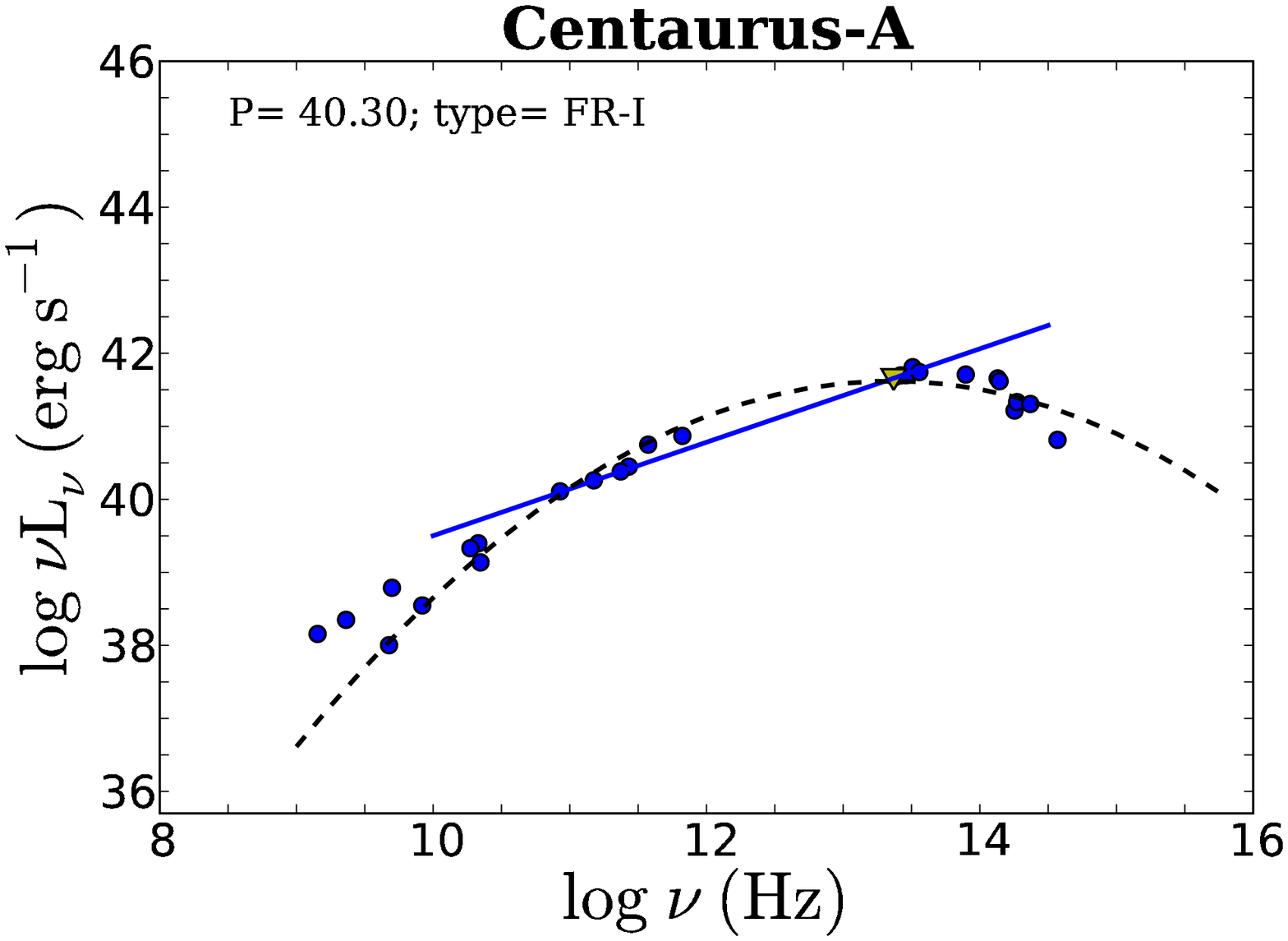} &
\includegraphics[width=6cm]{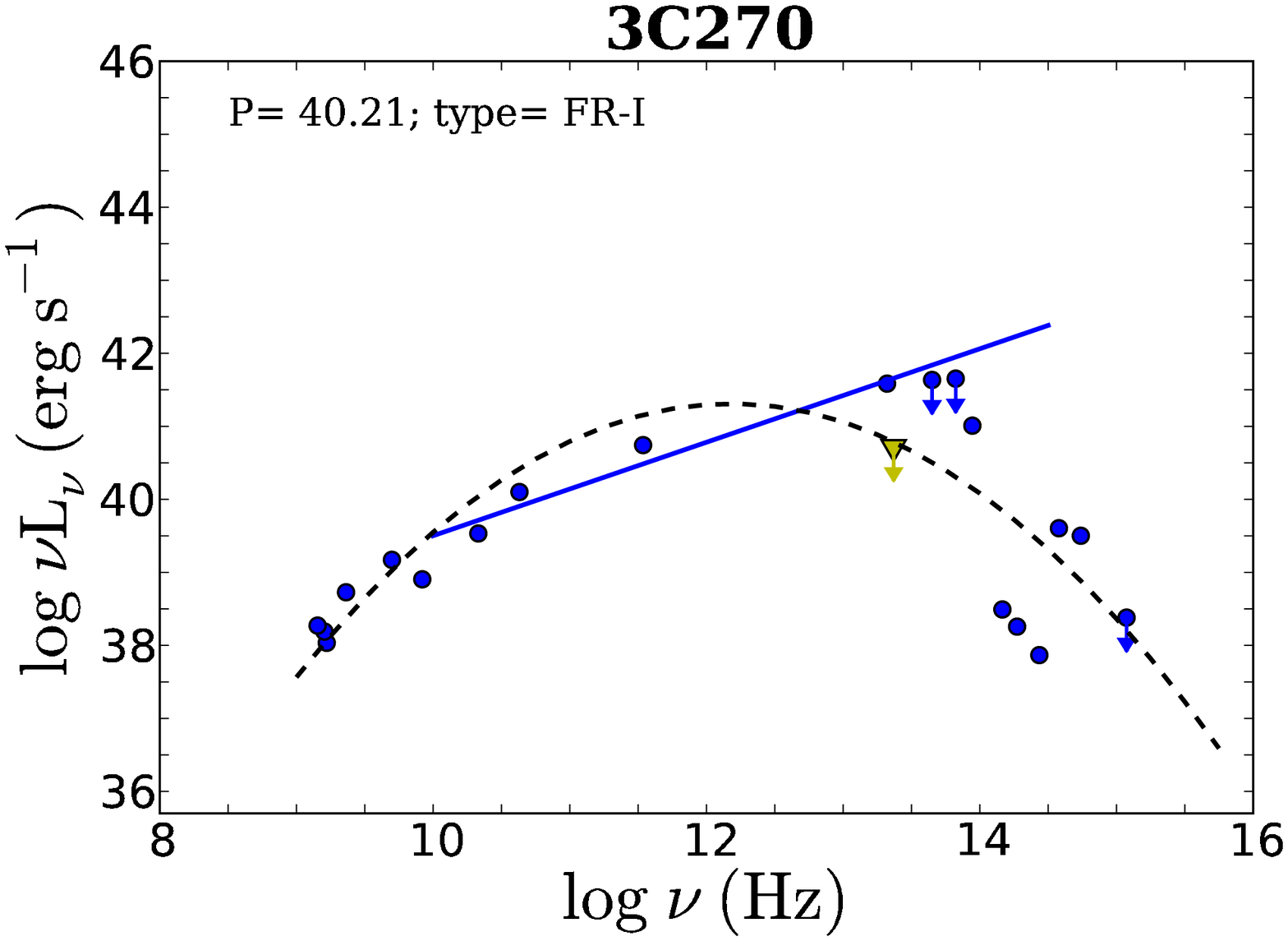} &
 \\
\end{array}
\end{displaymath}
\caption{Core spectral energy distributions of the FR-I radio galaxies ordered by decreasing total radio power from left to right. The high frequency radio core data are fitted by a straight power-law synchrotron spectrum $\nu L_{\nu} \propto \nu^{0.64}$, while the combined data points (filled circles) ranging from radio to UV frequencies, except the mid-infrared VISIR point (filled triangle), are fitted by a parabolic function.}
\end{figure*}

\begin{figure*}
\centering
\begin{displaymath}
\begin{array}{ccc}
\includegraphics[width=6cm]{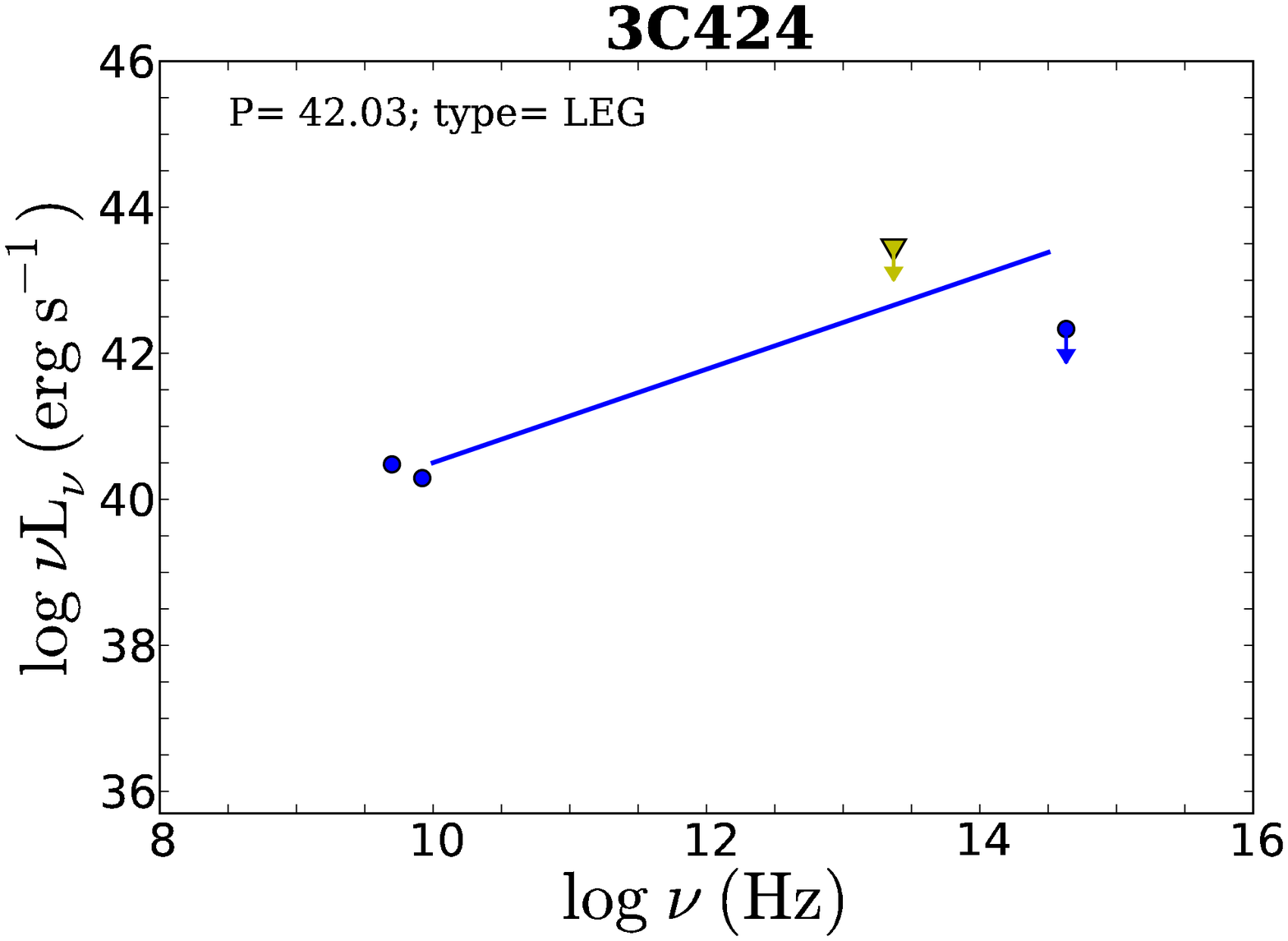} &
\includegraphics[width=6cm]{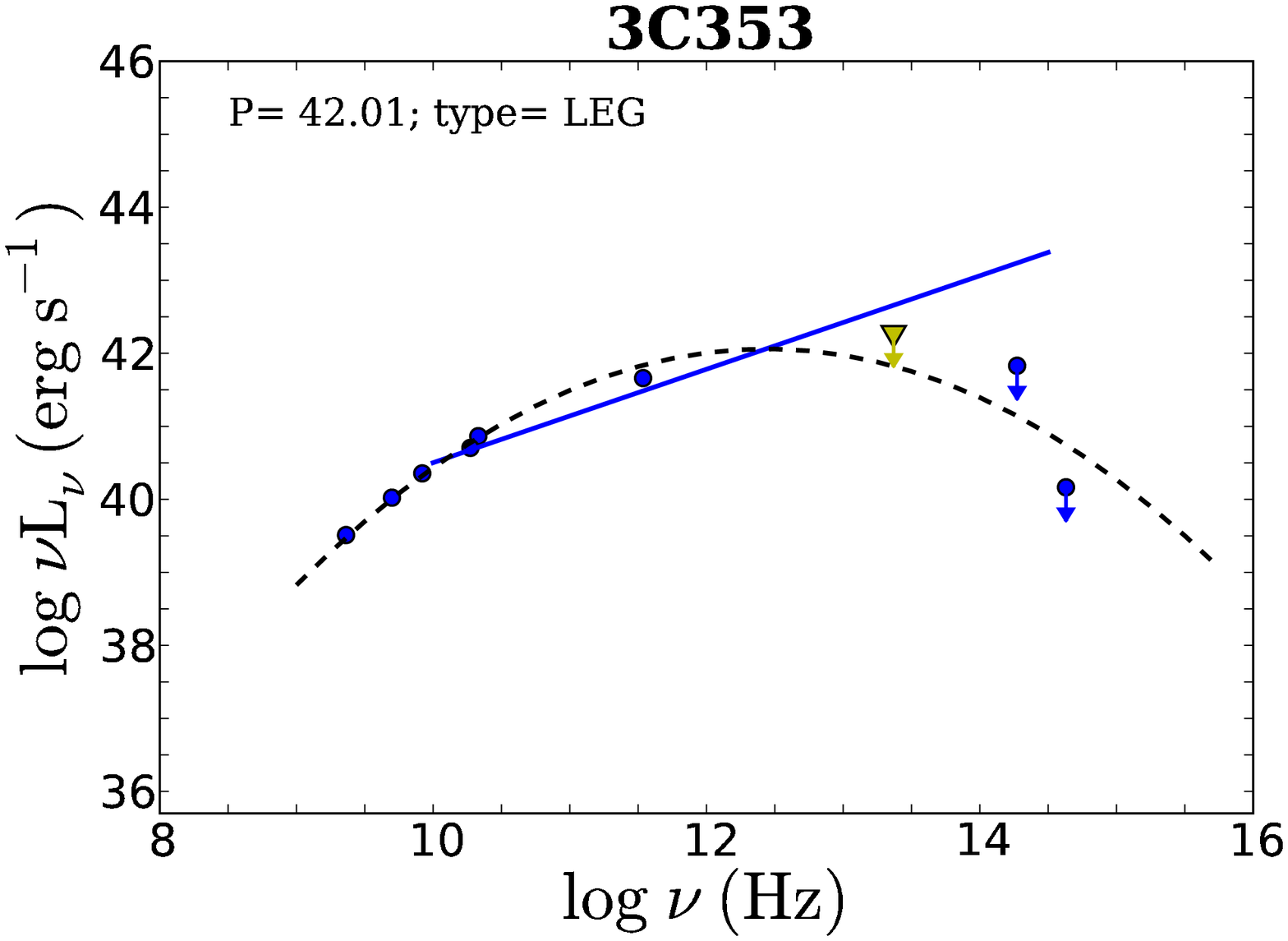} &
\includegraphics[width=6cm]{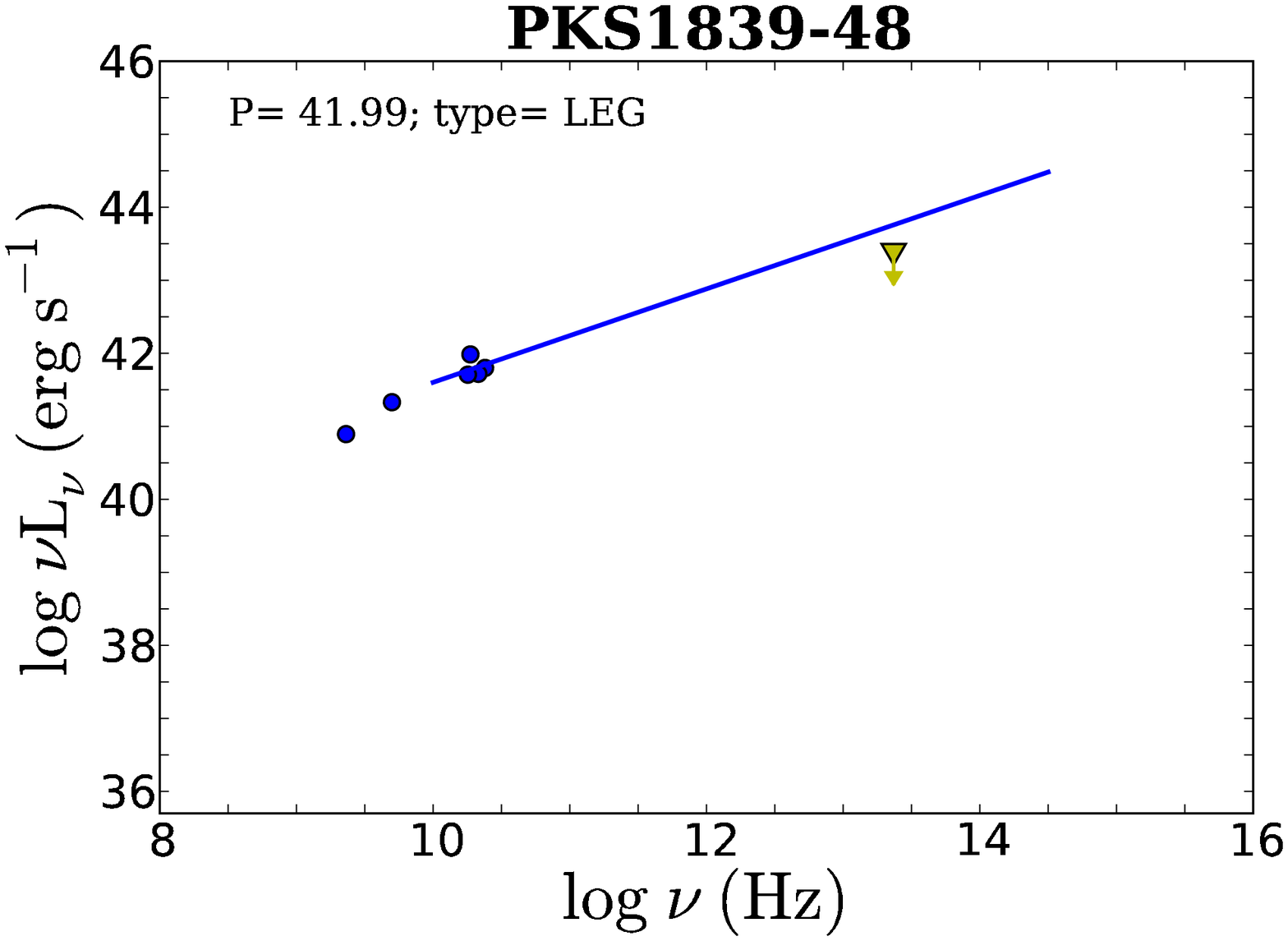} \\
\includegraphics[width=6cm]{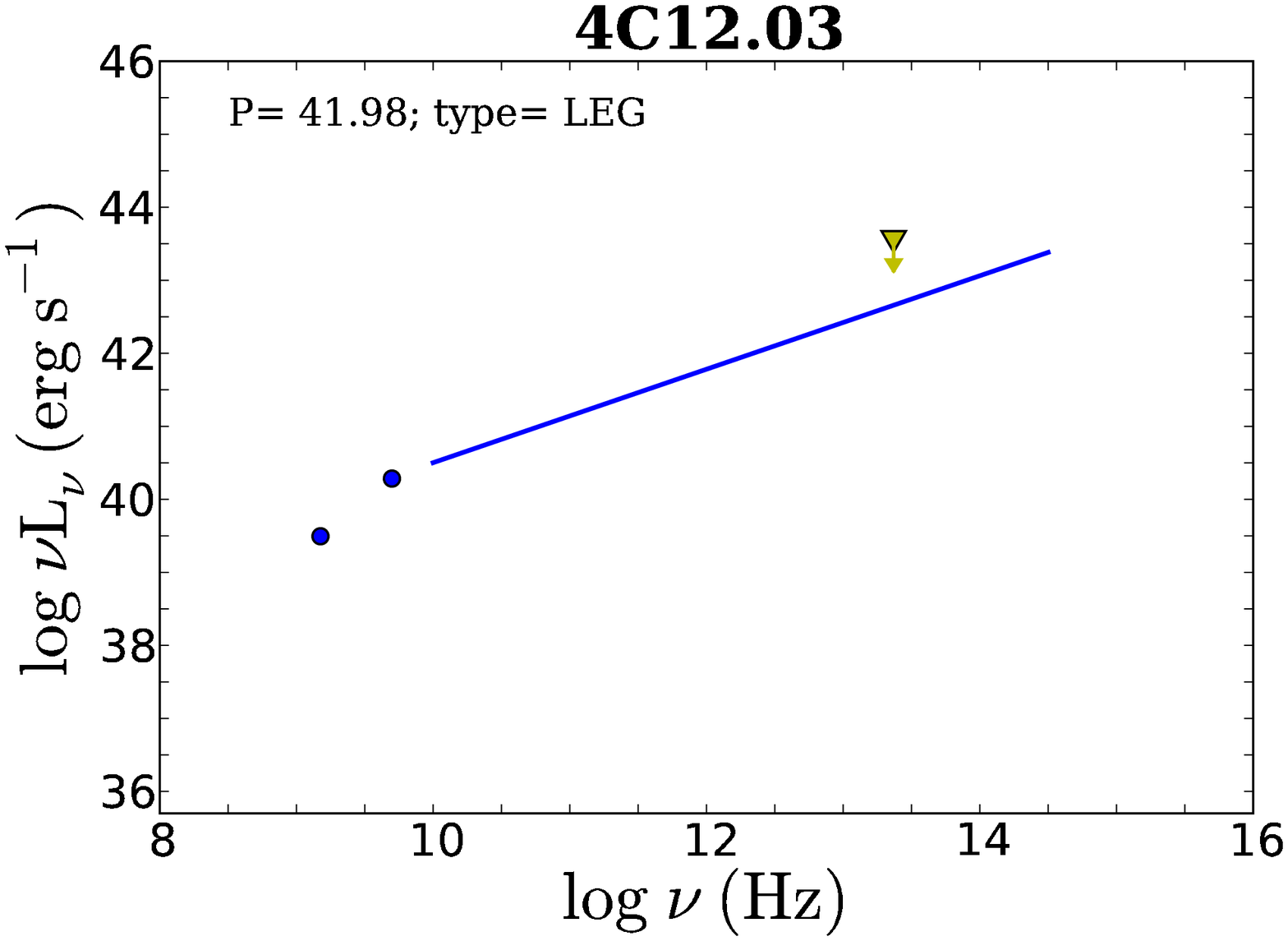} &
\includegraphics[width=6cm]{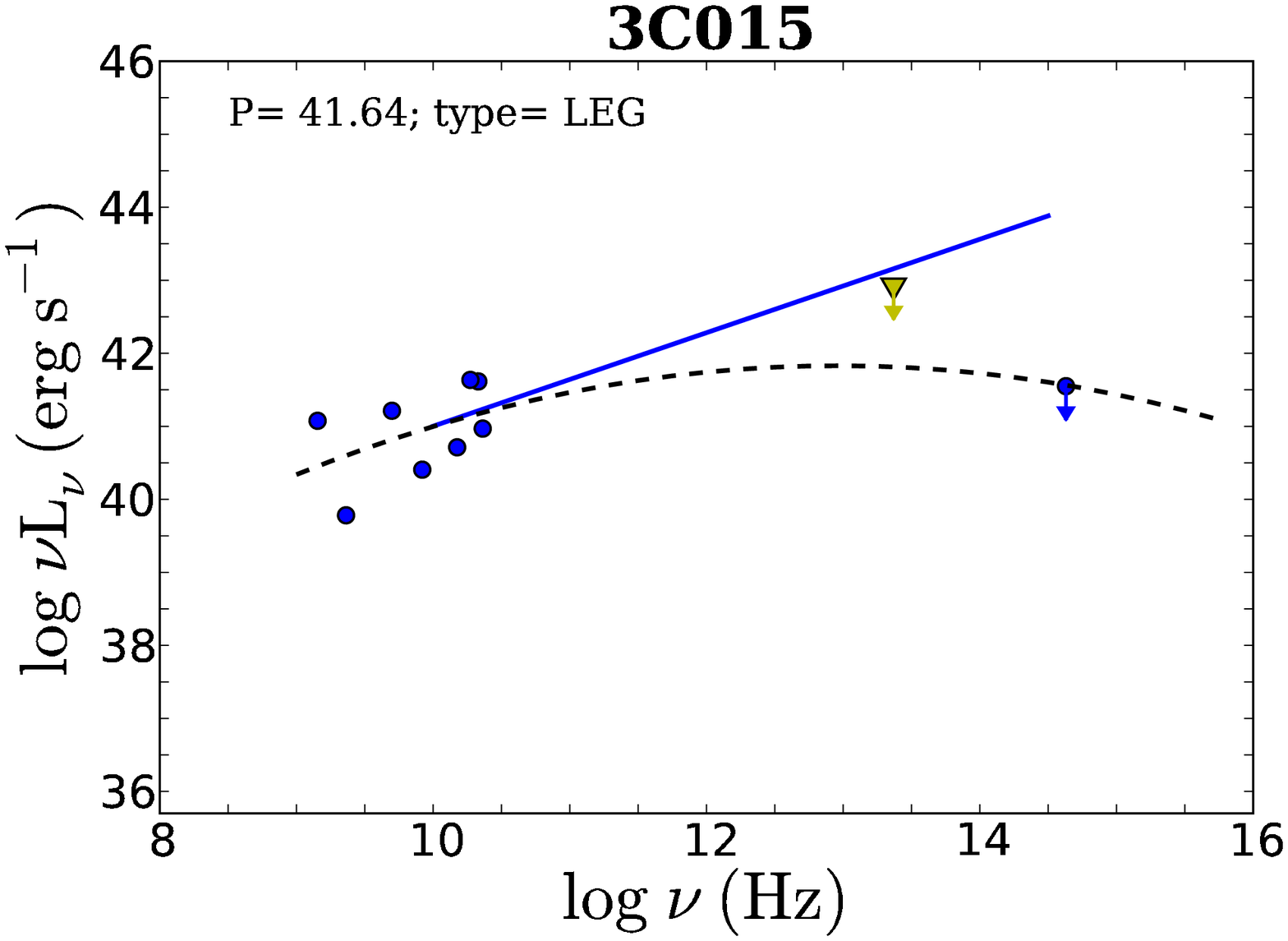} &
\includegraphics[width=6cm]{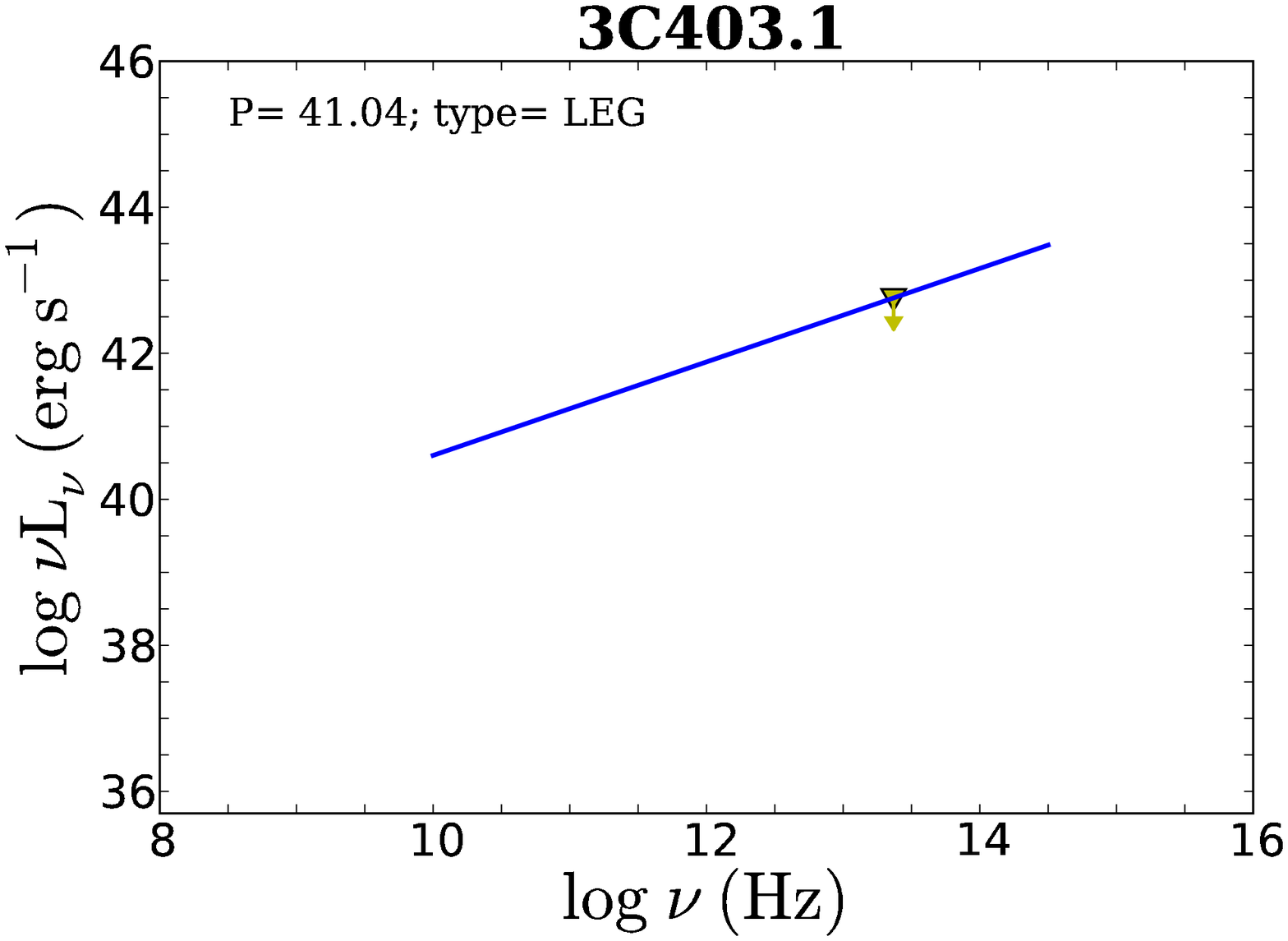} \\
\end{array}
\end{displaymath}
\caption{Same as Fig~\ref{allsedsFR-I}, but now for low-excitation (LEG) FR-II radio galaxies.}
\label{allsedsLEG}
\end{figure*}

\begin{figure*}
\centering
\begin{displaymath}
\begin{array}{ccc}
\includegraphics[width=6cm]{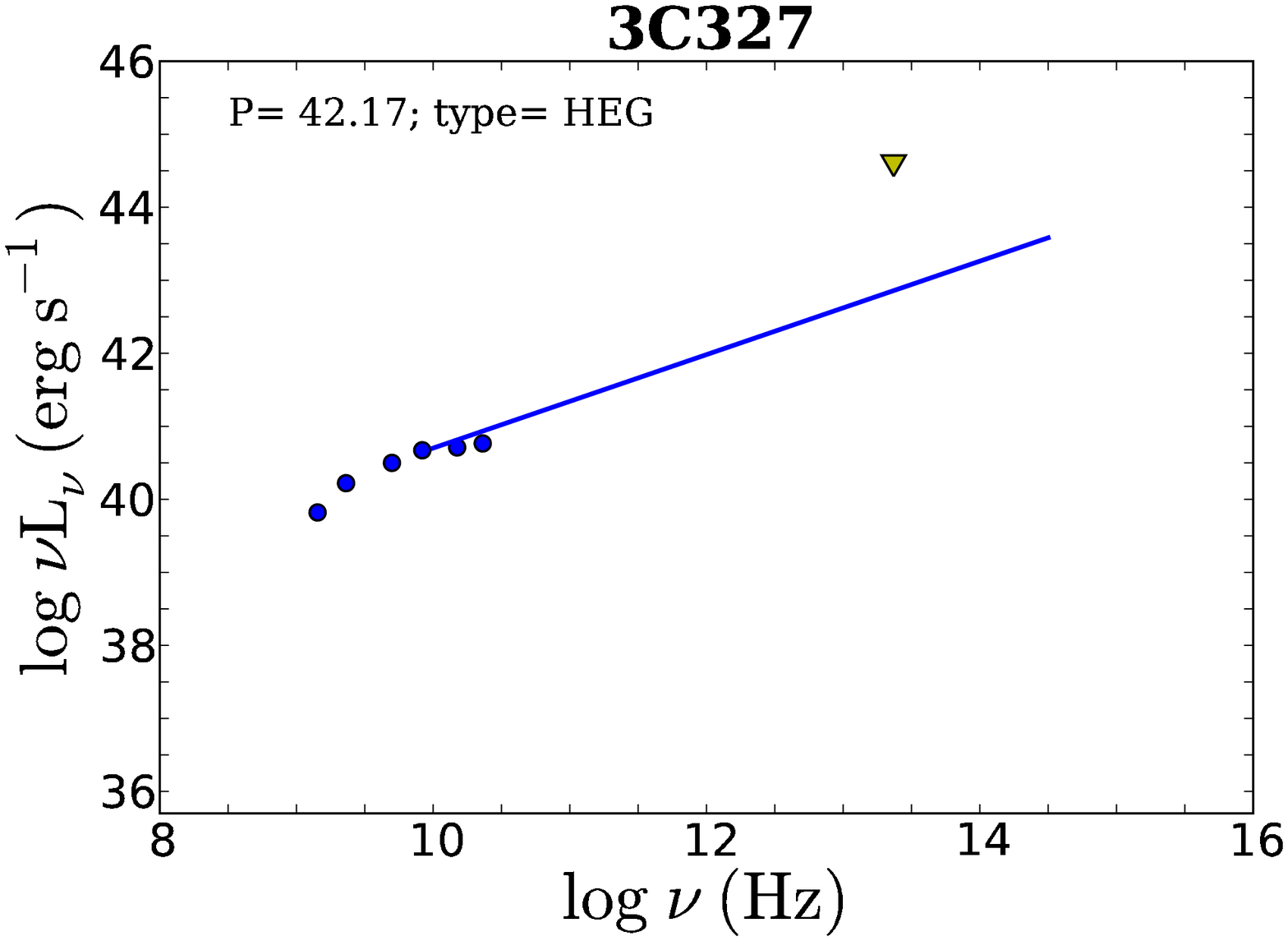} &
\includegraphics[width=6cm]{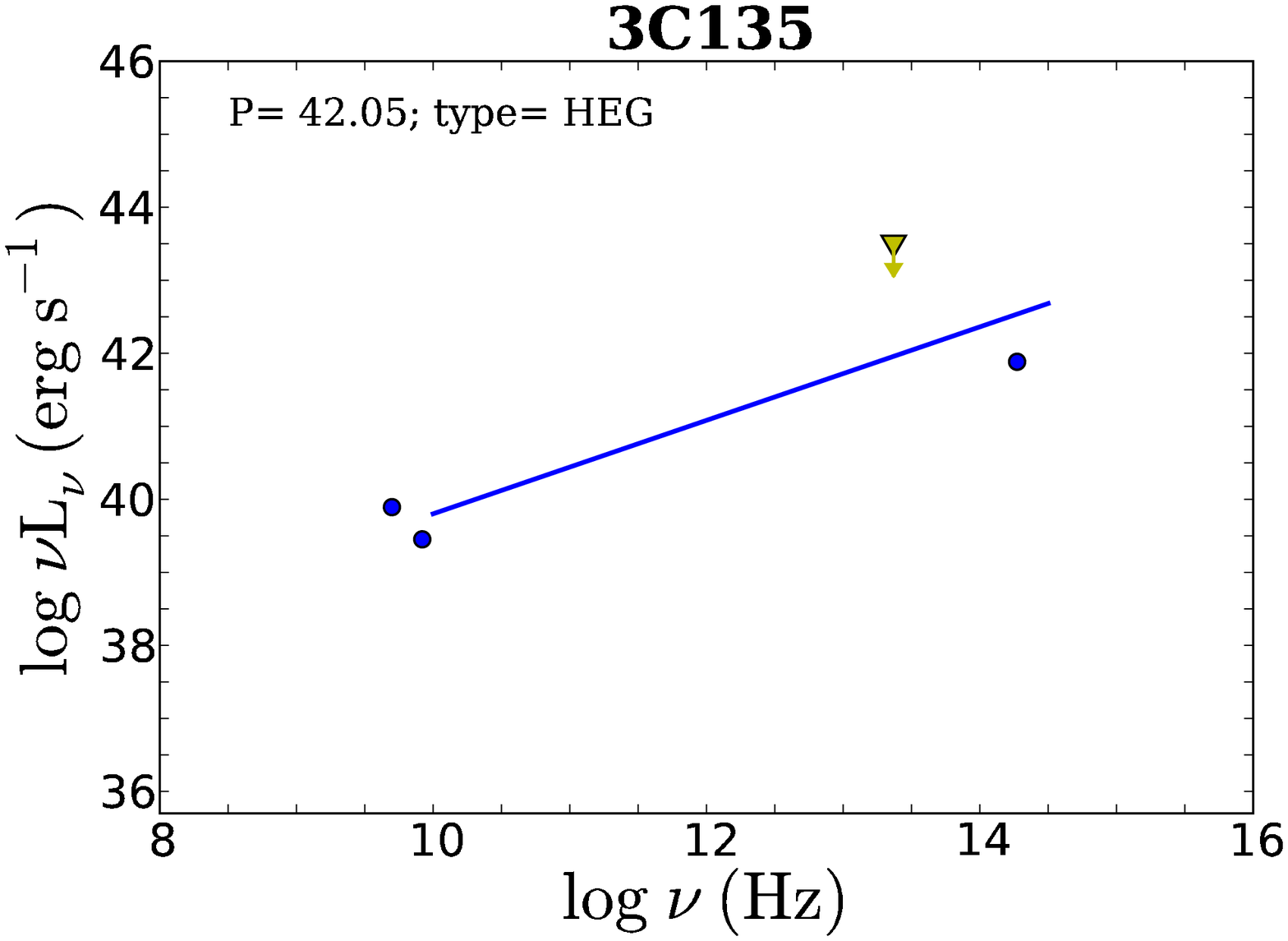} &
\includegraphics[width=6cm]{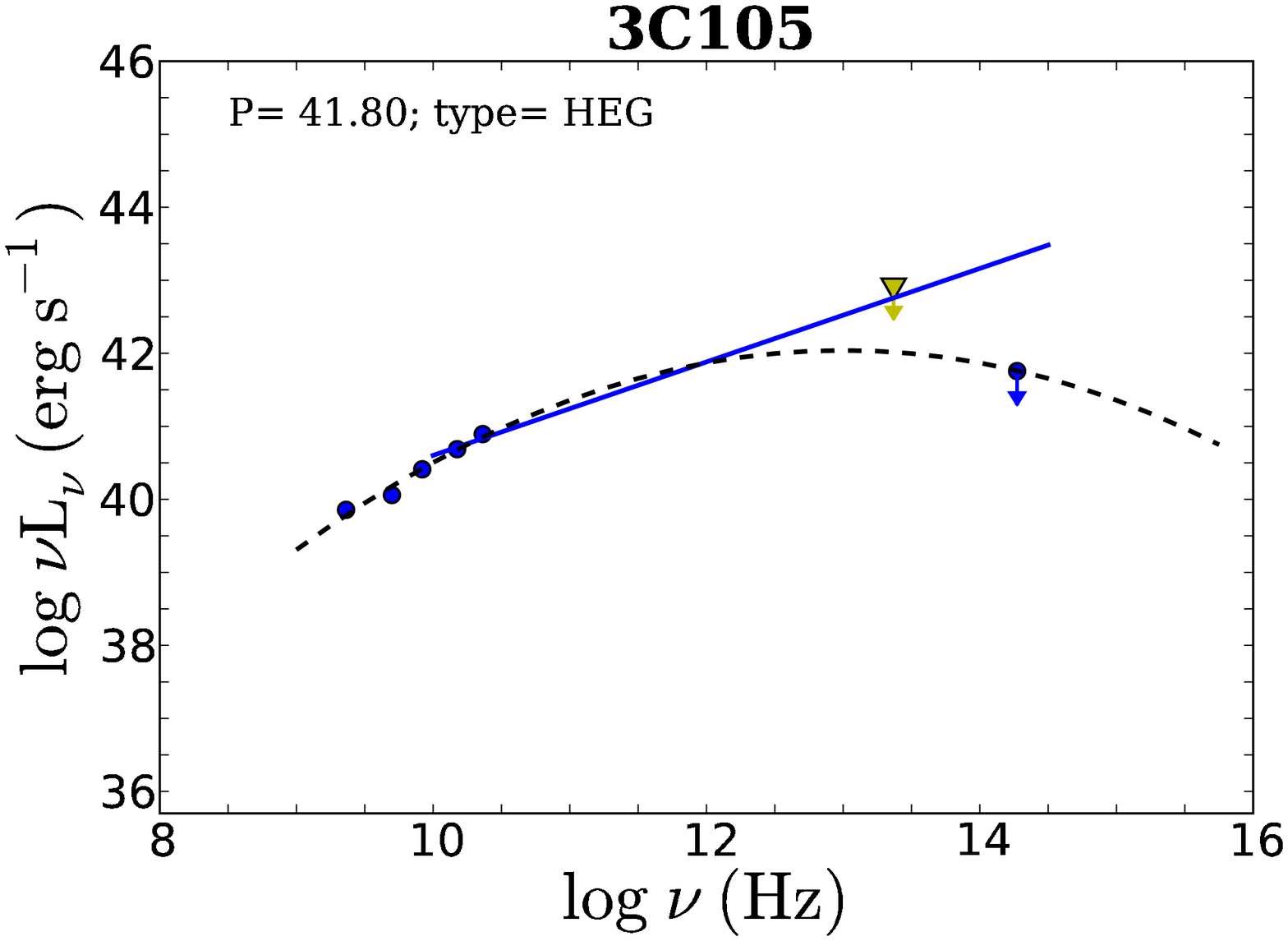} \\
\includegraphics[width=6cm]{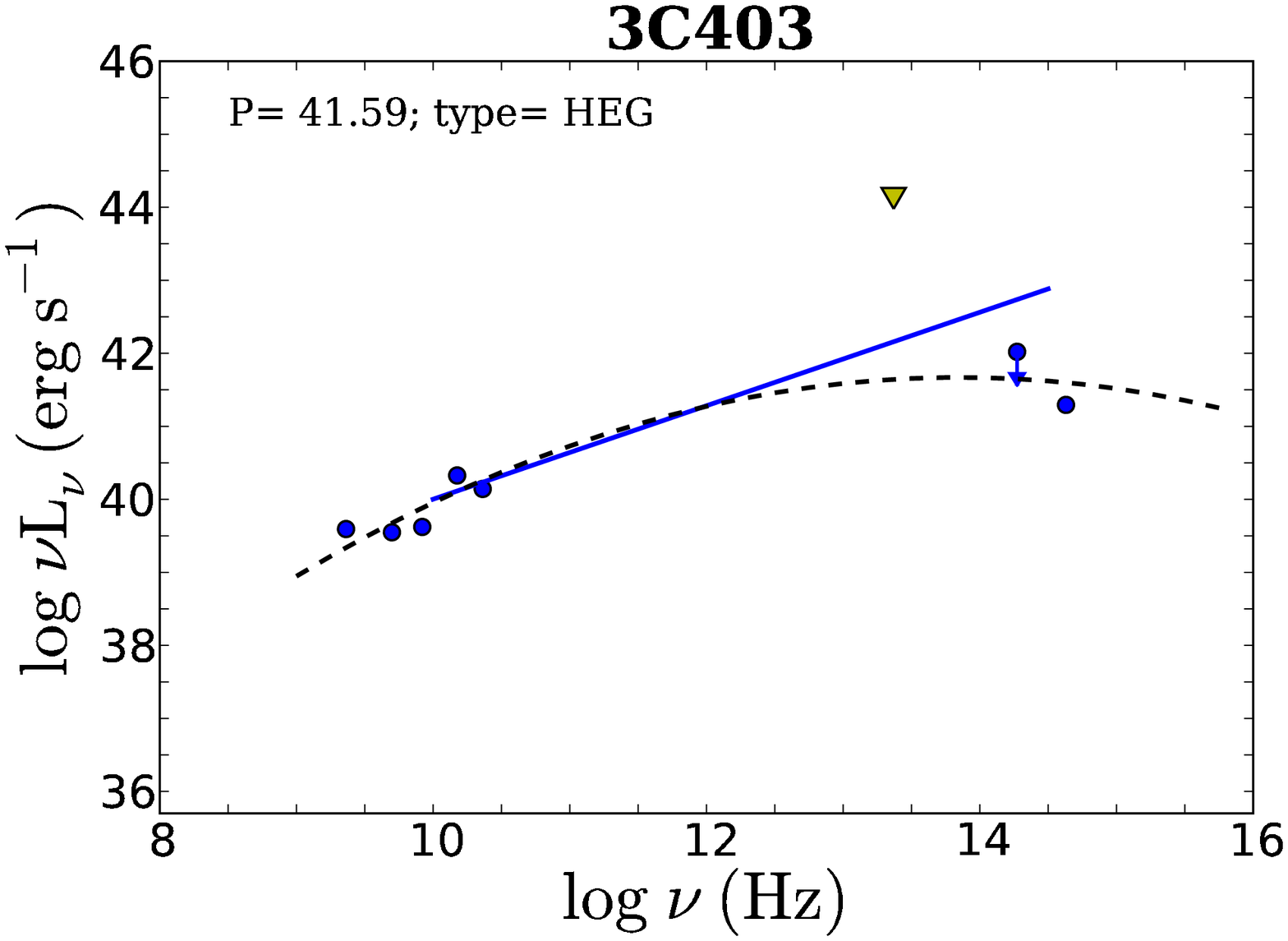} &
\includegraphics[width=6cm]{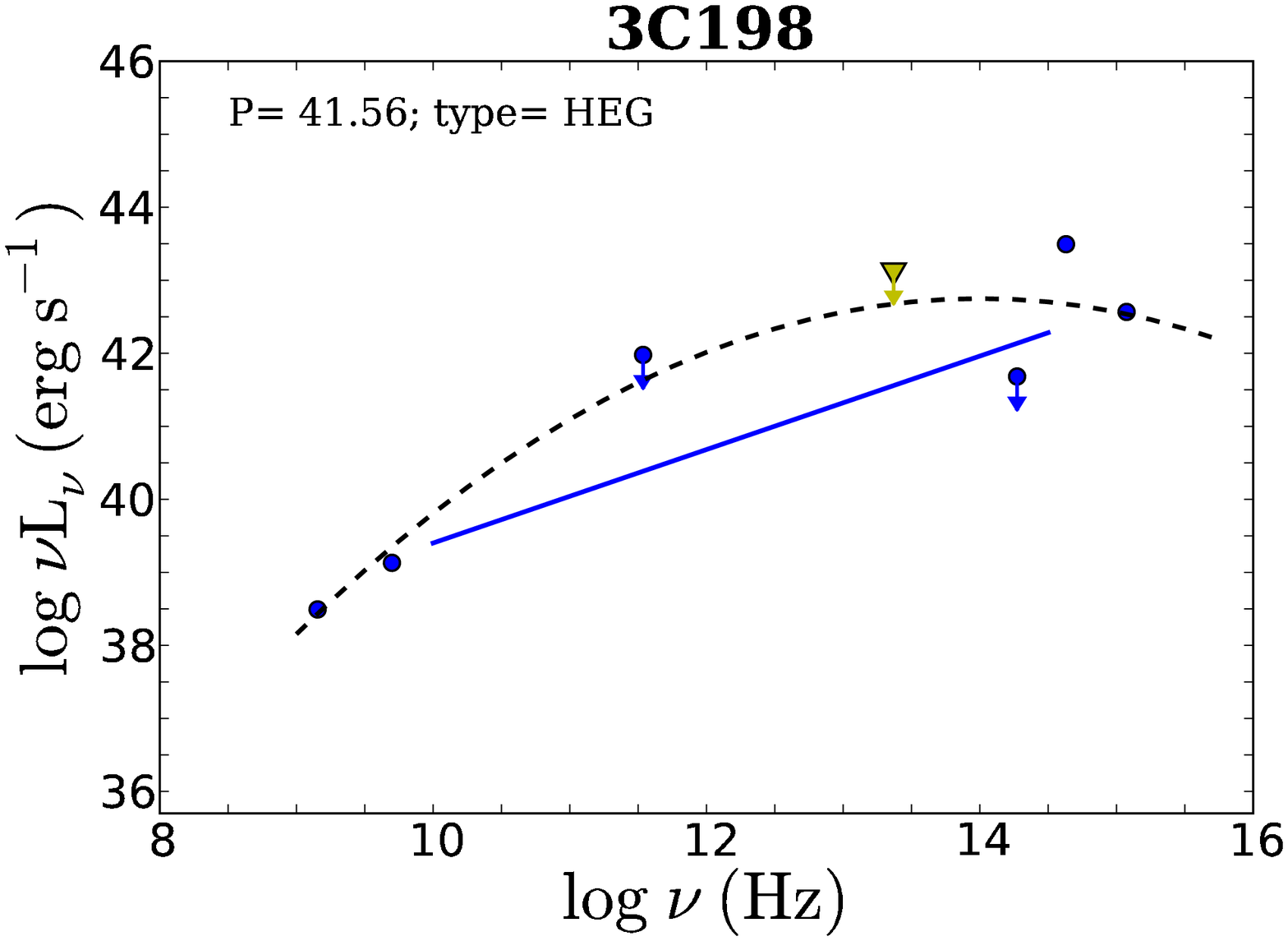} &
\includegraphics[width=6cm]{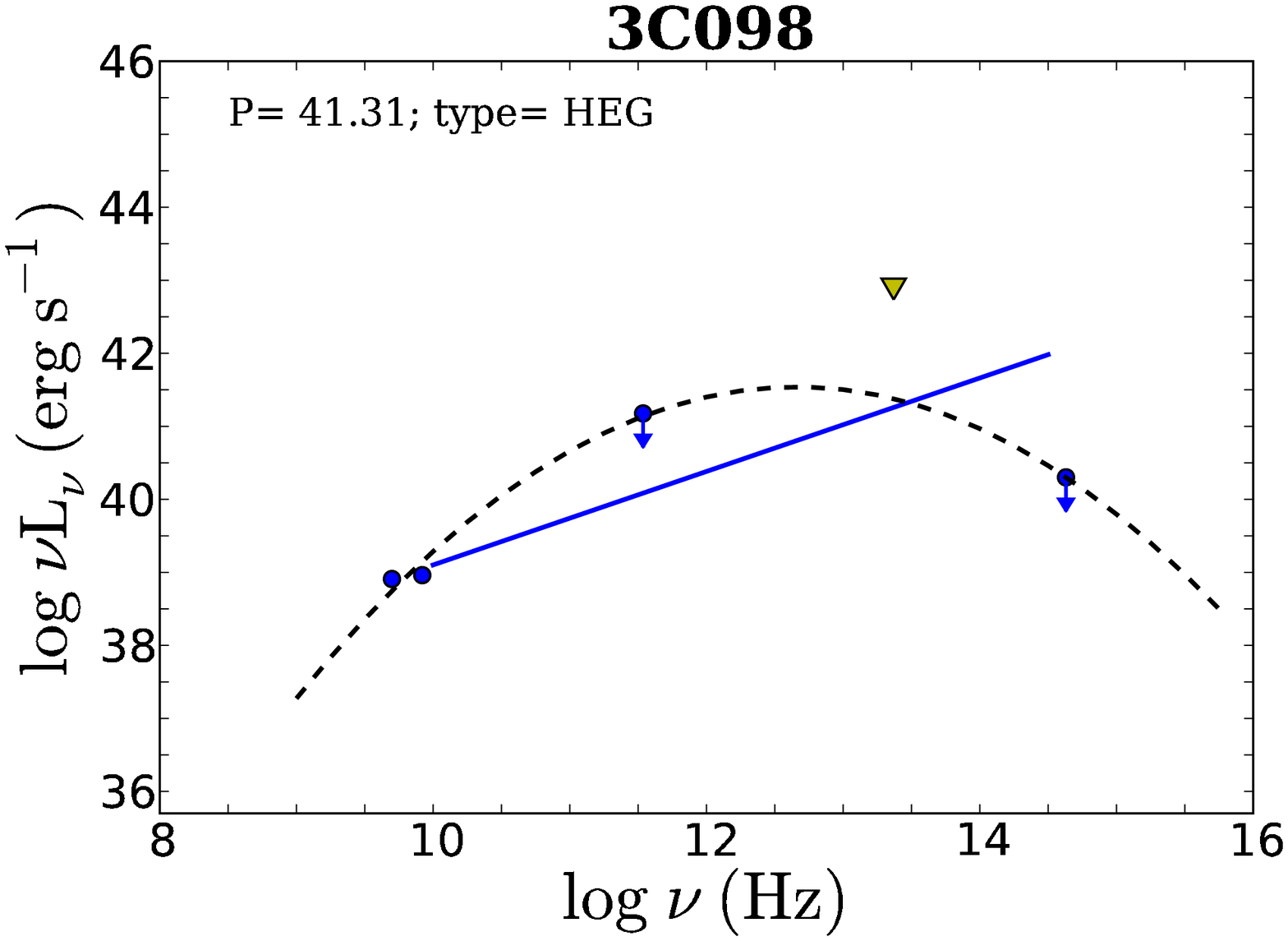} \\
\includegraphics[width=6cm]{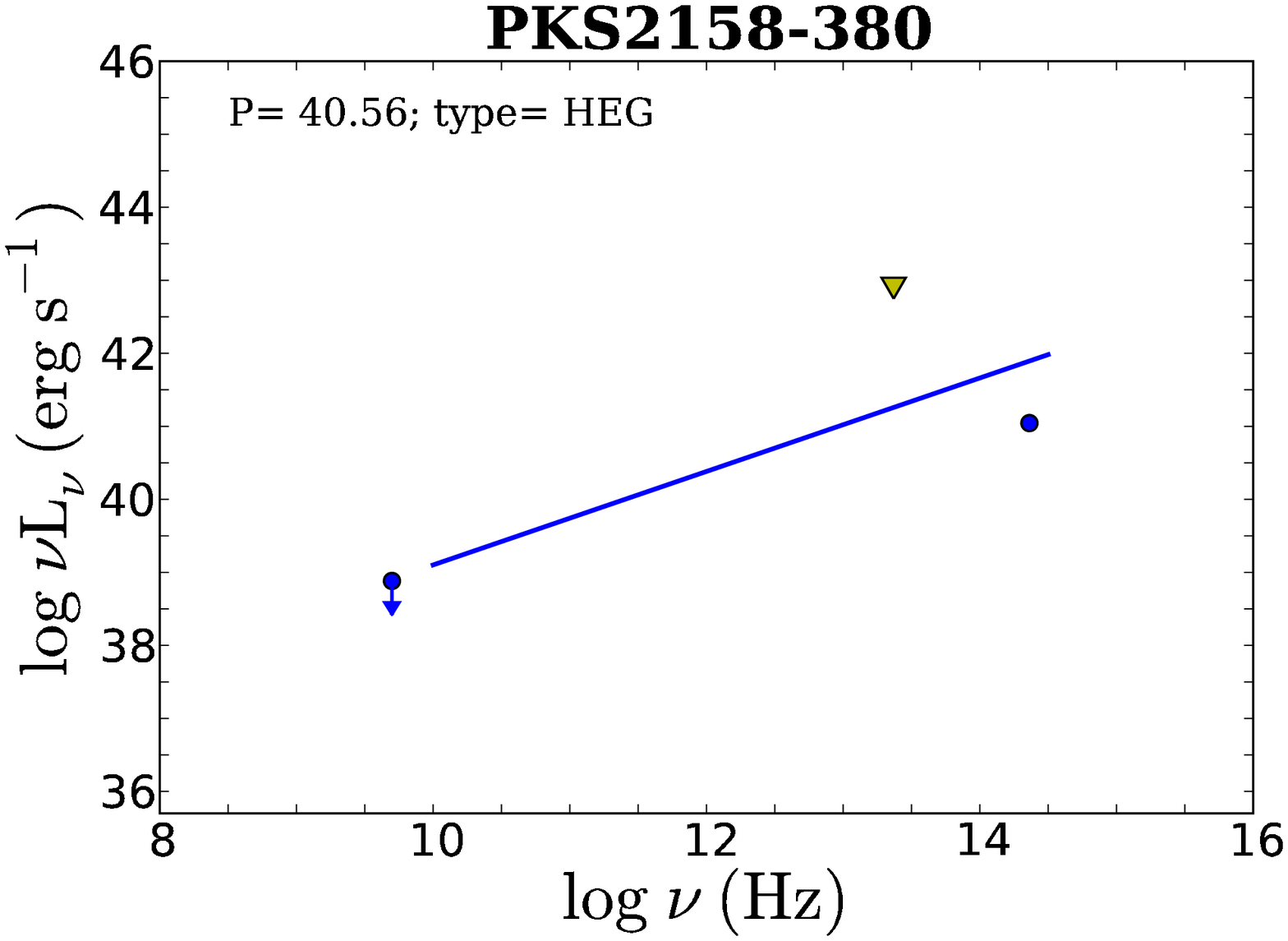} &
 &
 \\
\end{array}
\end{displaymath}
\caption{Same as Fig~\ref{allsedsFR-I}, but now for high-excitation (HEG) FR-II radio galaxies.}
\label{allsedsHEG}
\end{figure*}

\begin{figure*}
\centering
\begin{displaymath}
\begin{array}{ccc}
\includegraphics[width=6cm]{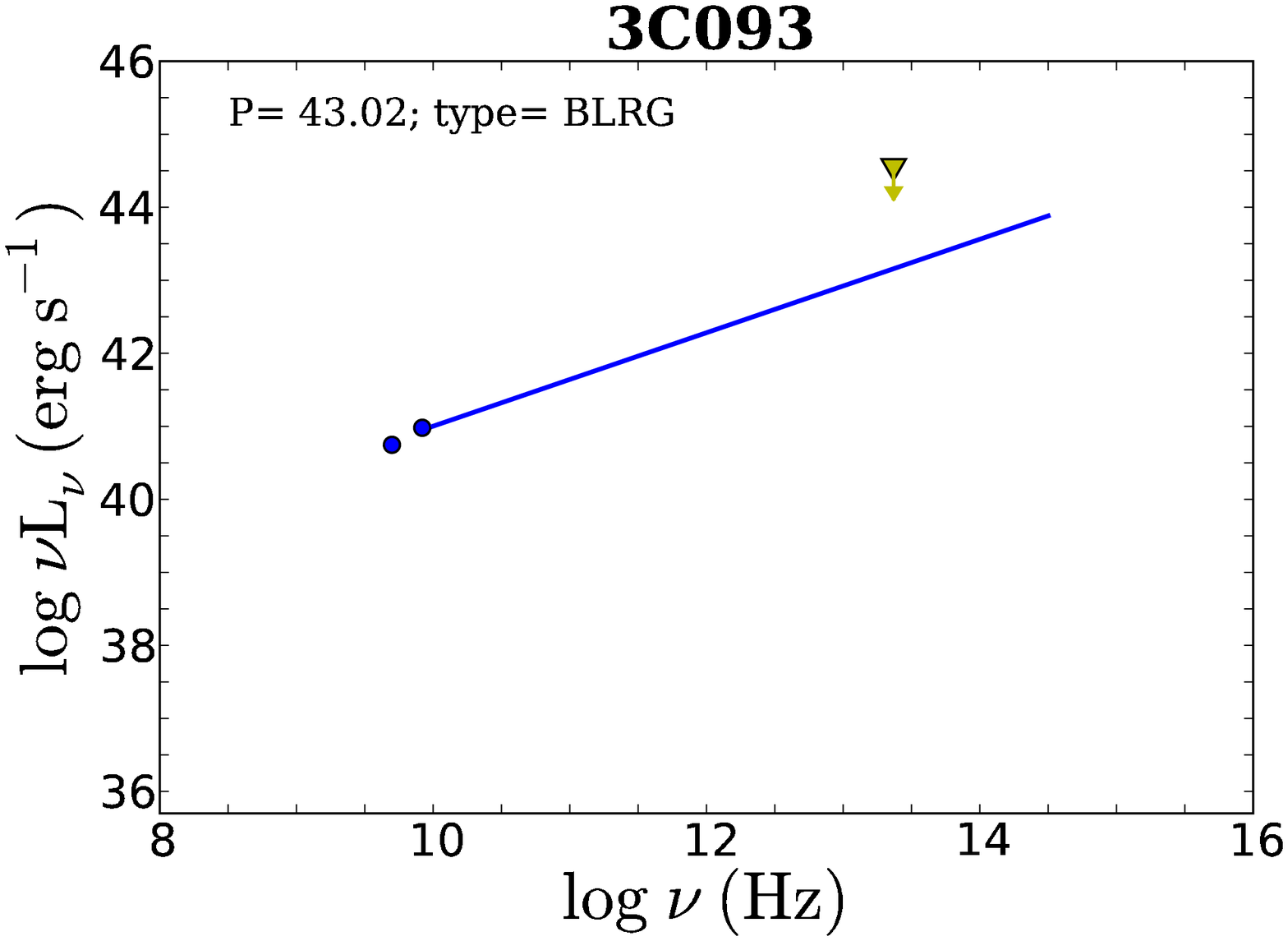} &
\includegraphics[width=6cm]{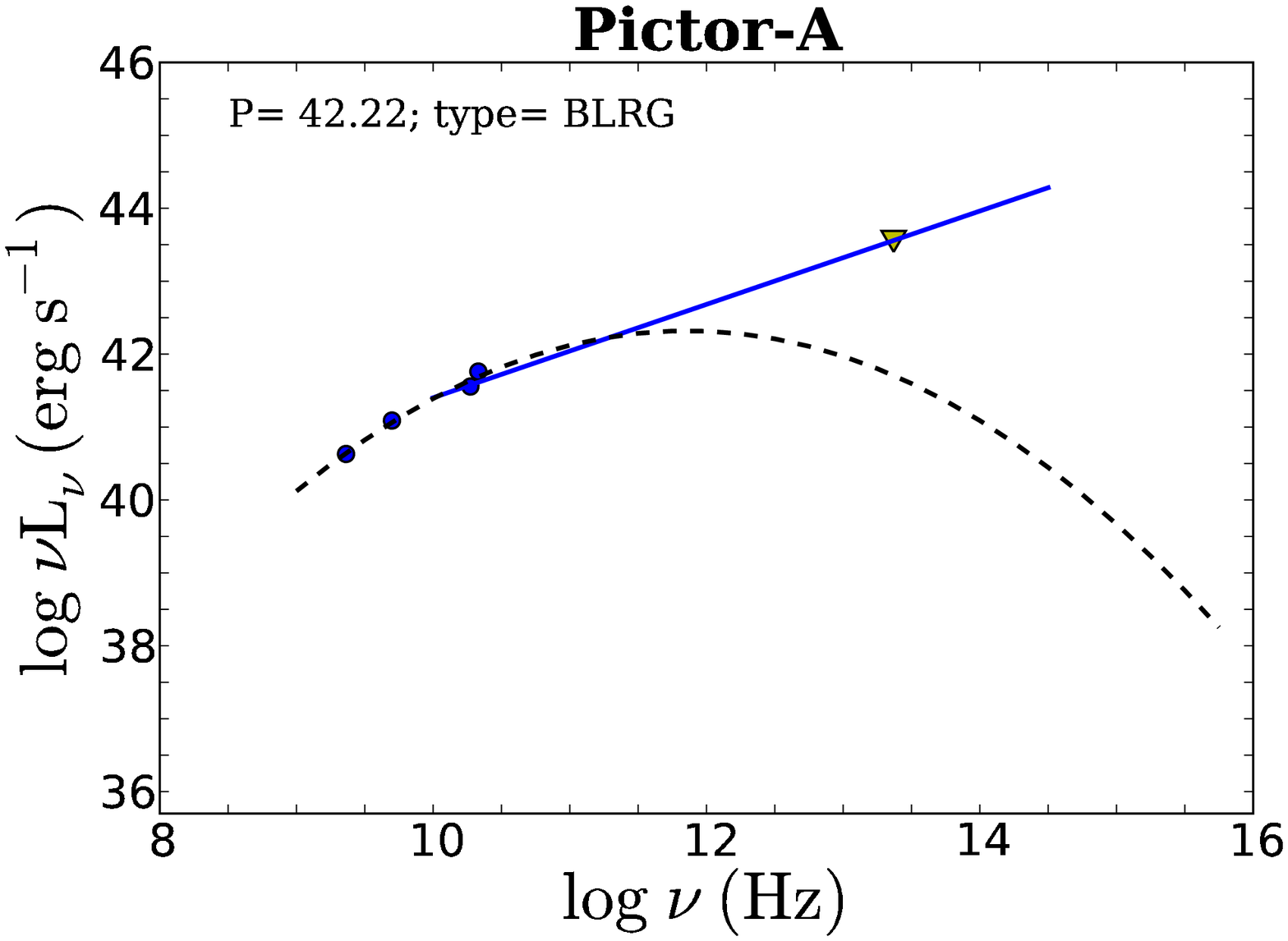} &
\includegraphics[width=6cm]{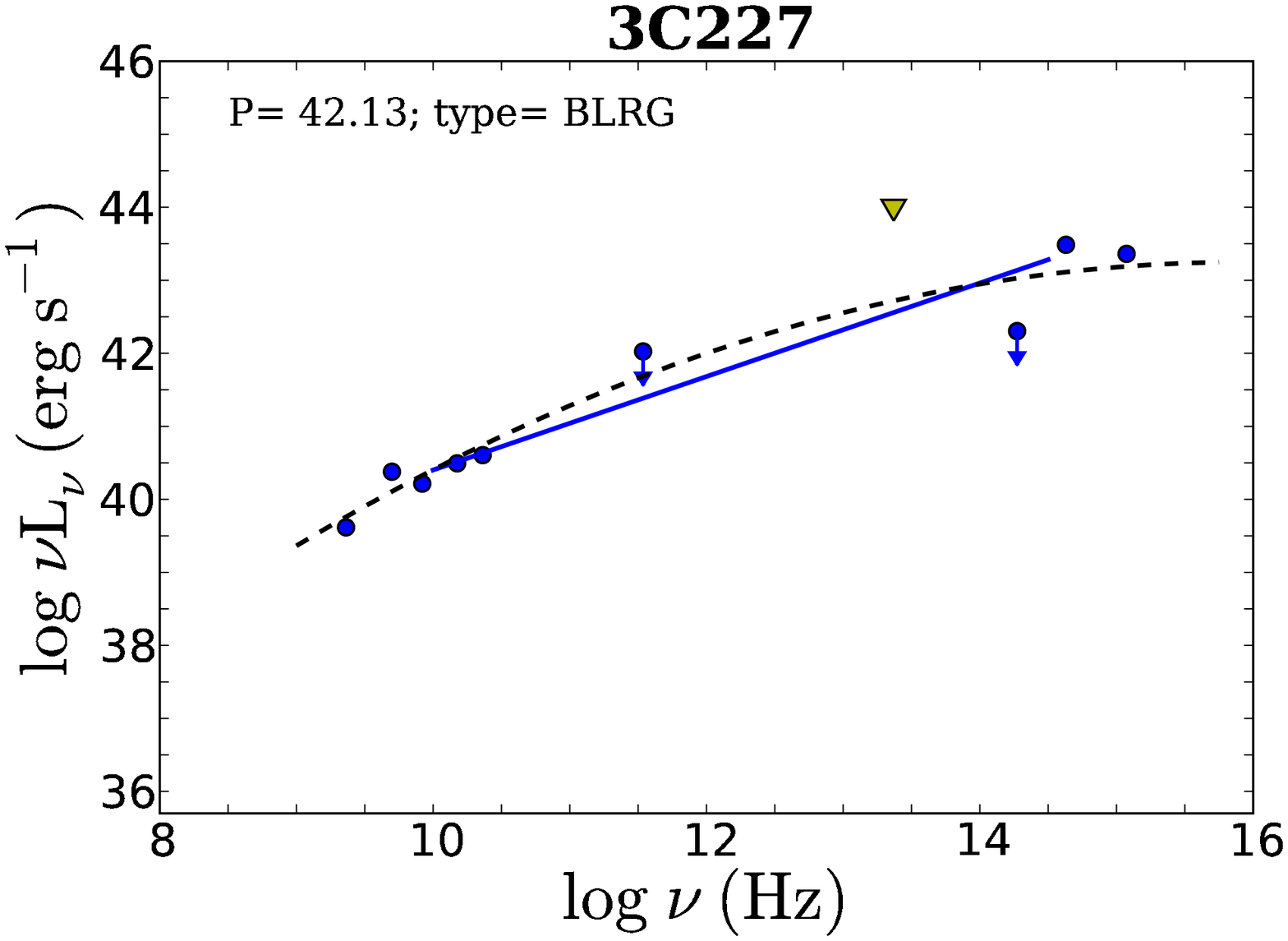} \\
\includegraphics[width=6cm]{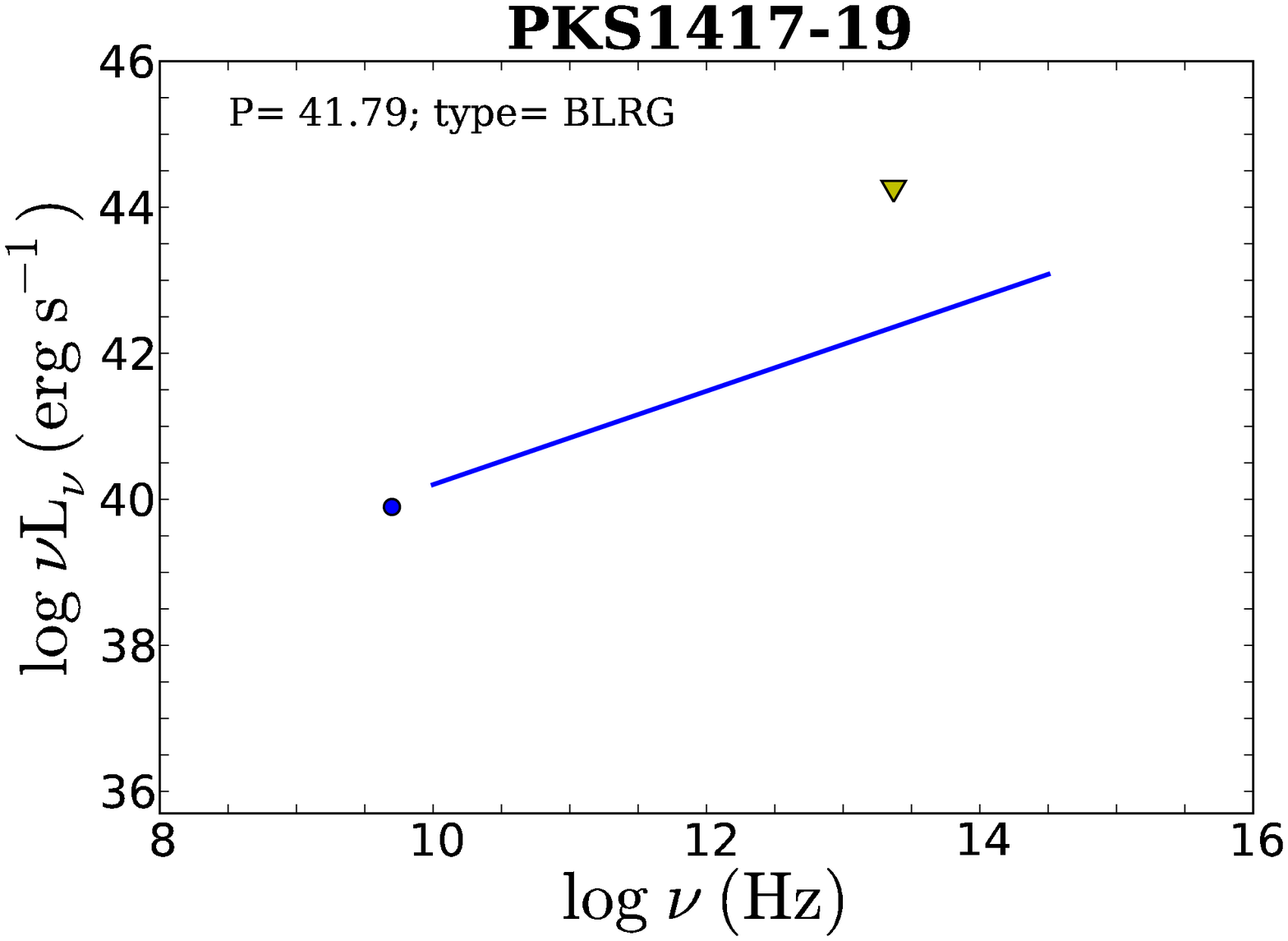} &
\includegraphics[width=6cm]{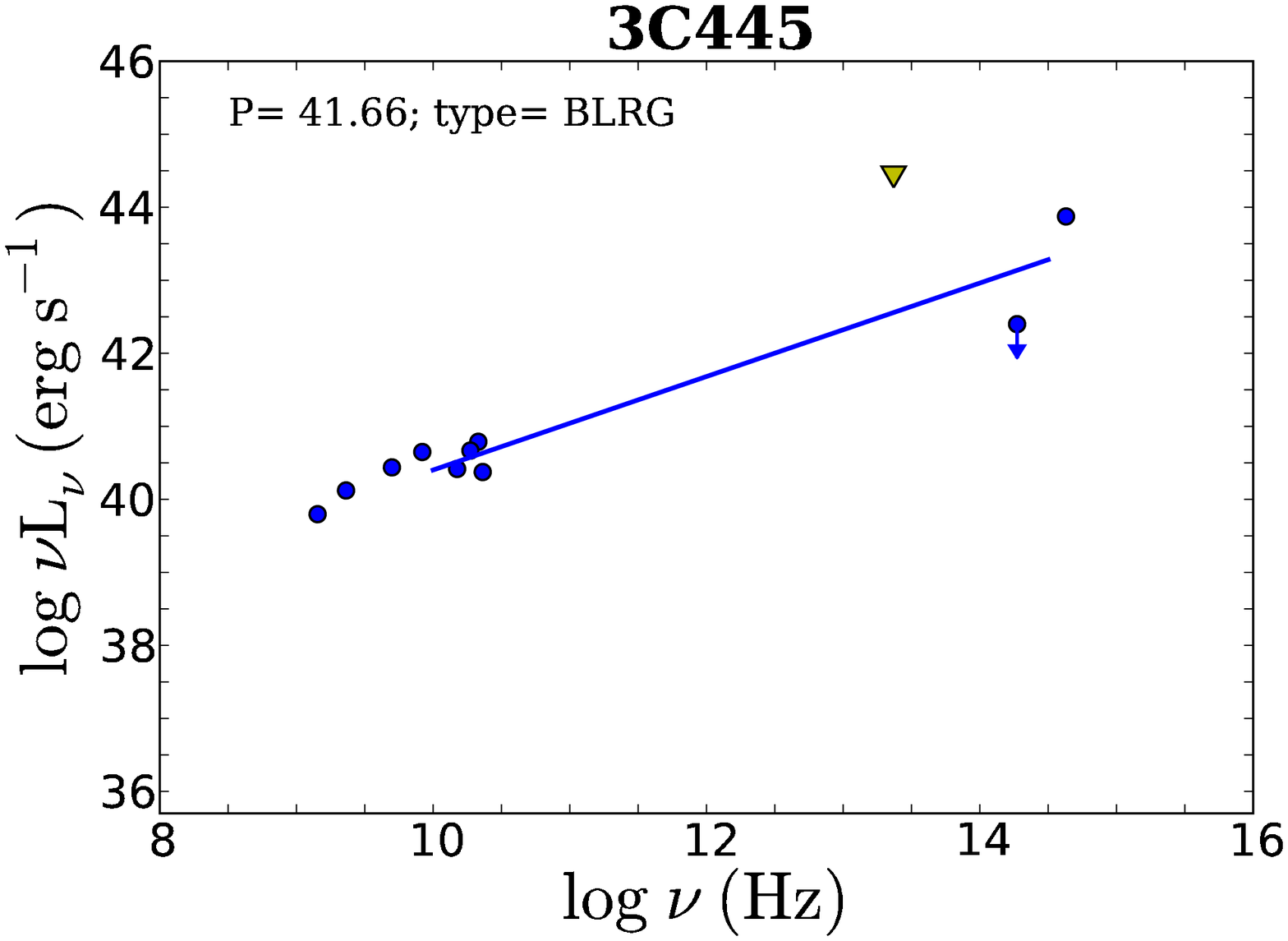} &
\includegraphics[width=6cm]{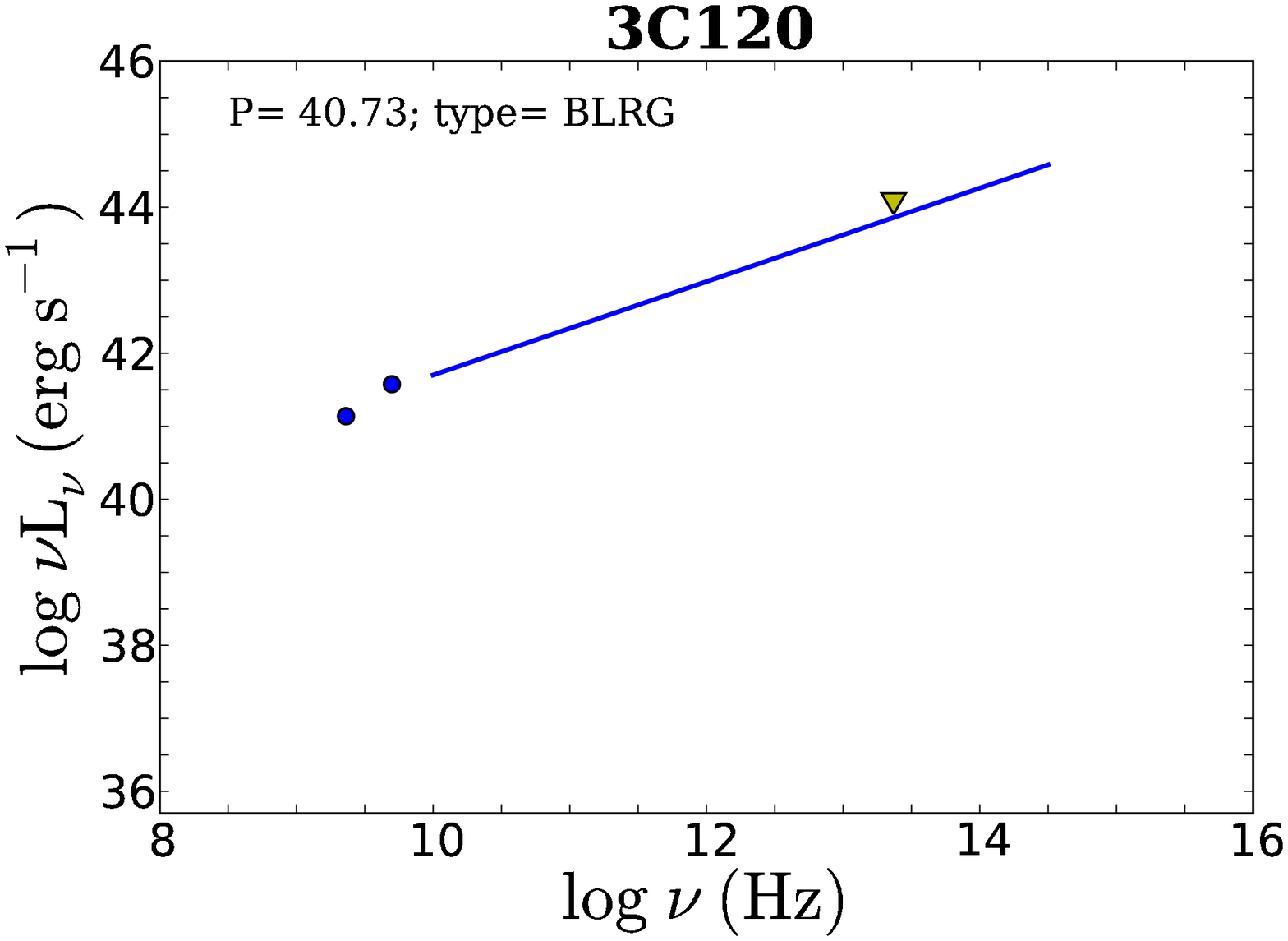} \\
\end{array}
\end{displaymath}
\caption{Same as Fig~\ref{allsedsFR-I}, but now for broad-line (BLRG) radio galaxies .}
\label{allsedsBLRG}
\end{figure*}

\section{Results}
\subsection{Dust morphology}
We detect an extended warm dust structure in the center of nearby FR-I Centaurus~A. For the FR-II radio galaxies, this is the case for none of the six LEGs, four out of seven HEGs, and five out of six BLRGs (see Fig.~\ref{psfsubtraction}). In the following subsections, we discuss the mid-infrared core morphologies and concentrations and the relation between the dust detected in the infrared and inferred from obscuration in the optical, for each type of radio galaxy.

\subsubsection{FR-I radio galaxies}
For the prototypical, close-by, FR-I source Centaurus~A / NGC5128, we find evidence of extended dust emission (Fig.~\ref{psfsubtraction}), surrounding the almost naked synchrotron core. As for \cite{2004A&A...414..123S}, but at a higher resolution, 0.33$\arcsec$ instead of 0.50$\arcsec$, and at a wavelength of $11.9~\mu\mathrm{m}$ instead of $10.4~\mu\mathrm{m}$, we find an unresolved core (FWHM~$\sim6$~parsec) showing some extended emission. We find that the extension is at an position angle of 100 degrees south-east of the nucleus. We interpret this emission as originating in a circumnuclear dust disk, of which the north-west emission is not visible at these wavelengths. The structure starts 2 parsec away from the center reaching outwards to 7 parsec and contributes $\sim6$ percent of the flux. Judging from the symmetry of the extended structure, the inclination of the disk must be almost edge-on. At even higher resolution, using the VLTI MIDI instrument, the core remains unresolved ($<~10$~mas) \citep{2007A&A...471..453M}, but an inner circumnuclear dust disk is resolved, which has a major axis that is probably orientated along a position angle of $127~\pm~9$ degrees, has a diameter of $\sim0.6$ parsec, and contributes $20$ percent of the flux at a wavelength of $8~\mu\mathrm{m}$ and $40$ percent at $13~\mu\mathrm{m}$. Hence, the inner and outer disk have comparable orientations. Together with other VISIR data of Cen~A, we are able to compile a N-band spectral energy distribution and compare it with low resolution spectra made by the Spitzer Infrared Spectrograph (IRS), \cite[see Fig. 10 in][]{2009A&A...495..137H}. The PAH emission feature at 11.3~$\mu$m is only visible at the lower Spitzer resolution. The N-band VISIR spectral energy distribution indicates pure continuum emission on the 10~parsec scale. 

From their optical and UV images, it is known that FR-Is 3C40 and 3C270 contain regular, large-scale circumnuclear dust disks \citep{1993Natur.364..213J, 1999ApJS..122...81M, 2002ApJS..139..411A}. We do not detect these disks with VISIR, indicating that they originate farther than 100~parsec from the center, or do not radiate. For the other FR-Is in our sample, all non-detected with VISIR, the nuclei do not show any obvious signs of dust structures in the optical. For FR-I 3C29, the optical nucleus is resolved at 0.1$\arcsec$ \citep{1999ApJS..122...81M}. 3C78's nucleus is unresolved, which is indicative of a compact core. An optical jet arising from the nucleus has been detected. 3C317's nucleus is likely unresolved and shows evidence of an extremely faint dust lane. Our VISIR non-detection confirms the lack of significant dust emission. The FR-I galaxies in our sample have weak total mid-infrared luminosities as well. Only the nearby galaxies Centaurus~A and Fornax~A were detected by IRAS (Table~\ref{tablevisir}). The core mid-infrared concentration of Cen~A is 10 percent and of Fornax~A, non-detected with VISIR, less than $4$ percent.

\subsubsection{No dust-enshrouded LEG/FR-IIs}
Unlike other radio galaxies where the jet radiation extends to optical wavelengths, FR-II/LEG 3C15 does not have a bright, compact, optical nucleus. It is understood to be enshrouded in dust \citep{1999ApJS..122...81M}. However, our mid-infrared non-detection provides evidence of a lack of an nuclear infrared excess above the synchrotron fit (Fig.~\ref{allsedsLEG}). 3C353 shows a peanut-shaped extended optical nucleus and it is suggested that it is bifurcated by a small-scale (300~parsec) dust lane or has a double nucleus \citep{1999ApJS..122...81M,2000ApJS..129...33D}. Our mid-infrared non-detection suggests the latter, although it could be that the dust lane does not radiate in the mid-infrared. LEG 3C403.1 also shows no signs of any dust structures \citep{1999ApJS..122...81M}. The LEGs also have weak total mid-infrared luminositites. None of the galaxies were detected with IRAS at mid-infrared wavelengths, suggesting that this class of active galaxies contains little warm dust. Our findings are in agreement with a Spitzer IRS spectrum study, from which it is understood that it is very unlikely that FR-II/LEGs can be heavily obscured AGN \citep{2009MNRAS.396.1929H}.

\subsubsection{Mid-infrared weak HEGs}
We detect two low-power sources, 3C98 and PKS2158$-$380, that are enshrouded in dust and show high-excitation lines. The nuclei of these sources are mid-infrared weak, as defined by their luminosities $\nu L_{11.85~\mu\mathrm{m}}<2\times10^{43}$~erg~s$^{-1}$. This dividing line is set at half the luminosity of the weakest BLRG in the mid-infrared of our sample, Pictor A. The two galaxies contain a radio-weak core and a total radio power comparable to FR-Is, but have a FR-II radio morphology. In 3C98, an extended dust structure contributing 10 percent of the flux, is seen at a position angle (PA) of 170~degrees outwards to 400~pc. The optical nucleus of this galaxy is not detected \citep{2002A&A...394..791C}, which is also indicative of a dusty nucleus. By comparing the nuclear mid-infrared radiation detected with VISIR to the total galaxy radiation with IRAS, the radiation is inferred to be entirely concentrated in the inner 400~pc (Table~\ref{tablevisir}). In PKS2158$-$380, an arc of extended emission, contributing only 4 percent of the VISIR flux, is detected 300~pc east of the nucleus. The mid-infrared core concentration of this galaxy is $>20$~percent.

\begin{figure*}
\centering
\begin{displaymath}
\begin{array}{cc}
\includegraphics[width=9.0cm]{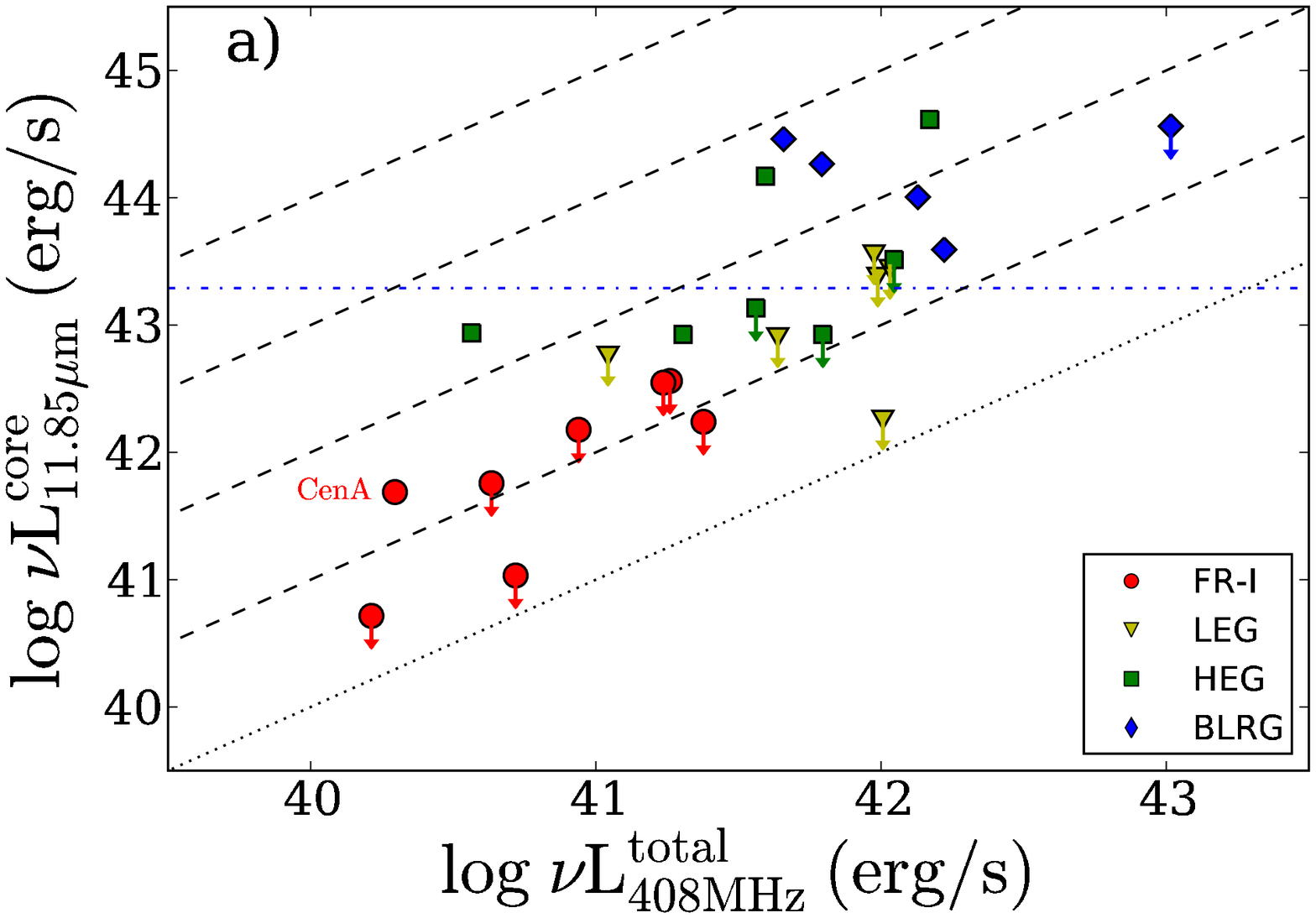} &
\includegraphics[width=9.0cm]{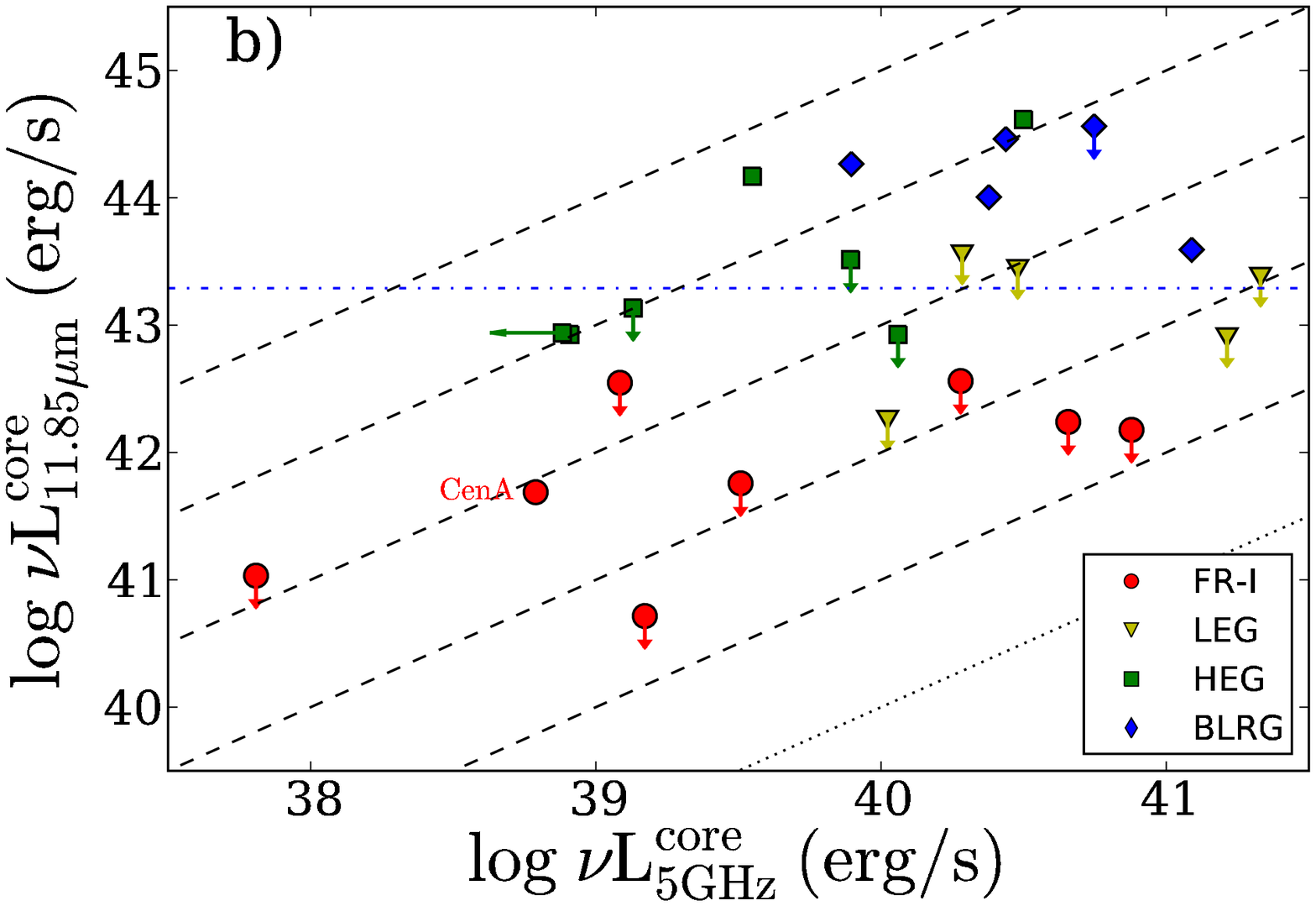} \\
\includegraphics[width=9.0cm]{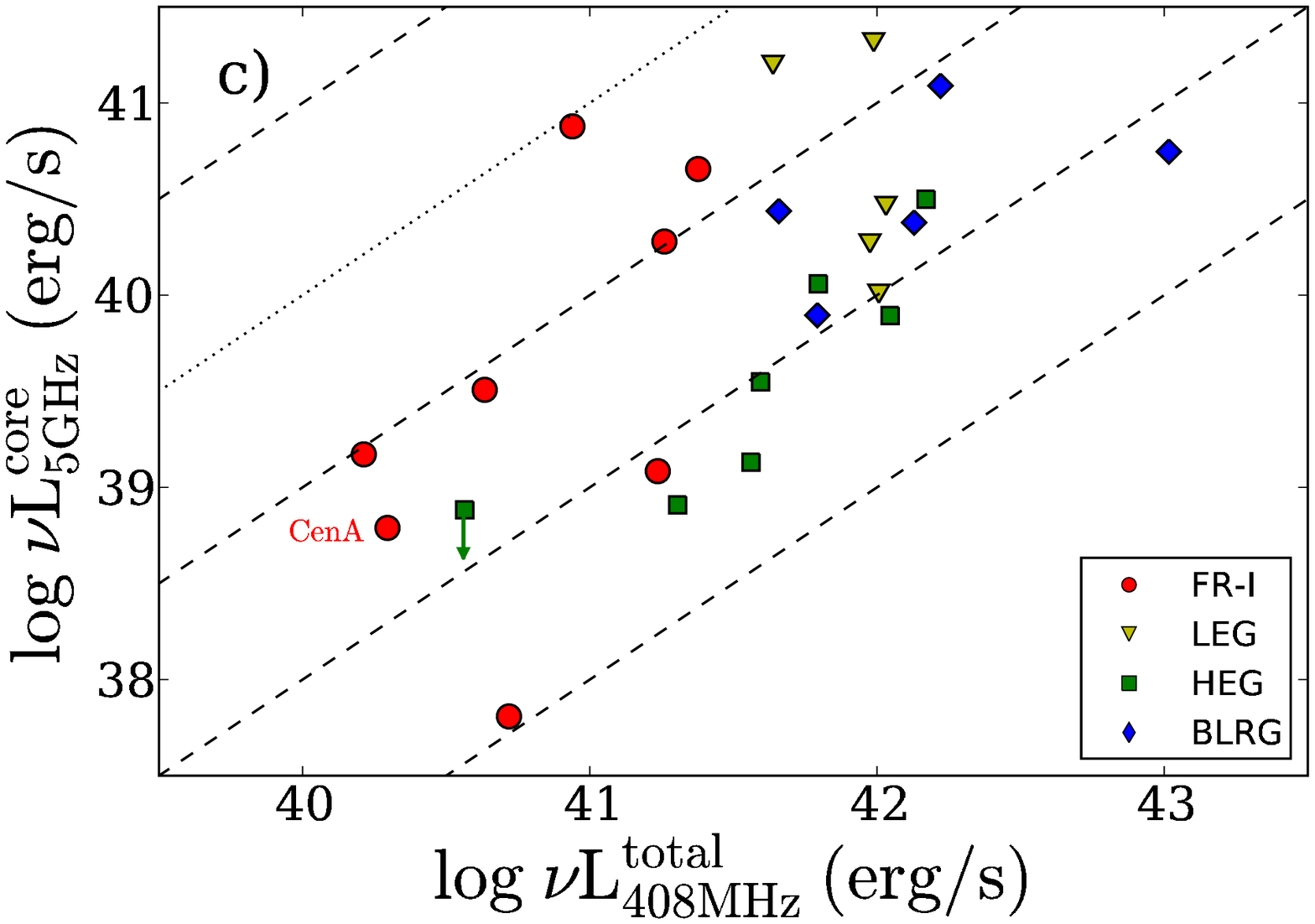} &
\includegraphics[width=9.0cm]{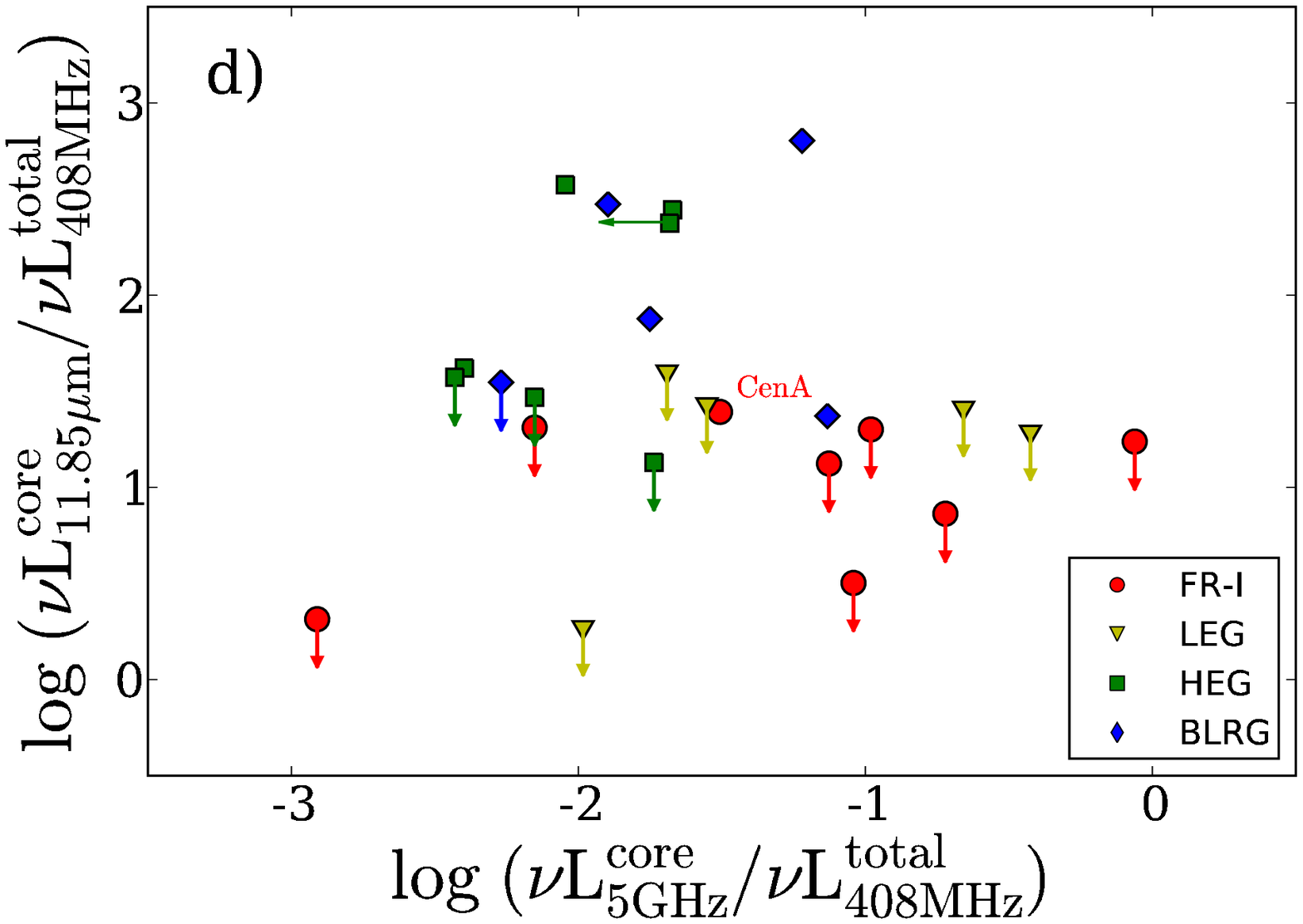} \\
\end{array}
\end{displaymath}
\caption{Luminosity comparison. \textit{a)} Total radio versus mid-infrared core luminosity. The HEGs show a large spread in the infrared-radio plane. The dot-dashed line, at $\nu L_{11.85~\mu\mathrm{m}}^{\mathrm{core}}=2\times10^{43}$~erg~s$^{-1}$, which is half the luminosity of BLRG Pictor~A, divides the sample into mid-IR strong, hidden quasars/BLRGs, and weak, likely inactive sources. The dotted line indicates a mid-IR core to total radio ratio of 1, and the dashed lines ratios of 10, 100, 1000, and 10\,000. \textit{b)} Radio core versus mid-infrared core luminosity. LEGs have high radio core powers. \textit{c)} Total versus core radio luminosity. FR-Is show a large scatter in core fraction, a measure of the orientation of the radio jet. \textit{d)} Mid-infrared core luminosity normalized by total radio power versus normalized radio core luminosity. The jets of the mid-infrared strong sources are oriented at intermediate angles with respect to the mid-infrared weak sources, which have a large spread in orientation.}
\label{mirradio}
\end{figure*}

\subsubsection{\label{hidden}Two hidden quasars among the HEGs}
Two sources among the high-excitation FR-II galaxies are as strong in the mid-infrared as the BLRGs, and therefore probably host hidden quasars. In the optical, these sources show multiple large dust-lanes 4-5~kpc in size. The mid-infrared strongest HEG, 3C327, shows a double dust disk north-west of the nucleus, and there is an indication of two elongated structures emerging from a possibly double nucleus \citep{2000ApJS..129...33D}. Our VISIR image (Fig.~\ref{psfsubtraction}) confirms the extended nature of the nucleus and we estimate its size to be 1.3~kpc in radius, with its major axis being along the north-west direction. This is rather large for a torus. It could well be that the dust is primarily heated by an extended large-scale emission region and that a torus is not present. Of the total mid-infrared radiation, 90 percent originates in this region. However, this source is very highly polarized in the UV \citep{2001ApJ...547..667K}, which is expected for a hidden quasar.

The nucleus of the second brightest HEG in the mid-infrared, 3C403, displays compact dust emission. Of the total mid-infrared radiation, 50 percent originates in the inner 700~pc. On large scales, this radio galaxy does not exhibit a standard two-sided jet but two pairs of jets. This x-shape is understood to be produced by the coalescence of two supermassive black holes at the center. At a two-percent level of the peak, 2~mJy, we detect the basis of the most recently formed jet. This is apparent from our PSF-subtracted mid-infrared image, which shows a mid-infrared jet at a position angle of 70 degrees extending from 120 to 240~parsec away from the nucleus. On a parsec scale, this jet is seen in the radio at a PA of 73 degrees, the same angle as observed for the jet on large-scales by \cite{2007A&A...475..497T}. This inner jet is coincident in location and velocity with three water-masers. The same authors do not detect central molecular gas emission and state that this component should be of size $<$200 pc. The same size limit can be placed on the dust emission.

\subsubsection{\label{extblrg}The extent of dust tori in BLRGs}
The mid-infrared radiation within the inner 300~parsec of BLRG 3C120 of FR-I type is 80~percent of the total emitted. As explained later, its nature is most likely non-thermal. Contributing 5 percent of the VISIR flux, we detect an extended structure 200-250~parsec south of 3C120 at a position angle of 180 degrees. This is probably the outer extent of the dust torus surrounding the nucleus, viewed at grazing incidence. We detect only the southern emission, and assume that the extended dust is symmetric but just not detectable at this wavelength. In this picture, the torus is compact in the inner few parsecs, and inclined away from the observer such that optically the weak quasar is not obscured. Extending to hundreds of parsecs, the density and flux of the torus decrease rapidly. The angle of the inner VLBI radio jet is directed 255 degrees in the north to east direction \citep{1998AJ....115.1295K}. This is perpendicular to the dust structure.

In Pictor~A, we detect an extended stucture 180-360~parsec in size at an position angle of 230~degrees, which also contributes 5 percent of the VISIR flux. The mid-infrared core concentration of this galaxy is 60~percent. The VLBI radio jet is oriented within its inner 5 parsec at an angle of 270 degrees, but after 7~parsec deflects to 310 degrees \citep{2000AJ....119.1695T} and has an inclination of $<$51 degrees. The outer edge of the central dust feature in 3C227 extends farther out, 360-540~parsec, at a PA of 30 degrees, contributing 6 percent of the flux. On larger scales, the jets are positioned at an angle of 89~degrees, while the direction of the inner jet is not known. We note that in large double-lobed sources the inner and outer jets are generally aligned. The mid-infrared core concentration of this galaxy is $>$10~percent. The extended emission in PKS1417$-$19 is probably mainly host galaxy dust, originating in star-forming regions. It extends to 2~kpc in a cone of 90~degrees from north to east and contributes 10 percent of the flux. The large-scale jet has a PA of 10~degrees \citep{1985ApJS...59..499A}.

The mid-infrared radiation from the inner 800~parsec of 3C445, the strongest mid-infrared BLRG, is equal to the total 12~$\mu$m radiation measured with IRAS. The outer edge of the torus in 3C445 is also the brightest in our sample of BLRGs. It emits 10 percent of the flux of the nucleus, extends out to about 350-460~parsecs, and is oriented at a position angle of 80~degrees. This is again orthogonal to the jet on large scales, which is oriented at an angle of 170~degrees \citep{1997MNRAS.291...20L}. Using other VISIR data of 3C445, we are able to construct a N-band spectral energy distribution and compare it with the low resolution IRS spectrum \citep[see Fig. 9 in][]{2009A&A...495..137H}. The IRS spectrum shows a weak and broad silicate feature, while the VISIR N-band spectrum rises slowly with wavelength and shows no sign of this silicate feature. From QSO IRS spectra, \cite{2008ApJ...679..101S} infer extended silicate emission features, which they interpret as being associated with cool dust in the narrow line region (NLR). For 3C445, this would mean that these clouds originate outside the VISIR beam, 200~parsec away from the center. The extended emission that we detect with VISIR is the tip of the iceberg.

Overall, we detect two trends. Firstly, in only one out of five cases, 3C227, the detected outer edges of the dust torus are not oriented perpendicularly to the inner and large-scale jets. Secondly, we detect the outer edges of the tori on one side only. The emission is probably asymmetric because of the optical thickness.

\begin{table*}
\caption{Median values of important physical parameters for the different radio galaxy classes.}
\label{tablemedian}
\centering
\begin{tabular}{lrccrcrrrcr}
\hline \hline
Class& $N$ & $z$ & $\log{\nu L^{\mathrm{core}}_{\mathrm{5GHz}}}$ & $\log{\nu L_{\mathrm{408MHz}}^{\mathrm{total}}}$ & $\log{R}$ &  $\log{\nu L^{\mathrm{core}}_{\mathrm{11.85}\mu\mathrm{m}}}$ & VISIR & $\log{L_{\rm{bol}}}$ & $\log{M_{\rm{BH}}}$ & $\log{\frac{L_{\rm{bol}}}{L_{\rm{Edd}}}}$\\
& & & & & & & Resolution & & & \\
& & & ($\mathrm{erg}~\mathrm{s}^{-1}$) & ($\mathrm{erg}~\mathrm{s}^{-1}$) & & ($\mathrm{erg}~\mathrm{s}^{-1}$) & (parsec) & ($\mathrm{erg}~\mathrm{s}^{-1}$) & ($M_{\odot}$) &  \\
(1) & (2) & (3) & (4) & (5) & (6) & (7) & (8) & (9) & (10) & (11) \\
\hline
All  &    27 & 0.05540 & 40.17 & 41.59 & $-$1.92 & 42.94    &   450 &  42.5    &   8.5 &  $-$4.1   \\
FR-I &     8 & 0.02332 & 39.34 & 40.83 & $-$1.48 & $<$41.97 &   180 &  41.8    &   8.8 &  $-$5.1   \\
LEG  &     6 & 0.09210 & 40.48 & 41.98 & $-$1.86 & $<$43.16 &   780 &  $<$42.3 &   8.3 & $<$ $-$4.2\\
HEG  &     7 & 0.08147 & 39.55 & 41.59 & $-$2.20 & $<$43.13 &   680 &  $<$42.5 &   8.5 &  $-$4.1   \\
BLRG &     6 & 0.07101 & 40.44 & 42.13 & $-$1.89 & 44.18    &   590 &  45.1    &   8.5 &  $-$1.7   \\
\hline
\end{tabular}
\end{table*}

\begin{table}
\caption{Fit parameters, estimated thermal excess, and ratio of ionization to mid-infrared emission strength.}
\label{tablefit}
\centering
\begin{tabular}{lcccr}
\hline \hline
Source & $A$ & $\chi^2$ & $\log{\mathrm{MIR}_\mathrm{excess}}$ & $\log{\frac{L_{[\ion{O}{III}]}}{L_{\mathrm{11.85}\mu\mathrm{m}}}}$ \\
       &     &          & ($\mathrm{erg}~\mathrm{s}^{-1}$) & \\
(1)    & (2) &       (3)&              (4)&(5) \\
\hline
4C12.03      &  $\dots$ &  $\dots$  & $<$1.2  & $>$$-$2.7  \\
3C015        &  $-$5.26 &  5.71     & $<$1.2  & $>$$-$2.5  \\
3C029        &  $-$2.05 &  0.14     & $<$0.5  & $>$$-$2.7  \\
3C040        &  $-$2.89 &  0.03     & $ $0.0  & $ $$\dots$  \\
3C078        &  $-$3.38 &  0.25     & $ $0.0  & $ $$\dots$  \\
Fornax~A     &  $-$2.50 &  3.96     & $<$1.0  & $ $$\dots$  \\
3C093        &  $\dots$ &  $\dots$  & $<$1.7  & $ $$\dots$  \\
3C098        &  $-$1.57 &  1.01     & $ $2.0  & $ $$-$2.0  \\
3C105        &  $-$2.93 &  0.12     & $<$1.0  & $>$$-$2.1  \\
3C120        &  $\dots$ &  $\dots$  & $ $0.4  & $ $$-$2.4  \\
3C135        &  $\dots$ &  $\dots$  & $<$1.8  & $>$$-$1.5  \\
Pictor~A     &  $-$1.86 &  0.19     & $ $0.2  & $ $$-$2.4  \\
3C198        &  $-$2.72 &  10.4     & $<$1.9  & $>$$-$2.0  \\
3C227        &  $-$6.18 &  2.51     & $ $1.7  & $ $$-$2.2  \\
3C270        &  $-$1.35 &  12.4     & $ $0.0  & $>$$-$1.8  \\
Centaurus~A  &  $-$1.89 &  0.50     & $ $0.0  & $ $$-$3.4  \\
PKS1417$-$19 &  $\dots$ &  $\dots$  & $ $2.2  & $ $$\dots$  \\
3C317        &  $-$2.53 &  0.69     & $ $0.0  & $>$$-$2.2   \\
3C327        &  $\dots$ &  $\dots$  & $ $2.0  & $ $$-$2.5  \\
3C353        &  $-$1.83 &  2.73     & $ $0.0  & $>$$-$3.0  \\
PKS1839$-$48 &  $\dots$ &  $\dots$  & $ $0.0  & $ $$\dots$  \\
3C403        &  $-$4.32 &  1.78     & $ $2.7  & $ $$-$2.6  \\
3C403.1      &  $\dots$ &  $\dots$  & $\dots$ & $ $$\dots$  \\
3C424        &  $\dots$ &  $\dots$  & $<$1.0  & $>$$-$2.8  \\
PKS2158$-$380&  $\dots$ &  $\dots$  & $ $2.0  & $ $$\dots$  \\
3C445        &  $\dots$ &  $\dots$  & $ $2.2  & $ $$-$2.5  \\
PKS2354$-$35 &  $\dots$ &  $\dots$  & $<$1.2  & $>$$-$2.2  \\
\hline
\end{tabular}
\begin{list}{}{}
\item[]\textit{Column (2)} The fitted curvature of the parabola. \textit{Column (3)} Goodness of fit parameter, reduced $\chi^2$-value, of the parabolic fit to the core spectral energy distribution data ranging from the radio to UV. \textit{Column (4)} Logarithm of the thermal excess above the synchrotron fit. \textit{Column (5)} Logarithm of $[\ion{O}{III}]$ ionization over mid-infrared emission strength.
\label{fitparam}
\end{list}
\end{table}

\begin{table}
\caption{Bolometric luminosities, black hole masses, and Eddington ratios.}
\label{table3}
\centering
\begin{tabular}{lrcccr}
\hline \hline
Source & $\log{L_{\rm{bol}}}$ & Ref. & $\log{M_{\rm{BH}}}$ & Ref. & $\log{\frac{L_{\rm{bol}}}{L_{\rm{Edd}}}}$ \\
&($\mathrm{erg}~\mathrm{s}^{-1}$)& & ($M_{\odot}$) & & \\
(1)&(2)&(3)&(4)&(5)&(6)\\
\hline
4C12.03      & $<$  42.0 & 6     &   8.5 & 9     & $<$  $-$4.6 \\
3C015        & $<$  42.8 & 1     &   8.8 & 4     & $<$  $-$4.1 \\
3C029        &      42.3 & 3     &   9.1 & 3     &      $-$4.9 \\
3C040        & $<$  41.5 & 3     &   7.9 & 2     & $<$  $-$4.5 \\
3C078        &      43.5 & 3     &   8.9 & 3     &      $-$3.5 \\
Fornax~A     & $<$  41.1 & 5     &   8.4 & 1     & $<$  $-$5.4 \\
3C093        &      45.3 & 3     &   9.5 & 6     &      $-$2.3 \\
3C098        & $<$  41.5 & 1     &   7.9 & 8     & $<$  $-$4.5 \\
3C105        &      43.6 & 5     &   7.8 & 10    &      $-$2.3 \\
3C120        &      45.3 & 4     &   7.4 & 8     &      $-$0.2 \\
3C135        &      43.2 & 1     &   8.5 & 7     &      $-$3.4 \\
Pictor~A     &      44.3 & 3     &   8.7 & 6     &      $-$2.5 \\
3C198        &      43.9 & 1     &   8.1 & 4     &      $-$2.3 \\
3C227        &      44.9 & 3     &   8.9 & 6     &      $-$2.1 \\
3C270        &      40.4 & 3     &   8.6 & 3     &      $-$6.3 \\
Centaurus~A  &      41.1 & 2     &   7.7 & 5     &      $-$4.7 \\
PKS1417$-$19 &      45.3 & 3     &   8.3 & 6     &      $-$1.1 \\
3C317        &      42.4 & 3     &   9.5 & 3     &      $-$5.2 \\
3C327        & $<$  42.4 & 1     &   9.1 & 4     & $<$  $-$4.8 \\
3C353        & $<$  41.4 & 1     &   7.6 & 4     & $<$  $-$4.3 \\
PKS1839$-$48 &      44.5 & 5     &   8.7 & 10    &      $-$2.3 \\
3C403        &      42.5 & 1     &   8.5 & 4     &      $-$4.1 \\
3C403.1      & $<$  42.0 & 6     &   8.1 & 10    & $<$  $-$4.2 \\
3C424        & $<$  42.6 & 5     &   7.8 & 7     & $<$  $-$3.3 \\
PKS2158$-$380& $<$  42.0 & 6     &   8.6 & 2     & $<$  $-$4.7 \\
3C445        &      44.8 & 3     &   8.0 & 6     &      $-$1.3 \\
PKS2354$-$35 & $<$  42.0 & 7     &   9.2 & 10    & $<$  $-$5.3 \\
\hline
\end{tabular}
\begin{list}{}{}
\item[]\textit{Columns (2-3)} Logarithm of bolometric luminosity from: 1=\cite{2004MNRAS.351..733M}, 2=\cite{2007A&A...471..453M}, 3=\cite{2007ApJ...658..815S}, 4=\cite{2002ApJ...579..530W}, 5=calculated from bolometric correction \citep{1994ApJS...95....1E, 2004MNRAS.351..733M} to optical nuclear luminosity \citep{1993MNRAS.263..999T, 2006A&A...447...97B}, 6=upper limit from spectral energy distribution fit in Fig.~\ref{allsedsLEG} and \ref{allsedsHEG}. \textit{Columns (4-5)} Logarithm of black hole mass from: 1=\cite{2006A&A...447...97B}, 2=\cite{2003A&A...399..869B}, 3=\cite{2004MNRAS.349.1419C}, 4=\cite{2004MNRAS.351..733M}, 5=\cite{2007ApJ...671.1329N}, 6=\cite{2007ApJ...658..815S}, 7=\cite{2005ApJ...631..762W}, 8=\cite{2002ApJ...579..530W}, 9=calculated from the velocity dispersion $\sigma$ \citep{2007A&A...470..531W} using the $M_{\mathrm{BH}}-\sigma$ correlation from \cite{2002ApJ...574..740T}, 10=calculated from B- or V-band absolute magnitudes \citep{1985MNRAS.216..173W, 1995ApJ...448..521Z} following the work of \cite{2004MNRAS.351..733M}. \textit{Column (6)} Efficiency expressed in the logarithm of the bolometric to Eddington luminosity ratio.
\label{masses}
\end{list}
\end{table}

\subsection{\label{nature}The nature of the mid-infrared emission}

We find that the sources in our sample span a range of three orders of magnitude in both core mid-infrared as well as in normalized radio core luminosity (Figs.~\ref{mirradio}a--d). These quantities are understood to reflect the relative importance of dust reradiating the accretion disk luminosity, assuming that all of the mid-infrared radiation traces the reradiated accretion disk luminosity, and jet beaming. Together with the core spectral energy distribution data (Figs.~\ref{allsedsFR-I}, \ref{allsedsLEG}, \ref{allsedsHEG}, and \ref{allsedsBLRG}), we can examine this for each of the four radio galaxy types.

The radiative properties, as indicated by the nuclear mid-infrared power, vary between radio galaxy types and with the amount of radio emission. We find that in terms of both total radio power and nuclear mid-infrared power, FR-Is are the weakest, FR-II LEGs and HEGs are intermediate, and BLRGs are the strongest sources. The HEGs exhibit the largest spread in the infrared-radio plane. Their cores are among the mid-infrared weak and strong sources (Fig.~\ref{mirradio}a). We choose the division line between mid-infrared strong, hidden quasars and BLRGs, and weak, likely inactive, sources at $\nu L_{11.85~\mu\mathrm{m}}^{\mathrm{core}}=2\times10^{43}$~erg~s$^{-1}$, which is half the luminosity of the weakest BLRG in the mid-infrared of our sample, Pictor A. This is four times lower than the dividing line that \cite{2006ApJ...647..161O} adopt using the same criterium. This difference is probably caused by our sample being at lower redshifts and our data, of higher resolution, shows more spread in core flux concentration. While we observed thirteen LEGs and HEGs in total, four have upper limits just above the dividing line between weak and strong sources. We can make statements about hidden quasars and inactive nuclei on the basis of nine sources. Of the LEGs and HEGs, two out of nine (22$\pm$14 percent) satisfy the criterion of being obscured quasars. These two, 3C327 and 3C403, are of the HEG type. They are as luminous in the mid-infrared as the BLRGs. The remaining 78$\pm$14 percent of the LEG and HEG FR-IIs and all of the FR-I radio galaxies are mid-infrared weak. All the mid-infrared weak sources have total radio powers of $\nu L_{408\rm{MHz}}^{\rm{total}}<10^{42}$~erg~s$^{-1}$, while the mid-infrared strong sources have radio powers of $\nu L_{408\rm{MHz}}^{\rm{total}}>4\times 10^{41}$~erg~s$^{-1}$.

We proceed by investigating whether the mid-infrared weak galaxy nuclei are purely non-thermal. Figure~\ref{mirradio}b displays the mid-infrared strength versus the radio core strength of the sample nuclei. The 8 FR-I nuclei exhibit a large range of radio core powers, while we find that all of the 6 FR-II LEGs have high radio core powers. The 3 mid-infrared weak HEGs in our sample have intermediate core radio powers. We find that the nuclei of FR-Is and LEGs can be up to three orders of magnitude more powerful in the mid-infrared than in the radio, as can be seen from the dashed lines. FR-I and LEG nuclei follow a tighter relation in the radio-optical plane -- they are one to two orders of magnitude more powerful in the optical than in the radio \citep{1999A&A...349...77C, 2000A&A...355..873C} -- indicating an at most small mid-infrared excess. The BLRGs and HEGs can be up to four orders of magnitude more powerful in the mid-infrared than in the radio, indicating a larger thermal excess.

To investigate the non-thermal and/or thermal nature of these galaxy nuclei further, we gathered from the literature core spectral energy data, ranging from the radio to the UV (Table~\ref{coresed}). We used two methods to fit the non-infrared core data, where a mid-infrared excess above the fit could indicate additional thermal radiation by reradiation by circumnuclear dust. Firstly, 15 out of 27 galaxies have a large enough core data set, permitting us to fit a parabolic function to all, except the mid-infrared, core data, of the form
\begin{equation}
\centering
\log{\nu L_{\nu}} = C + \frac{1}{2A}(\log{\nu}-B)^2 \,,
\label{parabola}
\end{equation} 
 where $C$ is the $\log{}$ of the peak power, $B$ is the $\log{}$ of the peak frequency, and $A$ represents the curvature of the parabola \citep[e.g.][]{1986ApJ...308...78L,2002A&A...381..389A,2009ApJ...701..891L}. The values of the curvatures and the goodness of fit, as indicated by the reduced $\chi^2$ parameter, are listed in Table~\ref{fitparam}. Secondly, we fit the high frequency radio core data of all the galaxies, with a straight power-law synchrotron spectrum $\nu L_{\nu} \propto \nu^{0.64}$, similar in slope to the core spectrum of Cen~A \citep{2007A&A...471..453M}. This slope is slightly steeper than for the general FR-I population (Fig~\ref{allsedsFR-I}). A comparison of the two methods of fits shows that this power-law fit is also a good indicator for a non-thermal and/or thermal nature of the galaxy nuclei (Figs.~\ref{allsedsFR-I}, \ref{allsedsLEG}, \ref{allsedsHEG} and \ref{allsedsBLRG}). We note that the true slope of the power-law fit is not the relevant factor, since a flatter slope would only increase the mid-infrared excess and not change our overall conclusions. So, when there is enough core data we use the parabolic fit, otherwise we use the power-law fit to determine the mid-infrared excess (Table~\ref{fitparam}). Another caveat of this method could be the host galaxy contribution, but unless in the case of very contrived geometries the host galaxy contribution to the mid-infrared emission in the 0.40\arcsec~apertures is not expected to differ significantly between the four subsamples.

All eight FR-I sources clearly lack a thermal excess (Fig.~\ref{allsedsFR-I}) but do exhibit three orders of magnitude of scatter in the strength of their non-thermal radiation. Half (50$\pm$18~percent) have the strength of Centaurus~A, while 3/8 (38$\pm$17~percent) are ten times brighter and 1/8 (13$\pm$12~percent) is ten times weaker. Four of the sources have published Spitzer IRS spectra \citep{2009A&A...495..137H, 2009ApJ...701..891L}, which support the non-thermal, PAH-free nature of the nuclei. An exception might be 3C270, for which an excess IRS mid-infrared radiation is argued to originate in 200K warm nuclear dust.

We find that two of the five LEG/FR-IIs, 3C353 and PKS1839$-$48, are definitely non-thermal in nature (Fig.~\ref{allsedsLEG}). The other three, 3C15, 3C424, and 4C12.03 have VISIR upper limits that could conceivably represent a small mid-infrared excess. On the basis of its Spitzer IRS spectrum, 3C15 and 3C424 are argued to host larger amounts of thermal nuclear emission \citep{2009ApJ...701..891L}. Out of the five LEGs, two LEGs are mid-infrared weak, while the other three are uncertain since they have upper limits above the dividing line between weak and strong sources. 3C353 has a non-thermal LEG nucleus with a synchrotron strength ten times brighter than Cen~A. The non-thermal nucleus of PKS1839$-$48, is even a hundred times brighter.

Of the four HEGs with mid-infrared weak galaxy nuclei, we find that 3C105 is non-thermal in nature, 3C198 has an uncertain nature, while 3C98 and PKS2158$-$380 definitely show a thermal excess of $\sim$100 times the synchrotron flux. 3C105 has a strong synchrotron core, ten times brighter than Cen~A, while the synchrotron cores with a thermal excess are comparable to that of Cen~A. Conform to expectation, the mid-infrared radiation of HEG 3C98, which shows a thermal excess, is entirely concentrated in the inner kpc (Sect.~\ref{hidden}). For the other mid-infrared weak HEG, showing a thermal excess, PKS2158$-$380, a lower limit of $>$20~percent on the mid-infrared core concentration can be set.

Most of the mid-infrared strong galaxy nuclei are thermal in nature. We find that HEG and BLRG nuclei are four orders of magnitude more powerful in the mid-infrared than in the radio (Fig.~\ref{mirradio}b). In the BLRG sample, 4/6 (67$\pm$19~percent) show a mid-infrared, thermal excess of $\sim$100 times the synchrotron flux, as is apparent from the synchrotron spectrum fits to the core radio data (Figs.~\ref{allsedsHEG}~and~\ref{allsedsBLRG}). As already noted in Sect.~\ref{extblrg} for one of these galaxies, 3C445, the mid-infrared radiation is entirely concentrated in the inner kpc. For 2/6 (33$\pm$19~percent) of the BLRGs, Pictor A and 3C120, the mid-infrared point is no stronger than the synchrotron spectrum. However, these two sources have non-thermal radiation luminosities a factor of a hundred higher than the FR-I radio galaxy Centaurus~A, which could indicate that these sources are beamed towards us. Indeed 3C120 shows large-scale beaming, the core being much brighter than the extended radio emission. For Pictor A, the core fraction is much less. Only one other source in our sample, LEG PKS1839$-$48, also exhibits this bright non-thermal radiation, and has the highest core fraction of all the LEGs. Beaming of the central parsec core jet might be the explanation. The mid-infrared core concentration values of Pictor~A and 3C120 are 60 and 80 percent, respectively. The two mid-infrared strong HEGs, 3C327 and 3C403, which presumably host hidden quasars as deduced from their thermal excess of $\sim$100 and 500 times the synchrotron flux, are found among sources with brighter synchrotron cores. As expected, the mid-infrared radiation of 3C327 is almost entirely concentrated within the inner 2~kpc, while for 3C403 half of the radiation originates in the inner kpc. 

We find that FR-Is show a large scatter in core fraction, which to some extent reflects the orientation of the radio jet, while the FR-IIs have a smaller spread, especially the HEGs and BLRGs (Fig.~\ref{mirradio}c). Mid-infrared core strength normalized by total radio power does not correlate with the core fraction. The mid-infrared strong sources have intermediate core fractions compared to FR-Is and LEGs, which both exhibit large spreads in core fraction (Fig.~\ref{mirradio}d). This similarity in spread could indicate that LEGs also form part of the parent population of variable active nuclei with a beamed component close to the line of sight. This might not be the case for the mid-infrared weak HEGs. Their core fraction is similar to the mid-infrared strong HEGs and BLRGs.

\begin{figure}
\centering
\includegraphics[width=8.8cm]{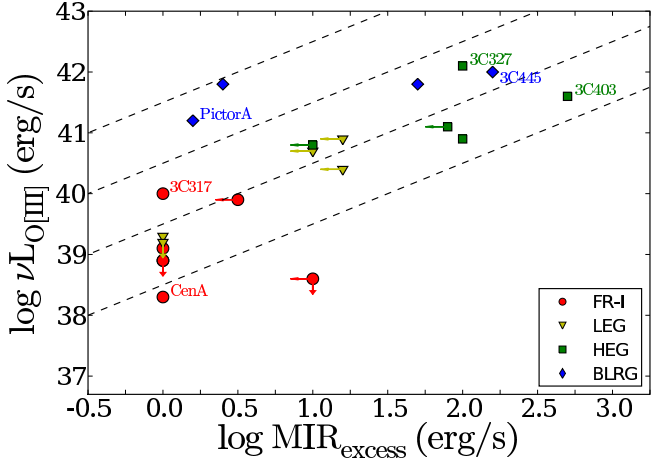}
\caption{Mid-infrared excess as determined from fits to the SEDs versus $[\ion{O}{III}]$ emission-line luminosity. Dashed lines indicate $\log$ ratios of -38, -37, -36, and -35}
\label{mirOIII}
\end{figure}

Normalising the $[\ion{O}{III}]$ emission-line luminosity by the mid-infrared strength, we find that this ratio does not vary much with radio galaxy type and is on average $\log{L_{[\ion{O}{III}]} /\/ L_{\mathrm{11.85}\mu\mathrm{m}}}=-2.5$ (Table~\ref{fitparam}). Similar values are found for Seyfert-1 and 2 galaxies \citep{2007A&A...473..369H}. This would suggest that both the $[\ion{O}{III}]$ and mid-infrared emission are fairly reliable indicators of the AGN luminosity, despite the $[\ion{O}{III}]$ emission being affected by dust obscuration \citep{1993Natur.362..326H}. A plot of the emission-line luminosity versus the mid-infrared excess determined from fits to the SEDs (Fig.~\ref{mirOIII}) does show two orders of magnitude scatter in its relation. However, we discover one remarkable trend, namely that the mid-infrared weak HEGs have ratios ($\log{L_{[\ion{O}{III}]} /\/ L_{\mathrm{11.85}\mu\mathrm{m}}}>-2.0$) that are higher than for mid-infrared strong HEGs (-2.6), and both  FR-Is and FR-II LEGs (-2.5). The explanation of this result could be simple: less dust produces less extinction towards the narrow line region.

\begin{figure}
\centering
\includegraphics[width=8.8cm]{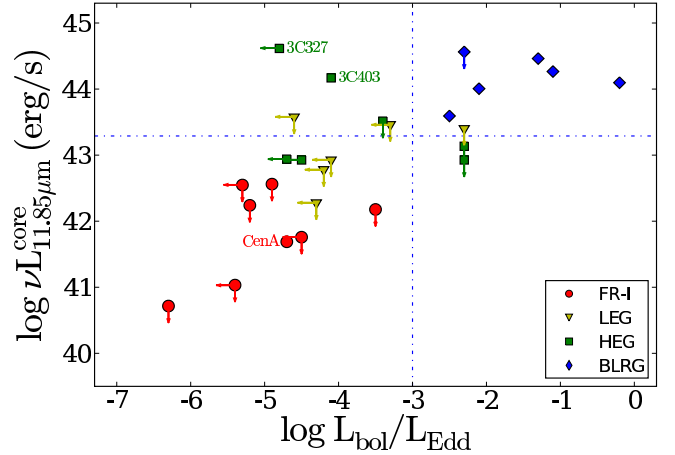}
\caption{Efficiency versus mid-infrared core luminosity. The HEGs are a mixed population with three types of nuclei: 1) low efficiency and dust torus, 2) low efficiency and weak dust torus, and 3) high efficiency and weak dust torus.}
\label{edd}
\end{figure}

\addtocounter{table}{1}

\section{\label{discussion}Discussion}
The inclusion of the mid-infrared points in the core spectral energy distributions has enabled us to discover that all the FR-I, all the FR-II LEG and a maximum of 3/7 (43$\pm$19~percent) of the HEG nuclei are dominated by pure synchrotron radiation and lack dust re-radiation of the accretion disk. We consider the possible physical origin of this behaviour. Theoretically, it is predicted that the torus disappears at low efficiencies, calculated to be the ratio of bolometric to Eddington luminosity, $L_{\rm{bol}}/L_{\rm{Edd}}<10^{-3}$ \citep{2008NewAR..52..274E} and that the available accretion energy is transformed into jet power. This idea is supported by our results for the BLRGs in our sample, for which the bolometric luminosity is higher and the extended dust emission reaches farther out. To investigate this further, we plot the efficiency versus mid-infrared core power for the sample objects (Fig.~\ref{edd}). The relevant data are compiled in Table~\ref{masses}. We note that many of the bolometric luminosities are determined using a bolometric correction $L_\mathrm{bol}=10L_\mathrm{B}$ to the optical nuclear emission, using quasar SEDs. This is argued to be justifiable, even for FR-I and LEG FR-IIs \citep{2004MNRAS.351..733M}, and is simply a rescaling of the blue continuum. Black hole masses are determined from velocity dispersions $\sigma$, using the relation of \cite{2002ApJ...574..740T}, or inferred from optical absolute magnitudes using $\log M_{\mathrm{BH}} = -1.58 - 0.488 M_{\rm{B}}$ \citep{2004MNRAS.351..733M}. All dust-tori-free FR-Is and all LEGs apart from one, PKS1839$-$48 with a likely beamed core, have low efficiencies. All BLRGs have high efficiencies. The HEGs are a mixed population with three types of nuclei: 1) objects with low efficiency with dust torus, 2) low efficiency and weak dust torus, and 3) high efficiency and weak dust torus. It will be interesting to investigate the host environment of a larger set of HEGs with VISIR data \citep[cf.][]{2008A&A...490..893T}.

Dust-tori-free FR-Is are the least active ($\log{L_{\rm{bol}}/L_{\rm{Edd}}}\sim-5.1$) but their energy is not necessarily transformed into jet power. The most actively accreting FR-I, 3C78, has a synchrotron core ten times more powerful than that of Centaurus~A. The FR-II LEGs of intermediate activity ($\log{L_{\rm{bol}}/L_{\rm{Edd}}}\sim-4.2$), which also lack significant dust reradiation, have stronger synchrotron cores than the FR-Is, indicating that the accretion power is transformed into jet power. Alternatively, the available energy is tapped with negligible accretion \citep{1984RvMP...56..255B}. On the other hand, as discussed in Sect.~\ref{nature}, the most active LEG PKS1839$-$48 possibly hosts a beamed central parsec core jet. The BLRGs, the most active accretors ($\log{L_{\rm{bol}}/L_{\rm{Edd}}}\sim-1.7$) in this sample indeed show thermal reradiation but have similar core jet powers as LEGs. They seem to host both dust tori and jets at the same time. The HEGs 3C105 and 3C198, with efficiencies as high as that of BLRGs lack or have weak dust tori. Among the HEGs, they are intermediate in total as well as in core radio and optical line emission strengths. The less active accretors, with tori that should have disappeared according to theory, are found in both the mid-infrared weak and strong regime. It is apparent that the mid-infrared strong sources, 3C327 and 3C403, have the most powerful radio lobes $\nu L_{408\rm{MHz}}^{\rm{total}}>4\times10^{41}$~erg~s$^{-1}$, cores and optical line emission strengths of the HEGs. The mid-infrared weak sources, 3C98 and PKS2158$-$380, are at the other end of this spectrum. In conclusion, our small sample study seems to indicate that the torus does not disappear at low efficiencies in HEGs with strong jets, but does so for FR-Is, LEGs, and mid-infrared weak HEGs.

The present study seems to indicate that the fraction of FR-IIs containing a dust torus covering a broad line region increases with redshift. This implies that the dust covering angle increases with redshift. The fraction of obscured radio galaxies between $z=0.5-1.0$ is $70\pm7$~percent, implying a viewing angle range of 0-45~degrees \citep{1989ApJ...336..606B} of quasars. A mid-infrared study of a sample with quasars and radio galaxies including lower redshift sources ($z=0.1-1.0$) seem to indicate a lower fraction of obscured radio galaxies of 50$\pm$15~percent \citep[]{2001A&A...372..719M, 2006ApJ...647..161O}. This would suggest that the unification viewing angle of quasars should be increased to $\sim60$~degrees, at least at lower luminosities. We note that since the subsamples presented here are small, the trends discovered are merely indicative. The drive to obtain a larger mid-infrared core sample study, especially of HEGs, is in place. Such a sample could play an important role in settling the question about the physical origin of dust tori.

\section{Summary and conclusions}
The main conclusions of our sub-arcsecond mid-infrared VISIR survey of the nuclei of 27 radio galaxies of various types are:
\begin{list}{}{}
\item[1.] On the basis of the core spectral energy distributions of our sample, we have found clear indications that many FR-I and several low-excitation FR-II and at maximum 43$\pm$19~percent of the high-excitation FR-II radio galaxies have non-thermal nuclei, but lack dust tori.
\item[2.] A circumnuclear dust disk is discovered, at a distance of 2-7 parsec from the center of Centaurus~A at a position angle of 100 degrees, which contributes 6 percent of the flux.
\item[3.] The extents of the dust tori around the naked nuclei of broad-line radio galaxies have been measured. In 4 out of 5 cases, these are directed perpendicular to the jet axis.
\item[4.] The warm dust radiation in objects showing a thermal excess is almost entirely concentrated in the inner kpc.
\item[5.] The number of mid-infrared weak sources among the FR-II radio galaxies, 78$\pm$14 percent, at redshifts $z\sim0.06$ is not too different from the 55$\pm$13 percent found by \cite{2006ApJ...647..161O} at $z<1.0$. Conversely, this indicates a hidden quasar fraction among $z\sim0.06$ FR-IIs of 22$\pm$14 percent, compared to 45$\pm$12 percent at redshifts $z<$1.0.
\item[6.] All FR-Is and all but one low-excitation FR-II galaxies contain nuclei with low efficiencies, $L_{\mathrm{bol}}$~$/L_{\mathrm{Edd}}<10^{-3}$. This suggests that the torus disappears at low efficiencies. High-excitation FR-II galaxies are a mixed population, having three types of nuclei: 1) low efficiency with dust torus, 2) low efficiency and weak dust torus, and 3) high efficiency and weak dust torus.
\item[7.] The present and other mid-infrared studies seem to indicate that roughly 50$\pm$15~percent of powerful radio galaxies at redshifts $z=0.1-1.0$ develop a nuclear broad line region with associated circumnuclear dust. This implies that the unification viewing angle range 0-45~degrees of quasars should be increased to $\sim$60~degrees, at least at lower luminosities.
\end{list}

\begin{acknowledgements}
The paper benefitted from comments by the referee, Dr.~R.~R.~J.~Antonucci. The observations, on which this article is based, were made with the VISIR instrument mounted at the ESO telescope Melipal on Paranal Observatory under programs 075.B-0621, 076.B-0194, 077.B-0135 and 078.B-0020. We are grateful to the staff at Paranal for assisting with the observations. We would like to thank Drs.~A.P.~Verhoeff and Prof.~Dr.~K.~Meisenheimer for observing on our behalf on respectively June 2nd and December 18th 2005. This research has made use of the NASA/IPAC Extragalactic Database (NED) which is operated by the Jet Propulsion Laboratory, California Institute of Technology, under contract with the National Aeronautics and Space Administration, and the VIZIER database at the CDS, Strasbourg, France.
\end{acknowledgements}

\bibliographystyle{aa}
\bibliography{12435ax.bib}

\Online

\appendix
\section{Core spectral energy distribution data}

{\scriptsize
\begin{center}
\tablefirsthead{%
  \hline \hline
  Filter & $\lambda$ & Flux & Obs. Date & Ref.\\
  &($\mu$m)&(mJy)&(yr,m,d)& \\
  \hline}
\tablehead{%
  \hline \hline
  Filter & $\lambda$ & Flux & Obs. Date & Ref.\\
  &($\mu$m)&(mJy)&(yr,m,d)& \\
  \hline}
\tabletail{%
\hline}
\tablelasttail{\hline}
\tablecaption{\label{coresed} Core spectral energy distribution data}
\begin{supertabular}{llrrc}
4C12.03      & & & & \\ 
\hline 
VISIR/SiC    & 1.3e+01 & $<$ 3.1e+00 & 2006,06,18   & $ $           \\ 
5GHz         & 6.0e+04 &     7.4e+00 & 1989,03      & 24$^{*}$     \\ 
1.5GHz       & 2.0e+05 &     4.0e+00 & 1989,03      & 24           \\ 
\hline 
3C015        & & & & \\ 
\hline 
F702W        & 7.0e-01 & $<$ 7.6e-03 & $<$1999,04   & 8            \\ 
VISIR/SiC    & 1.3e+01 & $<$ 3.3e+00 & 2006,06,18   & $ $           \\ 
22.4GHz      & 1.3e+04 &     3.7e+01 & 2006,08      & 12           \\ 
22GHz        & 1.4e+04 &     1.8e+02 & 2002,03      & 32           \\ 
18.5GHz      & 1.6e+04 &     2.1e+02 & 2002,03      & 32           \\ 
14.9GHz      & 2.0e+04 &     3.2e+01 & 2006,08      & 12           \\ 
8.35GHz      & 3.6e+04 &     2.8e+01 & 1989-1991    & 17           \\ 
5GHz         & 6.0e+04 &     3.0e+02 & 1989         & 29           \\ 
2.3GHz       & 1.3e+05 &     2.4e+01 & 1994         & 30           \\ 
1.4GHz       & 2.1e+05 &     7.6e+02 & 1993,03      & 2            \\ 
\hline 
3C029        & & & & \\ 
\hline 
Rosat/PSPC   & 1.2e-03 & $<$ 4.1e-05 & 1992         & 36           \\ 
F25SRF2      & 2.5e-01 &     1.1e-03 & 2000,06,08   & 9            \\ 
F702W        & 7.0e-01 &     9.5e-03 & 1995,01,12   & 7            \\ 
F160W        & 1.6e+00 & $<$ 7.1e-02 & 2004,12,04   & 26           \\ 
VISIR/SiC    & 1.3e+01 & $<$ 3.8e+00 & 2006,12,27   & $ $           \\ 
22GHz        & 1.4e+04 &     9.1e+01 & 2002,3       & 32           \\ 
18.5GHz      & 1.6e+04 &     1.1e+02 & 2002,3       & 32           \\ 
5GHz         & 6.0e+04 &     9.3e+01 & 1989         & 29           \\ 
2.3GHz       & 1.3e+05 &     5.5e+01 & 1994         & 30           \\ 
1.4GHz       & 2.1e+05 &     4.0e+01 & 1993-2003    & 2            \\ 
\hline 
3C040        & & & & \\ 
\hline 
VISIR/SiC    & 1.3e+01 & $<$ 3.8e+00 & 2006,12,27   & $ $           \\ 
22GHz        & 1.4e+04 &     1.5e+02 & 2002,3       & 32           \\ 
18.5GHz      & 1.6e+04 &     1.6e+02 & 2002,3       & 32           \\ 
5GHz         & 6.0e+04 &     1.0e+02 & 1989         & 29           \\ 
4.85GHz      & 6.2e+04 &     9.0e+01 & 1984         & 21           \\ 
2.3GHz       & 1.3e+05 &     7.9e+01 & 1994         & 30           \\ 
1.4GHz       & 2.1e+05 &     4.9e+01 & 93,03        & 2            \\ 
\hline 
3C078        & & & & \\ 
\hline 
Chandra      & 4.5e-04 &     1.3e-04 & $<$2005,01   & 1            \\ 
BeppoSax     & 1.2e-03 &     3.0e-04 & 1997,01,07   & 35           \\ 
F25QTZ       & 2.5e-01 &     5.7e-02 & 2000,03,15   & 9            \\ 
F555W        & 5.6e-01 &     3.2e-01 & 1996,09,16   & 7            \\ 
F702W        & 7.0e-01 &     3.9e-01 & 1994,08,17   & 7            \\ 
F28X50LP     & 7.2e-01 &     6.6e-01 & 2000,03,15   & 9            \\ 
VISIR/SiC    & 1.3e+01 & $<$ 4.1e+00 & 2006,12,27   & $ $           \\ 
345GHz       & 8.7e+02 &     2.8e+02 & 2001,02,11   & 31           \\ 
15GHz        & 2.0e+04 &     6.9e+02 & 1982,06,18   & 33           \\ 
5GHz2        & 6.0e+04 &     6.3e+02 & 1982,06,18   & 33           \\ 
5GHz         & 6.0e+04 &     9.6e+02 & 1989         & 29           \\ 
2.3GHz       & 1.3e+05 &     5.8e+02 & 1994         & 30           \\ 
1.5GHz       & 2.0e+05 &     7.5e+02 & 1982,06,18   & 33           \\ 
\hline 
Fornax~A     & & & & \\ 
\hline 
F175W        & 1.7e-01 &     6.3e-03 & 1992,02,01   & 14           \\ 
F480LP       & 5.1e-01 & $<$ 7.9e-03 & 1992,02,02   & 14           \\ 
F555W        & 5.6e-01 &     4.1e-02 & 1996-2006    & 25           \\ 
VISIR/SiC    & 1.3e+01 & $<$ 9.3e+00 & 2006,12,27   & $ $           \\ 
5GHz         & 6.0e+04 &     2.6e+01 & 1989         & 29           \\ 
2.3GHz       & 1.3e+05 & $<$ 4.0e+00 & 1994         & 30           \\ 
\hline 
3C093        & & & & \\ 
\hline 
VISIR/SiC    & 1.3e+01 & $<$ 5.3e+00 & 2006,12,27   & $ $           \\ 
8.4GHz       & 3.6e+04 &     3.9e+00 & 1991,12      & 4            \\ 
5GHz         & 6.0e+04 &     3.8e+00 & 1991,12      & 4$^{*}$      \\ 
\hline 
3C098        & & & & \\ 
\hline 
ROSAT        & 1.2e-03 &     5.8e-03 & $<$1999      & 18           \\ 
F702W        & 7.0e-01 & $<$ 2.6e-03 & $<$1999,04   & 8            \\ 
VISIR/SiC    & 1.3e+01 &     2.0e+01 & 2005,12,18   & $ $           \\ 
345GHz       & 8.7e+02 & $<$ 2.4e+01 & 2001,02,11   & 31           \\ 
8.35GHz      & 3.6e+04 &     6.1e+00 & 1989-1991    & 17           \\ 
5GHz         & 6.0e+04 &     9.0e+00 & 1985,01      & 16           \\ 
\hline 
3C105        & & & & \\ 
\hline 
F160W        & 1.6e+00 & $<$ 1.9e-02 & 2004,10,26   & 26           \\ 
VISIR/SiC    & 1.3e+01 & $<$ 2.2e+00 & 2005,12,18   & $ $           \\ 
22.4GHz      & 1.3e+04 &     2.1e+01 & 2006,08      & 12           \\ 
14.9GHz      & 2.0e+04 &     2.0e+01 & 2006,08      & 12           \\ 
8.35GHz      & 3.6e+04 &     1.9e+01 & 1989-1991    & 17           \\ 
5GHz         & 6.0e+04 &     1.4e+01 & 1989         & 29           \\ 
2.3GHz       & 1.3e+05 &     1.9e+01 & 1994         & 30           \\ 
\hline 
3C120        & & & & \\ 
\hline 
VISIR/SiC    & 1.3e+01 &     2.4e+02 & 2006,12,27   & $ $           \\ 
5GHz         & 6.0e+04 &     3.5e+03 & 1989         & 29           \\ 
2.3GHz       & 1.3e+05 &     2.7e+03 & 1994         & 30           \\ 
\hline 
3C135        & & & & \\ 
\hline 
F160W        & 1.6e+00 &     1.2e-02 & 2005,04,08   & 26           \\ 
VISIR/SiC    & 1.3e+01 & $<$ 4.1e+00 & 2006,12,27   & $ $           \\ 
8.35GHz      & 3.6e+04 &     1.0e+00 & 89,91        & 17           \\ 
5GHz         & 6.0e+04 &     4.6e+00 & $<$1993      & 37           \\ 
\hline 
Pictor~A     & & & & \\ 
\hline 
VISIR/SiC    & 1.3e+01 &     6.8e+01 & 2006,12,27   & $ $           \\ 
22GHz        & 1.4e+04 &     1.1e+03 & 2002,03      & 32           \\ 
18.5GHz      & 1.6e+04 &     7.8e+02 & 2002,03      & 32           \\ 
5GHz         & 6.0e+04 &     1.0e+03 & 1989         & 29           \\ 
2.3GHz       & 1.3e+05 &     7.5e+02 & 1994         & 30           \\ 
\hline 
3C198        & & & & \\ 
\hline 
F160W        & 1.6e+00 & $<$ 1.9e-02 & 2005,05,03   & 26           \\ 
F25SRF2      & 2.5e-01 &     2.3e-02 & 2000,03,15   & 9            \\ 
F702W        & 7.0e-01 &     5.4e-01 & $<$1999,4    & 8            \\ 
VISIR/SiC    & 1.3e+01 & $<$ 4.3e+00 & 2006,12,27   & $ $           \\ 
345GHz       & 8.7e+02 & $<$ 2.0e+01 & 2001,02,11   & 31           \\ 
5GHz         & 6.0e+04 &     2.0e+00 & 1993-2003    & 3$^{*}$      \\ 
1.4GHz       & 2.1e+05 &     1.6e+00 & 1993-2003    & 2            \\ 
\hline 
3C227        & & & & \\ 
\hline 
F25SRF2      & 2.5e-01 &     1.3e-01 & 2000,01,27   & 9            \\ 
F702W        & 7.0e-01 &     4.8e-01 & 1995,05,19   & 9            \\ 
F160W        & 1.6e+00 & $<$ 7.2e-02 & 2005,03,28   & 26           \\ 
VISIR/SiC    & 1.3e+01 &     2.9e+01 & 2006,12,27   & $ $           \\ 
345GHz       & 8.7e+02 & $<$ 2.1e+01 & 2001,02,11   & 31           \\ 
22.4GHz      & 1.3e+04 &     1.2e+01 & 2006,08      & 12           \\ 
14.9GHz      & 2.0e+04 &     1.4e+01 & 2006,08      & 12           \\ 
8.35GHz      & 3.6e+04 &     1.3e+01 & 1989-1991    & 17           \\ 
5GHz         & 6.0e+04 &     3.2e+01 & 1989         & 29           \\ 
2.3GHz       & 1.3e+05 &     1.2e+01 & 1994         & 30           \\ 
\hline 
3C270        & & & & \\ 
\hline 
Chandra      & 4.5e-04 &     6.6e-05 & $<$2005,01   & 1            \\ 
Chandra      & 2.0e-04 &     3.6e-05 & 2000,03,05   & 9            \\ 
F25SRF2      & 2.5e-01 & $<$ 2.1e-04 & 2000,03,05   & 9            \\ 
F547M        & 5.5e-01 &     6.0e-03 & 1994,12,13   & 7            \\ 
F791W        & 7.9e-01 &     1.1e-02 & 1994,12,13   & 7            \\ 
F110W        & 1.1e+00 &     2.8e-04 & 1998,04,23   & 6            \\ 
F160W        & 1.6e+00 &     1.0e-03 & 1998,04,23   & 6            \\ 
F205W        & 2.0e+00 &     2.2e-03 & 1998,04,23   & 6            \\ 
L            & 3.4e+00 &     1.2e+00 & 2000,03,21   & 31           \\ 
LW1          & 4.5e+00 & $<$ 7.0e+00 & 1996,07,02   & 31           \\ 
LW2          & 6.7e+00 & $<$ 1.0e+01 & 1996,07,02   & 31           \\ 
VISIR/SiC    & 1.3e+01 & $<$ 2.3e+00 & 2006,06,18   & $ $           \\ 
LW3          & 1.4e+01 &     1.9e+01 & 1996,07,02   & 31           \\ 
345GHz       & 8.7e+02 &     1.7e+02 & 2001,02,11   & 31           \\ 
43GHz        & 7.0e+03 &     3.0e+02 & 1997,09,07   & 22           \\ 
22GHz        & 1.4e+04 &     1.6e+02 & 1997,09,07   & 22           \\ 
8.4GHz       & 3.6e+04 &     1.0e+02 & 1995,04,01   & 23           \\ 
5GHz         & 6.0e+04 &     3.1e+02 & 1989         & 29           \\ 
2.3GHz       & 1.3e+05 &     2.4e+02 & 1994         & 30           \\ 
1.6GHz       & 1.9e+05 &     1.0e+02 & 1995,04,01   & 23           \\ 
1.4GHz       & 2.1e+05 &     1.4e+02 & 1993,03      & 2            \\ 
\hline 
Centaurus~A  & & & & \\ 
\hline 
WFPC2        & 8.1e-01 &     7.0e-03 & 1997.80      & 27           \\ 
NACO/1.28    & 1.3e+00 &     1.3e+00 & 2003.45      & 28           \\ 
F160W        & 1.6e+00 &     4.8e+00 & 1997.69      & 27           \\ 
NACO/1.67    & 1.7e+00 &     4.5e+00 & 2003.45      & 28           \\ 
NACO/2.15    & 2.1e+00 &     3.4e+01 & 2004.25      & 28           \\ 
NICMOS/2.22  & 2.2e+00 &     4.2e+01 & 1997.61      & 27           \\ 
NACO/3.8     & 3.8e+00 &     2.0e+02 & 2003.36      & 28           \\ 
MIDI/8.3     & 8.3e+00 &     4.7e+02 & 2005.28      & 28           \\ 
MIDI/9.3     & 9.3e+00 &     2.8e+02 & 2005.28      & 28           \\ 
MIDI/10.4    & 1.0e+01 &     2.5e+02 & 2005.28      & 28           \\ 
MIDI/11.4    & 1.1e+01 &     4.3e+02 & 2005.28      & 28           \\ 
MIDI/12.6    & 1.3e+01 &     6.2e+02 & 2005.28      & 28           \\ 
VISIR/SiC    & 1.3e+01 &     1.2e+03 & 2006,12,27   & $ $           \\ 
667GHz       & 4.5e+02 &     6.3e+03 & 1991.35      & 20           \\ 
375GHz       & 8.0e+02 &     8.5e+03 & 1991.35      & 20           \\ 
270GHz       & 1.1e+03 &     5.9e+03 & 2003.30      & 28           \\ 
235GHz       & 1.3e+03 &     5.8e+03 & 2003.18      & 28           \\ 
150GHz       & 2.0e+03 &     6.9e+03 & 2003.18      & 28           \\ 
90GHz        & 3.5e+03 &     8.6e+03 & 2003.18      & 28           \\ 
22.2GHz      & 1.4e+04 &     3.5e+03 & 1995.88      & 34           \\ 
22GHz        & 1.4e+04 &     6.6e+03 & 2002,03      & 32           \\ 
18.5GHz      & 1.6e+04 &     6.5e+03 & 2002,03      & 32           \\ 
8.4GHz       & 3.6e+04 &     2.4e+03 & 1996.22      & 34           \\ 
5GHz         & 6.0e+04 &     7.0e+03 & 1989         & 29           \\ 
4.8GHz       & 6.3e+04 &     1.2e+03 & 1993.13      & 34           \\ 
2.3GHz       & 1.3e+05 &     5.5e+03 & 1994         & 30           \\ 
1.4GHz       & 2.1e+05 &     5.7e+03 & 1993-2003    & 19           \\ 
\hline 
PKS1417$-$19   & & & & \\ 
\hline 
VISIR/SiC    & 1.3e+01 &     2.6e+01 & 2006,12,27   & $ $           \\ 
5GHz         & 6.0e+04 &     5.2e+00 & $<$1993      & 37           \\ 
\hline 
3C317        & & & & \\ 
\hline 
Chandra      & 4.5e-04 &     2.3e-05 & $<$2005,01   & 1            \\ 
ROSAT/PSPC   & 1.2e-03 &     4.6e-05 & 2000,09,03   & 36           \\ 
F210M        & 2.2e-01 &     7.7e-04 & 1994,03,05   & 9            \\ 
F555W        & 5.6e-01 &     1.5e-02 & 1997,10,18   & 7            \\ 
F702W        & 7.0e-01 &     2.0e-02 & 1994,03,05   & 9            \\ 
F814W        & 8.1e-01 &     2.1e-02 & 1997,10,18   & 7            \\ 
F160W        & 1.6e+00 &     1.2e-01 & 1998         & 36           \\ 
VISIR/SiC    & 1.3e+01 & $<$ 3.2e+00 & 2006,06,18   & $ $           \\ 
345GHz       & 8.7e+02 &     9.1e+01 & 2001,02,11   & 31           \\ 
5GHz         & 6.0e+04 &     3.9e+02 & 1989,09      & 29           \\ 
2.3GHz       & 1.3e+05 &     3.1e+02 & 1994         & 30           \\ 
1.4GHz       & 2.1e+05 &     5.0e+02 & $\dots$   & 2            \\ 
\hline 
3C327        & & & & \\ 
\hline 
VISIR/SiC    & 1.3e+01 &     7.8e+01 & 2005,07,02   & $ $           \\ 
22.4GHz      & 1.3e+04 &     1.1e+01 & 2006,08      & 12           \\ 
14.9GHz      & 2.0e+04 &     1.5e+01 & 2006,08      & 12           \\ 
8.35GHz      & 3.6e+04 &     2.5e+01 & 89,91        & 17           \\ 
5GHz         & 6.0e+04 &     2.8e+01 & 1989         & 29           \\ 
2.3GHz       & 1.3e+05 &     3.2e+01 & 1994         & 30           \\ 
1.4GHz       & 2.1e+05 &     2.1e+01 & 1993,03      & 2            \\ 
\hline 
3C353        & & & & \\ 
\hline 
F702W        & 7.0e-01 & $<$ 1.9e-03 & $<$1999,4    & 8            \\ 
F160W        & 1.6e+00 & $<$ 2.0e-01 & 2004,09,09   & 26           \\ 
VISIR/SiC    & 1.3e+01 & $<$ 4.5e+00 & 2005,07,02   & $ $           \\ 
345GHz       & 8.7e+02 &     7.4e+01 & 2001,02,11   & 31           \\ 
22GHz        & 1.4e+04 &     1.9e+02 & 2002,03      & 32           \\ 
18.5GHz      & 1.6e+04 &     1.5e+02 & 2002,03      & 32           \\ 
8.4GHz       & 3.6e+04 &     1.5e+02 & 1989-1991    & 17           \\ 
5GHz         & 6.0e+04 &     1.2e+02 & 1989         & 29           \\ 
2.3GHz       & 1.3e+05 &     7.8e+01 & 1994         & 30           \\ 
\hline 
PKS1839$-$48   & & & & \\ 
\hline 
VISIR/SiC    & 1.3e+01 & $<$ 4.1e+00 & 2006,06,18   & $ $           \\ 
24GHz        & 1.2e+04 &     1.0e+02 & 2006,07      & 12           \\ 
22GHz        & 1.4e+04 &     9.3e+01 & 2002,03      & 32           \\ 
18.5GHz      & 1.6e+04 &     2.0e+02 & 2002,03      & 32           \\ 
18GHz        & 1.7e+04 &     1.1e+02 & 2006,07      & 12           \\ 
5GHz         & 6.0e+04 &     1.6e+02 & 1991-1992    & 29           \\ 
2.3GHz       & 1.3e+05 &     1.3e+02 & 1994         & 30           \\ 
\hline 
3C403        & & & & \\ 
\hline 
F702W        & 7.0e-01 &     6.5e-03 & $<$1999,04   & 8            \\ 
F160W        & 1.6e+00 & $<$ 7.9e-02 & 2004,11,06   & 26           \\ 
VISIR/SiC    & 1.3e+01 &     8.9e+01 & 2006,06,18   & $ $           \\ 
22.4GHz      & 1.3e+04 &     8.5e+00 & 2006,08      & 12           \\ 
14.9GHz      & 2.0e+04 &     2.0e+01 & 2006,08      & 12           \\ 
8.35GHz      & 3.6e+04 &     7.1e+00 & 1989-1991    & 17           \\ 
5GHz         & 6.0e+04 &     1.0e+01 & 1989         & 29           \\ 
2.3GHz       & 1.3e+05 &     2.4e+01 & 1994         & 30           \\ 
\hline 
3C403.1      & & & & \\ 
\hline 
VISIR/SiC    & 1.3e+01 & $<$ 4.2e+00 & 2006,06,18   & $ $           \\ 
\hline 
3C424        & & & & \\ 
\hline 
F28X50LP     & 7.0e-01 & $<$ 1.5e-02 & 2000,03,15   & 9            \\ 
VISIR/SiC    & 1.3e+01 & $<$ 3.7e+00 & 2006,06,18   & $ $           \\ 
8.4GHz       & 3.6e+04 &     7.0e+00 & 1989-1991    & 17           \\ 
5GHz         & 6.0e+04 &     1.8e+01 & $ $   & 3            \\ 
\hline 
PKS2158$-$380  & & & & \\ 
\hline 
F130M        & 1.3e+00 &     2.2e-02 & 1991,05,07   & 5            \\ 
VISIR/SiC    & 1.3e+01 &     1.7e+01 & 2006,06,18   & $ $           \\ 
5GHz         & 6.0e+04 & $<$ 7.0e+00 & 1982         & 15           \\ 
\hline 
3C445        & & & & \\ 
\hline 
F702W        & 7.0e-01 &     2.8e+00 & $<$1999,04   & 8            \\ 
F160W        & 1.6e+00 & $<$ 2.1e-01 & 2005,06,24   & 26           \\ 
VISIR/SiC    & 1.3e+01 &     1.9e+02 & 2005,07,02   & $ $           \\ 
22.4GHz      & 1.3e+04 &     1.6e+01 & 2006,08      & 12           \\ 
22GHz        & 1.4e+04 &     4.5e+01 & 2002,03      & 32           \\ 
18.5GHz      & 1.6e+04 &     3.9e+01 & 2002,03      & 32           \\ 
14.9GHz      & 2.0e+04 &     2.7e+01 & 2006,08      & 12           \\ 
8.4GHz       & 3.6e+04 &     8.4e+01 & 1989-1991    & 17           \\ 
5GHz         & 6.0e+04 &     8.6e+01 & 1989         & 29           \\ 
2.3GHz       & 1.3e+05 &     9.0e+01 & 1994         & 30           \\ 
1.4GHz       & 2.1e+05 &     6.9e+01 & 1993-2003    & 2            \\ 
\hline 
PKS2354$-$35   & & & & \\ 
\hline 
VISIR/SiC    & 1.3e+01 & $<$ 3.1e+00 & 2006,06,18   & $ $           \\ 
5GHz         & 6.0e+04 &     5.0e+00 & $<$1981      & 13           \\ 
\end{supertabular}
\begin{list}{}{}
\item[]\textit{Column (5)} References from: 1=\cite{2006A&A...451...35B}, 2=\cite{2003yCat.8071....0B}, 3=\cite{1992MNRAS.256..186B}, 4=\cite{1994A&AS..105...91B}, 5=\cite{1996A&A...305..715B}, 6=\cite{2000MNRAS.318..493C}, 7=\cite{1999A&A...349...77C}, 8=\cite{2000A&A...355..873C}, 9=\cite{2002A&A...394..791C}, 10=\cite{2002ApJ...571..247C}, 11=\cite{2003ApJ...582..645C}, 12=\cite{2008ApJ...678..712D}, 13=\cite{1989MNRAS.236..737E}, 14=\cite{1994ApJ...434...67F}, 15=\cite{1982MNRAS.201..991F}, 16=\cite{1988A&A...199...73G}, 17=\cite{1998MNRAS.296..445H}, 18=\cite{2003MNRAS.338..176H}, 19=\cite{2007ApJ...670L..81H}, 20=\cite{1993MNRAS.260..844H}, 21=\cite{2006MNRAS.368..609J}, 22=\cite{2000ApJ...534..165J}, 23=\cite{1997ApJ...484..186J}, 24=\cite{2005AJ....129.2138L}, 25=\cite{1991AJ....102..537L}, 26=\cite{2006ApJS..164..307M}, 27=\cite{2000ApJ...528..276M}, 28=\cite{2007A&A...471..453M}, 29=\cite{1993MNRAS.263.1023M}, 30=\cite{1997MNRAS.284..541M}, 31=\cite{2003AJ....126.2677Q}, 32=\cite{2006A&A...445..465R}, 33=\cite{1986MNRAS.219..545S}, 34=\cite{1998AJ....115..960T}, 35=\cite{1999A&A...348..437T}, 36=\cite{2003A&A...403..889T} and 37=\cite{1995ApJ...448..521Z}.
\end{list}
\end{center}
}

\end{document}